%%%%%%%%%%%%%%%%%%%%%%%%%%%%%%%%%%%%%%%%%%%%%%%%%%%%%%%%%%%%%%%%%%%%%%%%%%%%%%
%%%%%%%%%%%%%%%%%%%%%%%%%%%%%%sx.tex%%%%%%%%%%%%%%%%%%%%%%%%%%%%%%%%%%%%%%
%%%%%%%%%%%%%%%%%%%%%%%%%%%%%%%%%%%%%%%%%%%%%%%%%%%%%%%%%%%%%%%%%%%%%%%%%%%%%%

%%%%%%%%%%%%%%%%%%  tex macros for preprints, cm version %%%%%%%%%%%%%%
%                     (P. Ginsparg, last updated 9/91)
%                if confused, type `b' in response to query 
%
%---------------------------------------------------------------------%
%% site dependent options: 
%% \unredoffs and \redoffs define horizontal and vertical offsets 
%% respectively for unreduced and reduced modes. \speclscape defines
%% the \special{} call that sets printer to landscape (sideways) mode.
%% from standard set below, leave uncommented as appropriate or redefine
%
%%% next 400dpi
%\def\unredoffs{} \def\redoffs{\voffset=-.31truein\hoffset=-.48truein}
%\def\speclscape{\special{landscape}}
%
%%% apple lw
%\def\unredoffs{} \def\redoffs{\voffset=-.31truein\hoffset=-.59truein}
%\def\speclscape{\special{ps: landscape}}
%
%%% qms lasergrafix:
%\def\unredoffs{} \def\redoffs{\voffset=-.4truein\hoffset=.125truein}
%\def\speclscape{\special{qms: landscape}}
%
%%% saclay A4 paper:
\def\unredoffs{\hoffset-.14truein\voffset-.2truein} 
 
%\def\speclscape{\special{landscape}}
%
%---------------------------------------------------------------------%
%
\newbox\leftpage \newdimen\fullhsize \newdimen\hstitle \newdimen\hsbody
\tolerance=1000\hfuzz=2pt
\catcode`\@=11 % This allows us to modify PLAIN macros.
%\def\bigans{b }
%\message{ big or little (b/l)? }\read-1 to\answ
%
%\ifx\answ\bigans\message{(This will come out unreduced.}
\magnification=1095\unredoffs\baselineskip=16pt plus 2pt minus 1pt
\hsbody=\hsize \hstitle=\hsize %take default values for unreduced format
%
%\else\message{(This will be reduced.} \let\l@r=L
%\magnification=800\baselineskip=16pt plus 1pt minus 0.5pt \vsize=7truein
%\redoffs \hstitle=8truein\hsbody=4.75truein\fullhsize=10truein\hsize=\hsbody
%
%\output={\ifnum\pageno=0 %%% This is the HUTP version
%  \shipout\vbox{\speclscape{\hsize\fullhsize\makeheadline}
%    \hbox to \fullhsize{\hfill\pagebody\hfill}}\advancepageno}
%  \else
% \almostshipout{\leftline{\vbox{\pagebody\makefootline}}}\advancepageno 
%  \fi}
%\def\almostshipout#1{%\if L\l@r \count1=1 \message{[\the\count0.\the\count1]}
%      \global\setbox\leftpage=#1 \global\let\l@r=R
% \else 
%\count1=2
%  \shipout\vbox{\speclscape{\hsize\fullhsize\makeheadline}
%      \hbox to\fullhsize{\box\leftpage\hfil#1}}  \global\let\l@r=L\fi}
%\fi
%---------------------------------------------------------------------
%
\newcount\yearltd\yearltd=\year\advance\yearltd by -1900

%
% 	restores pagenumbers
%
%       use following instead of \Date on the preliminary draft, 
%       puts date/time on each page in big mode, writes labels in margins

\def\draftmode{\message{ DRAFTMODE }\def\draftdate{{\rm preliminary draft:
\number\month/\number\day/\number\yearltd\ \ \hourmin}}%
\headline={\hfil\draftdate}\writelabels\baselineskip=16pt plus 2pt minus 2pt
 {\count255=\time\divide\count255 by 60 \xdef\hourmin{\number\count255}
  \multiply\count255 by-60\advance\count255 by\time
  \xdef\hourmin{\hourmin:\ifnum\count255<10 0\fi\the\count255}}}
%       use \nolabels to get rid of eqn, ref, and fig labels in draft mode
\def\nolabels{\def\wrlabeL##1{}\def\eqlabeL##1{}\def\reflabeL##1{}}
\def\writelabels{\def\wrlabeL##1{\leavevmode\vadjust{\rlap{\smash%
{\line{{\escapechar=` \hfill\rlap{\sevenrm\hskip.03in\string##1}}}}}}}%
\def\eqlabeL##1{{\escapechar-1\rlap{\sevenrm\hskip.05in\string##1}}}%
\def\reflabeL##1{\noexpand\llap{\noexpand\sevenrm\string\string\string##1}}}
\nolabels
%
% tagged sec numbers
\global\newcount\secno \global\secno=0
\global\newcount\meqno \global\meqno=1
\def\newsec#1{\global\advance\secno by1\message{(\the\secno. #1)}
%\ifx\answ\bigans \vfill\eject \else \bigbreak\bigskip \fi  %if desired
\global\subsecno=0\eqnres@t\noindent{\bf\the\secno. #1}
\writetoca{{\secsym} {#1}}\par\nobreak\medskip\nobreak}
\def\eqnres@t{\xdef\secsym{\the\secno.}\global\meqno=1\bigbreak\bigskip}
\def\sequentialequations{\def\eqnres@t{\bigbreak}}\xdef\secsym{}
\global\newcount\subsecno \global\subsecno=0
\def\subsec#1{\global\advance\subsecno by1\message{(\secsym\the\subsecno. #1)}
\ifnum\lastpenalty>9000\else\bigbreak\fi
\noindent{\bf\secsym\the\subsecno. #1}\writetoca{\string\quad 
{\secsym\the\subsecno.} {#1}}\par\nobreak\medskip\nobreak}
\def\appendix#1#2{\global\meqno=1\global\subsecno=0\xdef\secsym{\hbox{#1.}}
\bigbreak\bigskip\noindent{\bf Appendix #1. #2}\message{(#1. #2)}
\writetoca{Appendix {#1.} {#2}}\par\nobreak\medskip\nobreak}
%
%       \eqn\label{a+b=c}	gives displayed equation, numbered
%				consecutively within sections.
%     \eqnn and \eqna define labels in advance (of eqalign?)
%
\def\eqnn#1{\xdef #1{(\secsym\the\meqno)}\writedef{#1\leftbracket#1}%
\global\advance\meqno by1\wrlabeL#1}
\def\eqna#1{\xdef #1##1{\hbox{$(\secsym\the\meqno##1)$}}
\writedef{#1\numbersign1\leftbracket#1{\numbersign1}}%
\global\advance\meqno by1\wrlabeL{#1$\{\}$}}
\def\eqn#1#2{\xdef #1{(\secsym\the\meqno)}\writedef{#1\leftbracket#1}%
\global\advance\meqno by1$$#2\eqno#1\eqlabeL#1$$}
%
%			 footnotes
\newskip\footskip\footskip14pt plus 1pt minus 1pt %sets footnote baselineskip
\def\footnotefont{\ninepoint}\def\f@t#1{\footnotefont #1\@foot}
\def\f@@t{\baselineskip\footskip\bgroup\footnotefont\aftergroup\@foot\let\next}
\setbox\strutbox=\hbox{\vrule height9.5pt depth4.5pt width0pt}
\global\newcount\ftno \global\ftno=0
\def\foot{\global\advance\ftno by1\footnote{$^{\the\ftno}$}}
%
%say \footend to put footnotes at end
%will cause problems if \ref used inside \foot, instead use \nref before
\newwrite\ftfile   
\def\footend{\def\foot{\global\advance\ftno by1\chardef\wfile=\ftfile
$^{\the\ftno}$\ifnum\ftno=1\immediate\openout\ftfile=foots.tmp\fi%
\immediate\write\ftfile{\noexpand\smallskip%
\noexpand\item{f\the\ftno:\ }\pctsign}\findarg}%
\def\footatend{\vfill\eject\immediate\closeout\ftfile{\parindent=20pt
\centerline{\bf Footnotes}\nobreak\bigskip\input foots.tmp }}}
\def\footatend{}
%
%     \ref\label{text}
% generates a number, assigns it to \label, generates an entry.
% To list the refs on a separate page,  \listrefs
%
\global\newcount\refno \global\refno=1
\newwrite\rfile
\def\ref{[\the\refno]\nref}
\def\nref#1{\xdef#1{[\the\refno]}\writedef{#1\leftbracket#1}%
\ifnum\refno=1\immediate\openout\rfile=refs.tmp\fi
\global\advance\refno by1\chardef\wfile=\rfile\immediate
\write\rfile{\noexpand\item{#1\ }\reflabeL{#1\hskip.31in}\pctsign}\findarg}
%	horrible hack to sidestep tex \write limitation
\def\findarg#1#{\begingroup\obeylines\newlinechar=`\^^M\pass@rg}
{\obeylines\gdef\pass@rg#1{\writ@line\relax #1^^M\hbox{}^^M}%
\gdef\writ@line#1^^M{\expandafter\toks0\expandafter{\striprel@x #1}%
\edef\next{\the\toks0}\ifx\next\em@rk\let\next=\endgroup\else\ifx\next\empty%
\else\immediate\write\wfile{\the\toks0}\fi\let\next=\writ@line\fi\next\relax}}
\def\striprel@x#1{} \def\em@rk{\hbox{}} 
\def\lref{\begingroup\obeylines\lr@f}
\def\lr@f#1#2{\gdef#1{\ref#1{#2}}\endgroup\unskip}
\def\semi{;\hfil\break}
\def\addref#1{\immediate\write\rfile{\noexpand\item{}#1}} %now unnecessary
\def\startrefs#1{\immediate\openout\rfile=refs.tmp\refno=#1}
\def\xref{\expandafter\xr@f}\def\xr@f[#1]{#1}
\def\refs#1{\count255=1[\r@fs #1{\hbox{}}]}
\def\r@fs#1{\ifx\und@fined#1\message{reflabel \string#1 is undefined.}%
\nref#1{need to supply reference \string#1.}\fi%
\vphantom{\hphantom{#1}}\edef\next{#1}\ifx\next\em@rk\def\next{}%
\else\ifx\next#1\ifodd\count255\relax\xref#1\count255=0\fi%
\else#1\count255=1\fi\let\next=\r@fs\fi\next}
%

%
% this is ugly, but moore insists
\newwrite\ffile\global\newcount\figno \global\figno=1
\def\fig{fig.~\the\figno\nfig}
\def\nfig#1{\xdef#1{fig.~\the\figno}%
\writedef{#1\leftbracket fig.\noexpand~\the\figno}%
\ifnum\figno=1\immediate\openout\ffile=figs.tmp\fi\chardef\wfile=\ffile%
\immediate\write\ffile{\noexpand\medskip\noexpand\item{Fig.\ \the\figno. }
\reflabeL{#1\hskip.55in}\pctsign}\global\advance\figno by1\findarg}
\def\xfig{\expandafter\xf@g}\def\xf@g fig.\penalty\@M\ {}
\def\figs#1{figs.~\f@gs #1{\hbox{}}}
\def\f@gs#1{\edef\next{#1}\ifx\next\em@rk\def\next{}\else
\ifx\next#1\xfig #1\else#1\fi\let\next=\f@gs\fi\next}
\newwrite\lfile
{\escapechar-1\xdef\pctsign{\string\%}\xdef\leftbracket{\string\{}
\xdef\rightbracket{\string\}}\xdef\numbersign{\string\#}}

\def\writestop{\def\writestoppt{\immediate\write\lfile{\string\pageno%
\the\pageno\string\startrefs\leftbracket\the\refno\rightbracket%
\string\def\string\secsym\leftbracket\secsym\rightbracket%
\string\secno\the\secno\string\meqno\the\meqno}\immediate\closeout\lfile}}
\def\writestoppt{}\def\writedef#1{}
\def\seclab#1{\xdef #1{\the\secno}\writedef{#1\leftbracket#1}\wrlabeL{#1=#1}}
\def\subseclab#1{\xdef #1{\secsym\the\subsecno}%
\writedef{#1\leftbracket#1}\wrlabeL{#1=#1}}
\newwrite\tfile \def\writetoca#1{}
\def\leaderfill{\leaders\hbox to 1em{\hss.\hss}\hfill}
%	use this to write file with table of contents
\def\writetoc{\immediate\openout\tfile=toc.tmp 
   \def\writetoca##1{{\edef\next{\write\tfile{\noindent ##1 
   \string\leaderfill {\noexpand\number\pageno} \par}}\next}}}
%       and this lists table of contents on second pass

%
\catcode`\@=12 % at signs are no longer letters
%
%	Unpleasantness in calling in abstract and title fonts
\edef\tfontsize{\ifx\answ\bigans scaled\magstep3\else scaled\magstep4\fi}
 \tfontsize  \tfontsize
 \tfontsize \font\titlei=cmmi10 \tfontsize
\font\titleis=cmmi7 \tfontsize \font\titleiss=cmmi5 \tfontsize
\font\titlesy=cmsy10 \tfontsize \font\titlesys=cmsy7 \tfontsize
\font\titlesyss=cmsy5 \tfontsize  \tfontsize
\skewchar\titlei='177 \skewchar\titleis='177 \skewchar\titleiss='177
\skewchar\titlesy='60 \skewchar\titlesys='60 \skewchar\titlesyss='60
 \ifx\answ\bigans\else scaled\magstep1\fi
\ifx\answ\bigans\else

 \font\absi=cmmi10 scaled\magstep1
\font\absis=cmmi7 scaled\magstep1 \font\absiss=cmmi5 scaled\magstep1
\font\abssy=cmsy10 scaled\magstep1 \font\abssys=cmsy7 scaled\magstep1
\font\abssyss=cmsy5 scaled\magstep1 
\skewchar\absi='177 \skewchar\absis='177 \skewchar\absiss='177
\skewchar\abssy='60 \skewchar\abssys='60 \skewchar\abssyss='60
\fi
\font\ninerm=cmr9 \font\sixrm=cmr6 \font\ninei=cmmi9 \font\sixi=cmmi6 
\font\ninesy=cmsy9 \font\sixsy=cmsy6 \font\ninebf=cmbx9 
\font\nineit=cmti9 \font\ninesl=cmsl9 \skewchar\ninei='177
\skewchar\sixi='177 \skewchar\ninesy='60 \skewchar\sixsy='60 
\def\ninepoint{\def\rm{\fam0\ninerm}% switch to footnote font
\textfont0=\ninerm \scriptfont0=\sixrm \scriptscriptfont0=\fiverm
\textfont1=\ninei \scriptfont1=\sixi \scriptscriptfont1=\fivei
\textfont2=\ninesy \scriptfont2=\sixsy \scriptscriptfont2=\fivesy
\textfont\itfam=\ninei \def\it{\fam\itfam\nineit}\def\sl{\fam\slfam\ninesl}%
\textfont\bffam=\ninebf \def\bf{\fam\bffam\ninebf}\rm} 
%
%---------------------------------------------------------------------
%

\hyphenation{anom-aly anom-alies coun-ter-term coun-ter-terms}
\def\inv{^{\raise.15ex\hbox{${\scriptscriptstyle -}$}\kern-.05em 1}}

\def\Dsl{\,\raise.15ex\hbox{/}\mkern-13.5mu D} %this one can be subscripted
\def\dsl{\raise.15ex\hbox{/}\kern-.57em\partial}

 %pound sterling
\def\lspace{\ifx\answ\bigans{}\else\qquad\fi}
\def\lbspace{\ifx\answ\bigans{}\else\hskip-.2in\fi} % $$\lbspace...$$
\def\boxeqn#1{\vcenter{\vbox{\hrule\hbox{\vrule\kern3pt\vbox{\kern3pt
	\hbox{${\displaystyle #1}$}\kern3pt}\kern3pt\vrule}\hrule}}}
\def\mbox#1#2{\vcenter{\hrule \hbox{\vrule height#2in
		\kern#1in \vrule} \hrule}}  %e.g. \mbox{.1}{.1}
%	matters of taste
%\def\tilde{\widetilde} \def\bar{\overline} \def\hat{\widehat}
%
% some sample definitions
  %     curly letters

\def\darr#1{\raise1.5ex\hbox{$\leftrightarrow$}\mkern-16.5mu #1}
 %pound sterling

\def\half{{\textstyle{1\over2}}} %puts a small half in a displayed eqn
\def\fourninths{{\textstyle{4\over9}}} %ditto 4/9
\def\ninefourths{{\textstyle{9\over4}}} %ditto 9/4
\def\threeovertwo{{\textstyle{3\over2}}} %ditto 3/2
\def\roughly#1{\raise.3ex\hbox{$#1$\kern-.75em\lower1ex\hbox{$\sim$}}}

\def\p2inf{\mathrel{\mathop{\sim}\limits_{\scriptscriptstyle
{p^2 \rightarrow \infty }}}}
\def\kap2inf{\mathrel{\mathop{\sim}\limits_{\scriptscriptstyle
{\kappa \rightarrow \infty }}}}
\def\x2inf{\mathrel{\mathop{\sim}\limits_{\scriptscriptstyle
{x \rightarrow \infty }}}}
\def\Lam2inf{\mathrel{\mathop{\sim}\limits_{\scriptscriptstyle
{\Lambda \rightarrow \infty }}}}
\def\frac#1#2{{{#1}\over {#2}}}
\def\half{\hbox{${1\over 2}$}}

\def\Mev{{\rm MeV}}\def\Gev{{\rm GeV}}

\def\eigplus{\exp\biggl[\int_{\alpha_s(Q^2)}^{\alpha_s(Q_I^2)}
\!\!{\tilde\gamma^0_{gg}
(\alpha_s(q^2)/N) \over\alpha^2_s(q^2)}d\,\alpha_s(q^2)\biggr]}

\def\eigvecplus{g_0(N)+\fourninths f^S_{0}(N)}

\def\eigpluscor{\exp\biggl[\int_{\alpha_s(Q^2)}^{\alpha_s(Q_I^2)}\!\!
{\tilde\gamma^0_{gg}
(\alpha_s(q^2)/N) \over\alpha^2_s(q^2)} + {\tilde
\gamma^1_{gg}(\alpha_s(q^2)/N)+ \fourninths \tilde
\gamma^1_{fg}(\alpha_s(q^2)/N)\over\alpha_s(q^2)}
d\,\alpha_s(q^2)\biggr]}

\def\eigpluscorp{\exp\biggl[\int_{\alpha_s(Q^2)}^{\alpha_s(Q_I^2)}\!\!
{\tilde\Gamma^0_{LL}
(\alpha_s(q^2)/N) \over\alpha^2_s(q^2)} + {\tilde
\Gamma^1_{LL}(\alpha_s(q^2)/N)-\biggl({36-8N_f\over 27}\biggr)
 \tilde\Gamma^1_{2L}(\alpha_s(q^2)/N)\over \alpha_s(q^2)}
d\,\alpha_s(q^2)\biggr]}

\def\lneigcomiC{\int_{\alpha_s(Q^2)}^{\alpha_s(Q_I^2)}\!\!{\tilde
\gamma^1_{gg}(\alpha_s(q^2)/N)+ \fourninths \tilde
\gamma^1_{fg}(\alpha_s(q^2)/N)\over \alpha_s(q^2)}
-{d\over d\alpha_s(q^2)}\biggl(\ln\biggl({C^g_{L,1}(\alpha_s(q^2)/N)\over
C^g_{L,1,0}}\biggl)\biggl)d\,\alpha_s(q^2)}

\def\lsim{\mathrel{mathpalette\@versim<}}
\def\gsim{\mathrel{mathpalette\@versim>}}
\def\higherorder{\hbox{\rm higher order in $\alpha_s$ and/or $N$}}

\catcode`@=11 %This allows us to modify plain macros
\def\slash#1{\mathord{\mathpalette\c@ncel#1}}
 \def\c@ncel#1#2{\ooalign{$\hfil#1\mkern1mu/\hfil$\crcr$#1#2$}}
\def\lsim{\mathrel{\mathpalette\@versim<}}
\def\gsim{\mathrel{\mathpalette\@versim>}}
 \def\@versim#1#2{\lower0.2ex\vbox{\baselineskip\z@skip\lineskip\z@skip
       \lineskiplimit\z@\ialign{$\m@th#1\hfil##$\crcr#2\crcr\sim\crcr}}}
\catcode`@=12 %at signs are no longer letters

\def\PR{{\it Phys.~Rev.~}}
\def\PRL{{\it Phys.~Rev.~Lett.~}}
\def\NP{{\it Nucl.~Phys.~}}
\def\PL{{\it Phys.~Lett.~}}

\def\SJNP{{\it Sov.~Jour.~Nucl.~Phys.~}}
\def\ZP{{\it Zeit.~Phys.~}}

\def\vyp#1#2#3{#1 (#2) #3}

\def\Asl{\raise.15ex\hbox{/}\mkern-11.5mu A}
\def\psl{\lower.12ex\hbox{/}\mkern-9.5mu p}
\def\qsl{\lower.12ex\hbox{/}\mkern-9.5mu q}
\def\rsl{\lower.03ex\hbox{/}\mkern-9.5mu r}
\def\ksl{\raise.06ex\hbox{/}\mkern-9.5mu k}
\def\Mev{\hbox{MeV}}

%%%%%%%%%%%%%%%%%%%%%%%%%%%%%%%%%%%%%%%%%%%%%%%%%%%%%%%%%%%%%%%%%%%%%%%%%%%%%%%

\pageno=0\nopagenumbers\tolerance=10000\hfuzz=5pt
\line{\hfill}
\vskip 36pt
\centerline{\bf A Complete Leading--Order Renormalization--Scheme--Consistent}
\vskip 6pt
\centerline{\bf Calculation of Structure Functions.}
\vskip 36pt
\centerline{Robert~S.~Thorne}
\vskip 12pt
\centerline{\it Rutherford Appleton Laboratory,}
\centerline{\it Chilton, Didcot, Oxon., OX11 0QX, U.K.}
\vskip 0.9in
{\narrower\baselineskip 10pt
\centerline{\bf Abstract}
\medskip
We present consistently ordered calculations of the structure
functions ${\cal F}_2(x,Q^2)$ and ${\cal F}_L(x,Q^2)$, in different 
expansion schemes. After discussing the standard expansion  
in powers of $\alpha_s(Q^2)$
we consider a leading--order expansion in $\ln (1/x)$ and finally
an expansion which is leading order in both $\ln(1/x)$ and $\alpha_s(Q^2)$,
and which is the only really correct expansion scheme. 
Ordering the calculation in a renormalization--scheme--consistent manner, 
there is no factorization scheme dependence, and the calculational method
naturally includes to the ``physical anomalous dimensions'' of Catani. 
However, it imposes stronger constraints than just the use of these effective 
anomalous dimensions. A relationship between the small--$x$ 
forms of the inputs ${\cal F}_2(x,Q_I^2)$ and ${\cal F}_L(x,Q_I^2)$ 
is predicted. Analysis of a wide range of data for ${\cal F}_2(x,Q^2)$
is performed, and a very good global fit obtained, particularly for data at 
small $x$. The fit allows a prediction for ${\cal F}_L(x,Q^2)$ to be produced, 
which is smaller than those produced by the usual NLO--in--$\alpha_s(Q^2)$ 
fits to ${\cal F}_2(x,Q^2)$ and different in shape.}   

\vskip 0.7in
\line{\hfill}
\line{July 1997\hfill}
\vfill\eject
\footline={\hss\tenrm\folio\hss}

%%%%%%%%%%%%%%%%%%%%%%%%%%%%%%%%%%%%%%%%%%%%%%%%%%%%%%%%%%%%%%%%%%%%%%%%%%%%%%%

%%%%%%%%%%%%%%%%%%%%%%%%%%%%%%%%%%%%%%%%%%%%%%%%%%%%%%%%%%%%%%%%%%%%%%%%%%%%%%%

\newsec{Introduction.}

The recent measurements of ${\cal F}_2(x,Q^2)$ at HERA have provided data on
a structure function at far lower values of $x$ than any previous
experiments, and show that there is a marked rise in ${\cal F}_2(x,Q^2)$ at
very small $x$ down to rather low values of $Q^2$ \ref\hone{H1 collaboration,
\NP \vyp{B470}{1996}{3}.}\ref\zeus{ZEUS collaboration: M. Derrick {\it et al},
\ZP \vyp{C69}{1996}{607}\semi preprint DESY 96--076 (1996), 
\ZP C, in print.}. Indeed, the rise persists for values of
$Q^2$ as low as 1.5 $\Gev^2$.

The qualitative result of a steep rise at small $x$ 
was in conflict with standard methods used to fit
the data were based on the solution of the Altarelli--Parisi evolution
equation \ref\apeqn{G. Altarelli and G. Parisi, \NP \vyp{B126}{1977}{298}\semi
Yu.L. Dokshitzer, {\it Sov. Jour. JETP} \vyp{46}{1977}{641}\semi
L.N.Lipatov, \SJNP \vyp{20}{1975}{95}\semi V.N. Gribov and L.N. Lipatov,
\SJNP \vyp{15}{1972}{438}.} at the
two--loop level, using nonperturbative, flat (the Donnachie--Landshoff 
pomeron used to soft physics has behaviour $x^{-0.08}$
\ref\dlp{A. Donnachie and P.V. Landshoff, \NP \vyp{B244}{1984}{322};
\NP \vyp{B267}{1986}{690}.}, and we will take
steep to mean any powerlike behaviour steeper than this)
input parton distributions at starting
scales of $Q_I^2 \sim 4 \Gev^2$ (e.g. \ref\mrsi{A.D. Martin, R.G. Roberts and
W.J. Stirling, \PR \vyp{D47}{1993}{867}; \PL \vyp{B306}{1993}{145}.}).
This procedure results in an effectively steep
behaviour at small $x$ \ref\wilc{A. de R\'ujula {\it et al}, \PR 
\vyp{D10}{1974}{1649}.}, but only after a long evolution length, 
and therefore at values of $Q^2 \gg 4 \Gev^2$.
Thus, the data  led to optimism amongst those
advocating the use of the BFKL equation \ref\bfkl{L.N. Lipatov, \SJNP 
\vyp{23}{1976}{338}\semi E.A. Kuraev, L.N. Lipatov and V.S. Fadin {\it Sov.
Jour. JETP} \vyp{45}{1977}{199}\semi Ya. Balitskii and L.N. Lipatov, \SJNP
\vyp{28}{1978}{6}.}, which provides the
unintegrated gluon Green's function which includes the leading power of
$\ln (1/x)$ for any power of (fixed) $\alpha_s$. 
It was solved analytically in the asymptotic
limit $x \to 0$, or numerically for finite $x$, and predicted a
powerlike behaviour of $x^{-1-\lambda}$ for the gluon distribution function,
where $\lambda = 4 \ln 2 \bar \alpha_s$ and $\bar \alpha_s =(3/\pi)
\alpha_s$, i.e. $\lambda \sim 0.5$ if $\alpha_s \sim 0.2$. This was assumed to
lead to ${\cal F}_2(x,Q^2)$ behaving like $x^{-\lambda}$ and was
in rough qualitative agreement with the data, even if $\lambda$ was somewhat
high. It could also be seen as some justification for choosing
powerlike inputs (with $\lambda \sim 0.2-0.3$) for the parton
distributions (e.g. \ref\mrsii{A.D Martin, R.G. Roberts and W.J. Stirling,
\PL \vyp{B354}{1995}{155}\semi H.L. Lai {\it et al}, \PR  
\vyp{D51}{1995}{4763}.}), which could then enable a good fit to the 
data using the Altarelli--Parisi equation. 

However, it was also convincingly demonstrated that it was possible to
generate the observed steep behaviour by being a little less conservative
concerning the region in which perturbative evolution could be
applied. Gl\"uck, Reya and Vogt had in fact predicted a sharp rise 
in $F_2(x,Q^2)$ at small $x$, even for $Q^2 \sim 1\Gev^2$, by using 
two--loop evolution from roughly valence--like parton distributions at a 
starting scale of $Q_I^2 = 0.34 \Gev^2$ \ref\grv{M. Gl\"uck, E. Reya and 
A. Vogt, \ZP \vyp{C48}{1990}{471}; \ZP \vyp{C53}{1992}{127}; \PL 
\vyp{B306}{1993}{391}; \ZP \vyp{C67}{1995}{433}.}. Starting at a higher scale,
$Q_I^2 = 1 \Gev^2$, Ball and Forte were able to fit the small--$x$ data
using their double asymptotic scaling (DAS) formula \ref\bfi{R.D. Ball and
S. Forte, \PL \vyp{B335}{1994}{77}.}, which is a simple, but
very accurate, approximation to the solution of the
one--loop evolution equation with flat\foot{Parton inputs behaving like 
$x^{-1}$, as opposed to $x^{-1-\lambda}$, are referred to as flat, i.e. the 
parton density rescaled by $x$ is flat.} inputs and which is valid in the 
region of small $x$ ($x \lsim 0.01$). They also showed that
two--loop evolution with flat inputs starting at $Q_I^2 \approx 2
\Gev^2$ could fit the data available at that time very
well \ref\bfii{R.D. Ball and S. Forte, \PL \vyp{B336}{1994}{77}.}. 

This state of affairs left scope for argument about the real
underlying physics describing the small--$x$ behaviour of
${\cal F}_2(x,Q^2)$. Those who used Altarelli--Parisi evolution from small
scales could be accused of working in regions where perturbation
theory was questionable and, perhaps more importantly, of ignoring
terms of higher order in $\alpha_s$ (but also higher order in $\ln
(1/x))$, which seemed from the BFKL equation to have very important
effects. Conversely, those who used the BFKL approach could be accused
of ignoring all but the leading--$\ln (1/x)$ terms (and hence ignoring
the large--$x$ data) and also of working
in a less well--defined theoretical framework than the renormalization
group approach based on the factorization of collinear
singularities \ref\collfac{J.C. Collins, D.E. Soper and G. Sterman, in: 
Perturbative Quantum Chromodynamics, ed. A.H. Mueller (World Scientific,
Singapore, 1989), and references therein.}. Starting from an input for 
the parton
distributions with $\lambda \sim 0.25$ at values of $Q_I^2 \sim
4\Gev^2$ was taking the best
of both worlds. However, this lacked a real justification 
for the choice
of input, which was significantly steeper than that expected from 
non--perturbative physics, but rather smaller than that from the BFKL equation,
and also ignored potentially important $\ln (1/x)$ terms in
the evolution. 

A significant step forward in the investigation of small--$x$ structure 
functions was the development of the
$k_T$--factorization theorem \ref\kti{S. Catani, M. Ciafaloni and 
F. Hautmann, \PL \vyp{B242}{1990}{97}; \NP \vyp{B366}{1991}{135}; \PL
\vyp{B307}{1993}{147}.}\ref\ktii{J.C. Collins and R.K. Ellis, \NP 
\vyp{B360}{1991}{3}.}, the prescription for the way in
which an off--shell photon--gluon scattering amplitude can be
convoluted with the unintegrated gluon Green's function calculated
using the BFKL
equation to provide the small--$x$ structure functions themselves. Hence, it
enables one to find effective moment--space coefficient functions within 
the BFKL framework. Numerical calculations performed
using this method were able to match the available data in a 
qualitative manner \ref\akms{A.J. Askew, J. Kwieci\'nski, A.D. Martin and 
P.J. Sutton, \PR \vyp{D49}{1994}{4402}\semi A.J. Askew {\it et al}, \PL 
\vyp{B325}{1994}{212}.}, as did similar calculations 
\ref\ccfmphen{J. Kwieci\'nski, 
A.D. Martin and P.J. Sutton, \PR \vyp{D53}{6094}{1996}; \ZP  
\vyp{C71}{1996}{585}.} using a modification of the
BFKL equation, i.e. the CCFM equation \ref\ccfm{M. Ciafaloni, \NP  
\vyp{B296}{1988}{49}\semi S. Catani, F. Fiorani and G. Marchesini, \PL 
\vyp{B234}{1990}{339}; \NP \vyp{B336}{1990}{18}.}. 
The $k_T$--factorization formula was also shown to be very important
if one insisted on working within the rigorous framework of the 
traditional renormalization group approach \ref\cathaut{S. Catani and 
F. Hautmann, \PL \vyp{B315}{1993}{157}; \NP \vyp{B427}{1994}{475}.}. By
showing how $k_T$--factorization fits within the collinear
factorization framework, Catani and Hautmann were able to calculate
all the renormalization group anomalous dimensions to lowest
nontrivial order in $\alpha_s$ for each power of $\ln (1/x)$, and similarly
for a number of coefficient functions. 
They demonstrated
that within the renormalization group framework 
the Mellin transformations of the splitting functions, the anomalous
dimensions $\gamma^0_{gg}(N,Q^2)$
and $\gamma^0_{gf}(N,Q^2)$ were the same 
renormalization--scheme--independent expressions as the effective anomalous
dimensions \ref\jaro{T. Jaroszewicz, \PL \vyp{B116}{1982}{291}.} given
by the BFKL equation. i.e. 
\eqn\gammadef{\gamma^0_{gg}(\alpha_s(Q^2)/N) = 
\sum_{m=1}^{\infty} a_{0,m} \biggl(
{\bar \alpha_s(Q^2) \over N}\biggr)^m, \hskip 0.6in
\gamma^0_{gf}(\alpha_s(Q^2)/N)= {4 \over 9} \gamma^0_{gg}(\alpha_s(Q^2)/N),}
where $\gamma^0_{gg}(\alpha_s(Q^2)/N)$ is given by the iterative solution of 
\eqn\chieqn{1={\bar\alpha_s(Q^2) \over N} \chi(\gamma^0_{gg}),
\hskip 0.8in \chi(\gamma) = 2\psi(1) -\psi(\gamma) -\psi(1-\gamma).}
and the solution as a power series in $\bar \alpha_s(Q^2)/N$
exists for $\vert N\vert \geq \lambda(\bar
\alpha_s(Q^2))$. 
They also derived expressions for
$\gamma^1_{ff}(\alpha_s(Q^2)/N)$ and $\gamma^1_{fg}(\alpha_s(Q^2)/N)$ 
(series of the same form as in \gammadef\ but with an extra power of 
$\alpha_s$) in certain
factorization schemes ($\gamma^0_{ff}(\alpha_s(Q^2)/N)$ and
$\gamma^0_{fg}(\alpha_s(Q^2)/N)$ being zero), and also for
the coefficient functions 
$C^g_{L,1}(\alpha_s(Q^2)/N)$, $C^f_{L,1}(\alpha_s(Q^2)/N)$, 
$C^g_{2,1}(\alpha_s(Q^2)/N)$ and $C^f_{2,1}(\alpha_s(Q^2)/N)$ (all
zeroth--order quantities being zero except $C^f_{2,0}$, which is
unity). This facilitated calculations of structure functions within
the normal renormalization group framework, but including much of what
is often called BFKL physics (i.e. the leading--$\ln (1/x)$ terms),
and indeed, a number of calculations were performed using somewhat different 
approaches \ref\bfresum{R.D. Ball 
and S. Forte, \PL \vyp{B351}{1995}{313};
\PL \vyp{B358}{1995}{365}.}\nref\ehw{R.K. Ellis, F. Hautmann and B.R. Webber,
\PL \vyp{B348}{1995}{582}.}\nref\frt{J.R. Forshaw, R.G. Roberts and 
R.S. Thorne, \PL \vyp{B356}{1995}{79}.}--\ref\brv{J. Bl\"umlein, S. 
Riemersma and A. Vogt, Proc. of the International Workshop {\sl QCD
and QED in higher orders}, Rheinsburg, Germany, April, 1996, 
eds. J. Bl\"umlein, F. Jegerlehner and T. Riemann, \NP B (Proc. Suppl.),
51C (1996) p. 30; {\tt hep-ph/9607329}, 
Proc. of the International Workshop on Deep Inelastic Scattering,
Rome, April, 1996, in print.}, and in most cases 
comparisons with data made. It seemed that by including these terms
it was not possible to
improve upon the best fits for the small--$x$ data using one-- or
two--loop evolution from soft inputs \bfresum\frt.
Indeed, many ways of including them made the
fits significantly worse, and this seemed to be universally true if
the fits were more global, i.e. constrained by large--$x$ data \bfresum.
Also, it seemed that there was a very strong
dependence on the factorization scheme used to perform the
calculations when including the leading--$\ln (1/x)$ terms \bfresum\frt
\ref\sdis{S. Catani, \ZP \vyp{C70}{1996}{263}.}\ref\qzero{M. Ciafaloni,
\PL \vyp{B356}{1995}{74}.}, and a
number of new factorization schemes were invented. 

The high precision of the most recent HERA data constrains theory far more 
than previously, and has changed the above picture somewhat. The best 
recent global fits seem to come from those intermediate approaches which 
use NLO perturbation theory with a quite steep (and completely unexplained) 
input for the singlet quark with 
$\lambda \sim 0.2$ and a very $Q_I^2$-sensitive small--$x$ input for the 
gluon \ref\mrsiii{A.D. Martin, W.J. Stirling and R.G. Roberts, 
\PL \vyp{B387}{1996}{419}.}.
Fixed order perturbation theory using 
flat or valence--like inputs and low $Q_I^2$ fails
at the lowest $x$ values, and relatively steep inputs for
the singlet quark, i.e. $\lambda \gsim 0.2$, seem to be absolutely necessary 
\ref\steep{R.D. Ball and S. Forte, {\tt hep-ph/9607289},
Proc. of the International 
Workshop on Deep Inelastic Scattering, Rome, April, 1996, in print.}.
Approaches including the leading--$\ln (1/x)$ terms seemed to 
fail \ref\faili{S. Forte and R.D. Ball, {\tt hep-ph/9607291},
Proc. of the International Workshop 
on Deep Inelastic Scattering, Rome, April, 1996, in print.}\ref\failii{I. 
Bojak and M. Ernst, {\tt hep-ph/9609378}, preprint DO--TH 96/18, 
September 1996.} in practically all factorization schemes.  

In this paper we will take issue with the above conclusions. 
In particular we will
demonstrate that the apparent failure of approaches
using the leading--$\ln (1/x)$ terms, and certainly the factorization scheme
dependence, is due to incorrect methods of incorporating these terms.
We will show that the
correct leading--order, renormalization--scheme--consistent (RSC) calculation 
of the structure functions is quite naturally 
factorization--scheme--independent, and 
also naturally includes leading--$\ln (1/x)$ 
terms in a well--defined form as well as terms leading--order in 
$\alpha_s$. We will also show that this calculation is clearly preferred 
by data.\foot{We 
note that a very brief presentation of the complete RSC calculation 
of structure functions has already appeared in \ref\mylet{R.S. Thorne,
\PL {B392}{1997}{463}.}.} 

\medskip

This paper will be structured as follows. We begin by defining conventions 
and giving a brief outline of the different possible types of scheme
dependence in the calculation of structure functions. We 
then quickly review the work of Catani, who has already shown how to obtain
factorization--scheme--invariant results in the small--$x$ expansion by
writing evolution equations in terms of the physical quantities, the
structure functions and ``physical anomalous dimensions'', rather than 
in terms of parton densities and of the usual anomalous dimensions \ref\cat{S. 
Catani, talk at UK workshop on HERA physics, September 1995, unpublished;
{\tt hep-ph/9609263}, preprint DDF 248/4/96, April 1996; {\tt hep-ph/9608310},
Proc. of the International 
Workshop on Deep Inelastic Scattering, Rome, April, 1996, in print.}. 
We then give a description of the calculation, within moment space, 
of structure functions in various
expansion schemes using the normal parton language.
The first part of this, regarding the normal loop expansion, will highlight 
points usually not discussed, particularly the form of the inputs. 
The second part, 
discussing the leading--$\ln (1/x)$ expansion, will 
present the only correct way to perform this expansion. One finds that
by calculating physical quantities in a well--ordered manner the 
physical
anomalous dimensions automatically appear, but not in the form one would 
obtain by simply solving the evolution equations using these
anomalous dimensions at some given order. Moreover, there is a certain 
degree of predictive power
for the form of the inputs for the structure functions at small $x$.
We also explain why the standard solutions using the small--$x$
expansions are strongly factorization scheme dependent.
To conclude this section we present the argument that there is a unique
renormalization--scheme--consistent calculation of structure functions, 
which applies to both large
and small $x$. We present this calculation for both the currently 
academic case of the nonsinglet
structure functions and for the phenomenologically important case of the 
singlet structure functions. We then briefly discuss
how we move from moment space and
obtain our $x$--space solutions, and the qualitative form 
these solutions must take, i.e. our best attempt at predictions.
After this long theoretical presentation we consider the comparison with
experiment, fitting the data for ${\cal F}_2(x,Q^2)$ 
using the renormalization--scheme--consistent solutions,
and comparing to global fits using the normal loop expansion at NLO. 
We conclude 
that the full renormalization--scheme--consistent calculation gives the best 
global fit to structure function data, particularly at small $x$. 
Finally we investigate the phenomenological consequences for 
${\cal F}_L(x,Q^2)$, and also preliminary indications for the charm structure 
function.\foot{There is a rather longer version of this paper which includes 
much more detail of a pedagogical nature and more 
comparison with other approaches
\ref\longsx{R.S. Thorne, RAL-TR-96-065, {\tt hep-ph/9701241.}.}}  

\newsec{Scheme and Scale Choices.}

For simplicity we work in moment--space for much of this paper, i.e. 
define the moment--space structure functions by the Mellin transform, i.e.
\eqn\melltranssf{F(N,Q^2)= \int_0^1\,x^{N-1}{\cal F}(x,Q^2)dx.}
The moment--space coefficient function is defined similarly but, as
with the definition of the anomalous dimension, we define the moment
space expression for the parton distribution as the Mellin transform
of a rescaled parton density i.e 
\eqn\melltranspd{f(N,Q^2)= \int_0^1\,x^{N}{\rm f}(x,Q^2)dx.}
The most
general moment--space expression for a structure function is the sum of the
products of the expressions for hard scattering with a certain
parton (the coefficient functions) with the corresponding,
intrinsically nonperturbative parton distributions,\foot{We deal with the heavy
quark thresholds in a rather naive manner, i.e. the quarks are taken to be
massless, with a particular flavour becoming active only above a certain
$Q^2$.}
\eqn\strcfunc{F(N,Q^2)= \sum_a
C^a(N,\alpha_s(\mu_R^2),Q^2/\mu_F^2, \mu_R^2/\mu_F^2)
f_a(N,\alpha_s(\mu_R^2),\mu_R^2/\mu_F^2),}
where $\mu_F$ is a factorization scale separating the ultraviolet physics
from the soft infrared physics. The coupling $\alpha_s(\mu_R^2)$ satisfies 
the renormalization group equation,
\eqn\runcoup{{d \,\alpha (\mu_R^2) \over d 
\ln(\mu_R^2)}=-\sum_{n=0}^{\infty}b_n\alpha^{n+2}(\mu^2)\equiv 
-\beta(\alpha(\mu_R^2)),}
where $\mu_R$ is the renormalization scale, the scale at which the coupling 
is defined. 

Two of the choices left open in the above expressions are 
common to all perturbative calculations in quantum field theory:
the choice of renormalization scheme (the choice of which expansion
parameter is to be used, and consequently the form of the perturbative
expansion), and subsequently the
choice of renormalization scale. Conventionally 
the $\overline{\hbox{\rm MS}}$ renormalization scheme is used, along with
the simple choice $\mu_R = \mu_F$, and indeed, the data seem to
favour this choice of scale \ref\mrsiv{A.D. Martin, R.G. Roberts and 
W.J. Stirling, \PL\vyp{B266}{1991}{173}.}. Hence, we take this simple choice. 
The remaining ambiguities are due to the particular problems in calculating 
quantities in QCD, i.e separating the physical quantity into the
perturbative coefficient function and the intrinsically
nonperturbative parton distribution. We have the freedom
of choosing the factorization scale $\mu_F$. As with renormalization scheme
dependence this does not affect the all--orders calculation and invariance 
under this choice of scale leads to the evolution equation
\eqn\apeqni{{d \, f_a(\mu_F^2) \over d \ln \mu_F^2}= \sum_b
\gamma_{ab}(\alpha_s(\mu_F^2)) f_b(\mu_F^2).}
It is desirable to choose $\mu_F^2$
to be large in order to make the expansion parameter
$\alpha_s(\mu_F^2)$ small, and to sum large logarithms in $Q^2$, and 
$\mu_F^2$ is
nearly always chosen to be equal to the hard scattering scale $Q^2$. We 
shall also make this simple choice.

This leaves us with our defining equations
\eqn\strcfuncii{F(N,Q^2)= 
\sum_a C^a(N,\alpha_s(Q^2)) f_a(N,Q^2)\hskip 0.3in \hbox{and}
\hskip 0.3in {d\, f_a(Q^2) \over d \ln Q^2}=
 \sum_b\gamma_{ab}(\alpha_s(Q^2))  f_b(Q^2),}
or being more careful there are two independent structure functions 
$F_2(N,Q^2)$ and $F_L(N,Q^2)$. In general we may write
\eqn\srucdef{F_i(N,Q^2) =
{1\over N_f}\biggl(\sum_{j=1}^{N_f}e^2_j\biggr)F^S_i(N,Q^2) +
F^{NS}_{i}(N,Q^2) \hskip 0.5in (i=2,L),}
where the singlet and nonsinglet structure functions are defined by 
\eqn\fsing{F^S_i(N,Q^2) = C^f_i(N,\alpha_s)f^S(N,Q^2) +
C^g_i(N,\alpha_s)g(N,Q^2),}
\eqn\fnonsing{F^{NS}_i(N,Q^2) = C^{NS}_i(N,\alpha_s)\sum_{j=1}^{N_f}
e_j^2 f^{NS}_{q_j}(N,Q^2),}
where $N_f$ is the number of active quark flavours, $f^S(N,Q^2)$ and
$f^{NS}_{q_j}(N,Q^2)$ are the singlet and nonsinglet quark
distribution functions respectively, and $g(N,Q^2)$ is the gluon distribution.
The equations for the nonsinglet distributions are ordinary
differential equations,
\eqn\nonsingevol{{d \, f^{NS}_{q_j}(N,Q^2) \over d \ln Q^2} =
\gamma_{NS}(N,\alpha_s)f_{q_j}^{NS}(N,Q^2),}
while those for the singlet sector are coupled
\eqn\singevol{ {d\over d \ln Q^2} \pmatrix{f^S(N,Q^2)\cr g(N,Q^2)\cr} =
\pmatrix{\gamma_{ff}(N,\alpha_s) &\gamma_{fg}(N,\alpha_s)\cr 
\gamma_{gf}(N,\alpha_s) & \gamma_{gg}(N,\alpha_s)\cr}
\pmatrix{f^S(N,Q^2)\cr g(N,Q^2)\cr}.}
\medskip 

This still leaves us one more freedom in our calculation,
i.e. how we choose to remove the infrared divergences from the bare 
coefficient functions and hence how we define our parton distributions. 
Starting from any particular choice for the definition of parton 
distributions one may always choose a new set of
parton distributions by an invertible transformation
\eqn\transpart{\breve f_a(N,Q^2) =
\sum_b U_{ab}(N,\alpha_s) f_b(N,Q^2),}
where $ U_{ab}(N,\alpha_s)$ has a power series expansion in
$\alpha_s$ such that $U_{ab}(N,\alpha_s)=\delta_{ab}+ {\cal
O}(\alpha_s)$. The structure functions will clearly be unchanged
as long as the coefficient functions obey the transformation rule
\eqn\coefftrans{\breve C^a(N,\alpha_s) =
(U^T)^{-1}_{ab}(N,\alpha_s)C^b(N,\alpha_s).}
By substituting $f_a(N,Q^2) =
\sum_b U^{-1}_{ab}(N,\alpha_s) \breve f_b(N,Q^2)$ into \apeqni\ we
easily find that 
the new parton densities evolve according to the standard evolution
equations but with the new anomalous dimensions
\eqn\transgamma{\breve \gamma_{ab}(N, \alpha_s) = \sum_c\sum_d 
U_{a,c}(N,\alpha_s)\gamma_{cd}(N,\alpha_s)U^{-1}_{db}(N,\alpha_s) +
\sum_c \beta(\alpha_s){\partial   U_{ac}(N,\alpha_s) \over \partial \alpha_s}
 U^{-1}_{cb}(N,\alpha_s).}

The transformation defined above is called a change of factorization 
scheme.\foot{The matrix $U$ must 
obey a number of conditions in order that physical
requirements on the parton distributions are maintained, e.g. flavour
and charge conjugation invariance, fermion number conservation and
longitudinal momentum conservation (see for example the second of \cathaut).
However, none of these needs to be
satisfied simply in order to keep the structure functions unchanged.} 
However, 
it is important to realize that, unlike the changes in renormalization
scheme, a change in factorization scheme leaves the expression for the
structure functions unchanged not only to all orders, but 
order by order in $\alpha_s$.
{\it Calculations performed carefully at a given well--defined order in
one scheme will lead to precisely the same results for the
structure functions as those in another scheme.} 
Also, the coupling constant to 
be used depends only on the ultraviolet renormalization. We will illustrate
this point in \S 4. However, we will
first discuss Catani's recent proposal for the construction of
factorization--scheme--independent structure functions. 

\newsec{Evolution Equations for Structure Functions.}

It is, as Catani noticed \cat, very simple to obtain 
factorization--scheme--independent factorization--scheme--independent 
effective anomalous
dimensions governing the evolution of the structure functions
In order to obtain these effective  anomalous dimensions all one
has to do is eliminate the parton densities from the equations
\strcfuncii\ and \apeqni. In order to demonstrate this,
let us first consider the simple case of the nonsinglet
structure function $F_2^{NS}(N,Q^2)$. 
Multiplying both sides of \nonsingevol\ by
$\sum_{j=1}^{N_f}e_j^2$ we can clearly write 
\eqn\nonsingfsi{{{d   \over 
d \ln Q^2} \Bigl({F_2^{NS}(N,Q^2)\over C^{NS}_2(N,\alpha_s)}\Bigr)
=\gamma_{NS}(N,\alpha_s){F_2^{NS}(N,Q^2) \over 
C^{NS}_2(N,\alpha_s)},}}
which becomes the factorization--scheme--independent equation
\eqn\nonsingfsi{{{d \, F_2^{NS}(N,Q^2) \over 
d \ln Q^2} =\Gamma_{2,NS}(N,\alpha_s)F_2^{NS}(N,Q^2),}}
where $\Gamma_{2,NS}(N,\alpha_s)=
\gamma_{NS}(N,\alpha_s)+d\,\ln(C^{NS}_2(N,\alpha_s))/d\ln Q^2$.
Therefore, we have
an effective anomalous dimension governing the evolution of each of
the nonsinglet structure functions, and clearly
$\Gamma_i^{NS}(N,\alpha_s)$ must be a factorization--scheme--independent
quantity (and is in principle measurable). The solution to this
equation is trivial:
\eqn\nonsingfsisol{{F_2^{NS}(N,Q^2)  
 =F_2^{NS}(N,Q_I^2)\exp\biggl[\int_{\ln Q_I^2}^{\ln Q^2}
\Gamma_{2,NS}(N,\alpha_s)d\ln q^2\biggr].}}

The situation for the singlet structure functions is more
complicated. However, using
\fsing\ for $i=2,L$ we may solve for the parton densities in terms of
the two structure functions and the coefficient
functions. Substituting these into \singevol\ we then obtain the
coupled evolution equations
\eqn\singevolfsi{ {d\over d \ln Q^2} \pmatrix{F^S_2(N,Q^2)\cr 
F^S_L(N,Q^2)\cr} =
\pmatrix{\check\Gamma_{22}(N,\alpha_s) &\check\Gamma_{2L}(N,\alpha_s)\cr 
\check\Gamma_{L2}(N,\alpha_s) & \check\Gamma_{LL}(N,\alpha_s)\cr}
\pmatrix{F^S_2(N,Q^2)\cr F^S_L(N,Q^2)\cr}.}
The expressions for the physical anomalous dimensions,
$\check\Gamma_{22}(N,\alpha_s)$, $\check\Gamma_{2L}(N,\alpha_s)$,
etc., are straightforward to derive 
in terms of anomalous dimensions and coefficient functions in any
particular factorization scheme using
the above procedure, but result in rather cumbersome expressions.
It is simplest first to
define a factorization scheme such that $F^S_2(N,Q^2)= f^S(N,Q^2)$,
i.e. $C^f_2(N,\alpha_s)=1$, $C^g_2(N,\alpha_s)=0$ (this is generally
known as a DIS type scheme \ref\dis{G. Altarelli, R.K. Ellis and 
G. Martinelli, \NP \vyp{B157}{1979}{461}.}\foot{We call it a DIS 
``type'' scheme
because satisfying the above requirement still leaves freedom in how we
may define the gluon density, and thus we are still considering a
family of schemes.}). In terms of the coefficient functions and
anomalous dimensions in this type of scheme we have
\eqn\physanom{\eqalign{\check\Gamma_{22}(N,\alpha_s)&=
\gamma_{ff}(N,\alpha_s)
-{C^f_L(N,\alpha_s)\over
C^g_L(N,\alpha_s)}\gamma_{fg}(N,\alpha_s),\cr
\check\Gamma_{2L}(N,\alpha_s)&=
{\gamma_{fg}(N,\alpha_s)\over C^g_L(N,\alpha_s)},\cr
\check\Gamma_{L,2}(N,\alpha_s)&= C^g_L(N,\alpha_s)\gamma_{gf}(N,\alpha_s)-
C^f_2(N,\alpha_s)\gamma_{gg}(N,\alpha_s)+{d
C^f_L(N,\alpha_s)\over d \ln
Q^2}+C^f_L(N,\alpha_s)\gamma_{ff}(N,\alpha_s)\cr
&-C^f_L(N,\alpha_s){d \ln(C^g_L(N,\alpha_s))\over
d\ln Q^2} -{(C^f_L(N,\alpha_s))^2\over C^g_L(N,\alpha_s)}
\gamma_{fg}(N,\alpha_s),\cr
\check\Gamma_{LL}(N,\alpha_s)&=\gamma_{gg}(N,\alpha_s) + 
{d \ln(C^g_L(N,\alpha_s))\over d \ln
Q^2}+ {C^f_L(N,\alpha_s)\over C^g_L(N,\alpha_s)}\gamma_{fg}(N,\alpha_s).\cr}}

Before
going any further let us remark that we believe there is a 
(purely technical) problem
with the above expression. As is well known $F_L(N,Q^2)$ starts at an order 
of $\alpha_s$ higher than
$F_2(N,Q^2)$. Because of this there is an intrinsic asymmetry in the
above definitions, with $\check\Gamma_{2L}(N,\alpha_s)$ beginning at zeroth
order in $\alpha_s$, $\check\Gamma_{22}(N,\alpha_s)$ and 
$\check\Gamma_{LL}(N,\alpha_s)$
beginning at first order, and $\Gamma_{L2}(N,\alpha_s)$ beginning at
second order. This asymmetry leads to the results obtained by solving these 
equations order by order being different that those using the normal parton 
picture even in the loop expansion.
A trivial modification of Catani's approach is therefore to accept 
that $F_L(N,Q^2)$ contains an extra power of
$\alpha_s$, and to define the new structure function $\hat
F_L(N,Q^2)=F_L(N,Q^2)/(\alpha_s/(2\pi))$. The longitudinal coefficient 
functions are likewise changed to $\hat
C^a_L(N,\alpha_s)=C^a_L(N,\alpha_s)/(\alpha_s/(2\pi))$, and the
singlet evolution equations become 
\eqn\singevolfsimod{ {d\over d \ln Q^2} \pmatrix{F^S_2(N,Q^2)\cr 
\hat F^S_L(N,Q^2)\cr} =
\pmatrix{\Gamma_{22}(N,\alpha_s) &\Gamma_{2L}(N,\alpha_s)\cr 
\Gamma_{L2}(N,\alpha_s) & \Gamma_{LL}(N,\alpha_s)\cr}
\pmatrix{F^S_2(N,Q^2)\cr \hat F^S_L(N,Q^2)\cr},}
where the $\Gamma(N,\alpha_s)$'s are defined precisely as in \physanom, but
in terms of of $\hat C^a_L(N,\alpha_s)$ rather than $C^a_L(N,\alpha_s)$. 
This procedure
restores the symmetry between the physical anomalous dimensions, and
makes the order--by--order--in--$\alpha_s$ calculations 
essentially the same as when using
evolution of parton distributions. It is, of course, trivial to obtain the
physical $F_L(N,Q^2)$ from $\hat F_L(N,Q^2)$.\foot{Similarly, 
it is also 
best to work with the rescaled nonsinglet structure function $\hat
F^{NS}_L(N,Q^2)$.}  

Having made our redefinition of the quantities with which we work, we
now have a direct relationship between possible calculations using the
evolution equations for structure functions and the solutions using the parton
densities. At present the parton anomalous
dimensions and coefficient functions are known to order $\alpha_s^2$. It 
is easy to see that this allows us to derive each of the $\Gamma$'s to order
$\alpha_s^2$. Similarly, from the known expansions of the parton
anomalous dimensions and coefficient functions in the form
$\alpha_s^n\sum_{m=1-n}^{\infty}a_m(\alpha_s/N)^m$ \cathaut, we can calculate 
$\Gamma_{LL}^0(N,\alpha_s)$ and $\Gamma^0_{L2}(N,\alpha_s)$,
$\Gamma_{2L}^0(N,\alpha_s)$ and $\Gamma_{22}^0(N,\alpha_s)$ 
(where both are trivially zero), and $\Gamma_{2L}^1(N,\alpha_s)$ 
and $\Gamma_{22}^1(N,\alpha_s)$.
This is the same order as
for the parton anomalous dimensions, with the longitudinal anomalous
dimensions having similar structure to the gluon anomalous dimensions and the 
$\Gamma_{2a}(N,\alpha_s)$'s having similar structure to the quark 
anomalous dimensions. The resulting expressions are
relatively simple, being  
\eqn\physanomval{\eqalign{\Gamma^1_{22}(\alpha_s/N)&= 
-{1\over(2\pi)}(\hat C^{f}_{L,1,0}-
\fourninths \hat C^{g}_{L,1,0})
(\threeovertwo \gamma^0_{gg}(\alpha_s/N)+
\sum_{n=0}^{\infty}(\gamma^0_{gg}(\alpha_s/N))^n)\cr
&-\fourninths\gamma^{1,0}_{fg}(N,\alpha_s),\cr
\Gamma^1_{2L}(\alpha_s/N)&={1\over(2\pi)}(\threeovertwo 
\gamma^0_{gg}(\alpha_s/N)+
\sum_{n=0}^{\infty}(\gamma^0_{gg}(\alpha_s/N))^n),\cr
\Gamma^0_{L,2}(N,\alpha_s)&= -(\hat C^{f}_{L,1,0}-
\fourninths \hat C^{1,0}_{L,g})\gamma^0_{gg}(\alpha_s/N),\cr
\Gamma^0_{LL}(\alpha_s/N)&=\gamma^0_{gg}(\alpha_s/N),\cr}}
where $\gamma^{1,0}_{fg}$, $C^{f}_{L,1,0}$ and
$C^{g}_{L,1,0}$ are the one--loop contributions to 
$\gamma^{1}_{fg}(\alpha_s/N)$, $C^{f}_{L,1}(\alpha_s/N)$ and
$C^{g}_{L,1}(\alpha_s/N)$ respectively. Each of these
anomalous dimensions is renormalization scheme invariant as well as
factorization scheme invariant, as we would expect for leading--order 
physical quantities.   

Solving the evolution equations for the structure functions using any
subset of the currently known physical anomalous dimensions guarantees a
result which is factorization scheme independent. However, there is
considerable freedom in how we may solve the
equations. We could, for example, simply put all of the anomalous
dimensions currently known into \singevolfsimod\ and then 
find the whole solution. 
Alternatively, we could solve using just the order $\alpha_s$ anomalous
dimensions and then try to perturb about this solution in an ordered
manner. These two approaches would lead to rather different answers, but 
both would be factorization scheme independent. The
problem of obtaining a correctly ordered solution for the structure
function will be discussed in detail in the next section. We will
initially use the familiar parton distributions and coefficient
functions, and show that, even when
using this approach, if we solve producing a well--defined expansion 
for the structure functions, we constrain the form of the solution rigidly
and thus automatically avoid the problem of factorization scheme dependence.
In fact,
not only do we obtain the physical anomalous dimensions, but also find 
precisely how we should use them.

\newsec{Ordered Calculations of Structure Functions.}

There are in principle many different expansion methods one may use when
obtaining solutions for the structure functions. The standard one is
simply solving order by order in
$\alpha_s$. But there is also the expansion in leading powers of $\ln
(1/x)$ for given powers in $\alpha_s$ (or equivalently in powers of 
$N^{-1}$ in moment space), as we
have already discussed. One can also combine the two expansion
methods, and indeed, we will later argue that this is the only correct 
thing to do.
Nevertheless, we will begin by outlining the procedure for making
a well--ordered
calculation of structure functions using the standard loop
expansion. Although this is well known, we feel it is worth presenting
it briefly, and making some points which are not usually highlighted,
especially concerning the role of the starting scale. 
We may then discuss the more complicated cases
of the expansion in leading powers of $\ln (1/x)$ and the combined expansion. 

\subsec{Loop Expansion.}

We begin by introducing some new notation.
In order to solve the evolution equations for the parton densities
order by order in $\alpha_s$ and
hence obtain expressions for $F^{NS}_i(N,Q^2)$ and $F^S_i(N,Q^2)$ we
make use of equation \runcoup\ to rewrite the evolution equation for the
nonsinglet parton density as
\eqn\nonsingevolal{\alpha^2_s(Q^2) {d \, f^{NS}_{q_j}(N,Q^2) \over d 
\alpha_s(Q^2)} =- \tilde \gamma_{NS}(N,\alpha_s)f_{q_j}^{NS}(N,Q^2),}
where $\tilde \gamma_{NS}(N,\alpha_s) =
\alpha^2_s(Q^2)\gamma_{NS}(N,\alpha_s)/\beta(\alpha_s)$, and similarly for
the singlet distributions,
with similar definitions for $\tilde\gamma_{ff}(N,\alpha_s)$ etc. as for
$\tilde \gamma_{NS}(N,\alpha_s)$. Each of the $\tilde \gamma$'s may now be
written as
\eqn\gamtilde{\tilde \gamma(N,\alpha_s) = \sum_{n=0}^{\infty}\tilde
\gamma^{n,l}(N) \, \alpha_s^{n+1},}
where where the super--subscript $n,l$ denotes the $(n+1)$--loop quantity and
we have an analogous definition for the normal anomalous dimensions.
 
Using our definition, the evolution equations may be solved order by order in
$\alpha_s$. When doing this it is necessary to choose a starting
scale $Q_I^2$ for the perturbative evolution of the parton distributions 
(or equivalently a starting value of the coupling $\alpha_s(Q_I^2)$),
and specify input parton distributions at this scale.
Let us discuss the choice of this scale briefly.
$Q_I^2$ must clearly be chosen to be large enough that perturbative
evolution should be reliable, i.e. $\alpha_s(Q^2)\lsim 0.3$ 
and also such that ``higher twist'' ($\Lambda_{QCD}^2/Q^2$) corrections 
should be
very small. Traditionally evolution only takes place up from this starting 
scale, for the simple reason of convenience, and also because some form of the 
inputs has been expected at low starting scales. Until a few years ago
the requirements described above led to a choice of 
$Q_I^2 \approx 4 \Gev^2$. In the past couple of years this value has 
often been chosen to be rather
lower, due to the apparent success of evolution from lower starting scales,
and also because much of the interesting small--$x$ data is now at $Q^2
\leq 4 \Gev^2$. These choices have been accompanied by guesses for the form of
the inputs at low starting scales, e.g valence--like \grv, or
flat \mrsi\bfi. 
We make no assumptions of the above sort about the value of $Q_I^2$. We
require it to be high enough to be in the perturbative regime and to 
avoid higher twists, but acknowledge that  
there is no reason why $Q_I^2$ cannot be chosen to be
quite large, and evolution away from the starting scale 
performed both up and down in $Q^2$. Taking this open--minded approach we 
also assume that only perturbative effects can lead to deviations 
from soft behaviour of the structure functions, and also 
demand that the form of our well--ordered expressions for the structure 
functions is as insensitive as possible to this choice, thus making 
the choice of $Q_I^2$ as open as possible. 

\medskip

We begin by solving for the nonsinglet parton distributions, which are 
an easily understandable model. In this case the solution is
particularly simple. Integrating both sides of \nonsingevolal\ we
obtain
\eqn\nonsingsol{f^{NS}_{q_j}(N,Q^2)=\biggl[\sum_{k=0}^{\infty}\alpha_s^k(Q_I^2)
f^{NS}_{q_j,k}(N, Q_I^2)\biggr]
\biggl({\alpha_s(Q_I^2)\over\alpha_s(Q^2)}\biggr)^{\tilde \gamma^{0,l}_{NS}(N)}
\exp\biggl[\sum_{n=1}^{\infty}{(\alpha^n_s(Q_I^2)-\alpha_s^n(Q^2))\over
n}\tilde\gamma_{NS}^{n,l}(N)\biggr].}
Perhaps unconventionally, we explicitly express the input 
$f^{NS}_{q_j}(N,Q_I^2)$ as a power series in $\alpha_s(Q_I^2)$.
This is necessary because changes in the
starting scale $Q_I^2$ lead to $\alpha_s(Q_I^2)$--dependent changes in 
the expression for the evolution term, 
which must be compensated for by $\alpha_s(Q_I^2)$--dependent
changes in the starting distribution in order to leave 
the whole expression for
the parton distributions unchanged, as required.
Let us examine this briefly by looking at the change of the 
lowest--order piece of \nonsingsol, i.e.
\eqn\nonsingsolzero{f^{NS}_{q_j,0}(N,Q^2)=f^{NS}_{q_j,0}(N,Q_I^2)
\biggl({\alpha_s(Q_I^2)\over\alpha_s(Q^2)}\biggr)^{\tilde 
\gamma^{0,l}_{NS}(N)},}
under a change in starting scale, $Q_I^2 \to  (1+\delta) Q_I^2$,
where $\delta$ is some constant. We may regain expressions in
terms of $Q_I^2$ by expanding the coupling constant in the form 
\eqn\inscalchange{\alpha_s((1+\delta)Q_I^2)= \alpha_s(Q_I^2)-
\delta b_0\alpha_s^2(Q_I^2) + {\cal O}(\alpha_s^3(Q_I^2)).}
Under this change in the input coupling constant, the evolution term in 
\nonsingsolzero\ undergoes a change
\eqn\zeroevolcha{\Delta \biggl({\alpha_s(Q_I^2)\over\alpha_s(Q^2)}
\biggr)^{\tilde \gamma^{0,l}_{NS}(N)} = -\alpha_s(Q_I^2)
\delta b_0 \tilde \gamma^{0,l}_{NS}(N)
\biggl({\alpha_s(Q_I^2)\over\alpha_s(Q^2)}
\biggr)^{\tilde \gamma^{0,l}_{NS}(N)} + \hbox{\rm higher order in  
$\alpha_s(Q_I^2)$}.} 
This change in the parton distribution due to the variation in the leading 
term can be countered, up to higher orders, by a change in the order 
$\alpha_s(Q_I^2)$ input of the form 
\eqn\chainput{\Delta f^{NS}_{q_j,1}(N,Q_I^2)=\delta b_0 \tilde 
\gamma^{0,l}_{NS}(N) f^{NS}_{q_j,0}(N,Q_I^2).}
Hence, changes in the evolution
term due to a change in $Q_I^2$ begin at first order in
$\alpha_s(Q_I^2)$, and are therefore absorbed by terms in the input 
at first order and beyond. The zeroth--order input is insensitive
to such changes and is $Q_I^2$--independent: $f^{NS}_{q_j,0}(N,Q_I^2)\equiv
f^{NS}_{q_j,0}(N)$. 
Higher--order changes in the parton distributions due to changes in  
$Q_I^2$ are all accounted 
for by changes in the higher--order inputs which are equal to functions of 
$N$ dependent 
only on the anomalous dimensions. Therefore, these higher--order inputs 
consist of perturbative parts
multiplying the fundamentally nonperturbative $f^{NS}_{q_j,0}(N)$. 
As we will soon discuss, there are other constraints to be satisfied, e.g.
factorization scheme independence of the input for the structure function, 
and this slightly complicates the above picture, but does not change the
main conclusions. 

Thus, it is necessary to express the input as a power series in 
$\alpha_s(Q_I^2)$ in order to make the parton distribution 
$Q_I^2$--independent \foot{Another reason for explicitly writing the input 
as a power series in $\alpha_s(Q_I^2)$ is that
it makes little sense to demand that the starting distribution should
not have a perturbative expansion, unless there is
something special about a particular factorization scheme. 
Any transformation from a scheme in which the starting parton
distribution is purely zeroth order in $\alpha_s(Q_I^2)$ will lead to a
nonperturbative starting distribution multiplied by a power series expansion 
in $\alpha_s(Q_I^2)$.}, but only one intrinsically
nonperturbative input which is $Q_I^2$-independent is needed. 
Usually in analyses of structure functions the parton inputs are 
taken to be a single $\alpha_s(Q_I^2)$--independent function which is 
implicitly allowed to be $Q_I^2$--dependent. 
Phenomenologically, this is normally much the same,
but we stress the formally correct expression for the input here
since it is rather important when constructing properly ordered solutions,
and leads to some predictive power, especially in the small $x$ limit.  
 
Substituting our solution
for the parton distribution into \fnonsing, we obtain the general
expression for the nonsinglet structure functions,
\eqn\fnonsingsol{\eqalign{F^{NS}_i(N,Q^2)=
\biggl[\Bigl(\delta_{i,2} +\sum_{m=1}^{\infty}
C^{NS}_{i,m,l}(N)&\alpha_s^m(Q^2)\Bigr)\sum_{j=1}^{N_f}e_j^2
\sum_{k=0}^{\infty}\alpha_s^k(Q_I^2)
f^{NS}_{q_j,k}(N, Q_I^2)\biggr]\cr
&\biggl({\alpha_s(Q_I^2)\over\alpha_s(Q^2)}\biggr)^{\tilde 
\gamma_{NS}^{0,l}(N)}
\exp\biggl[\sum_{n=1}^{\infty}{(\alpha^n_s(Q_I^2)-\alpha_s^n(Q^2))\over
n}\tilde\gamma_{NS}^{n,l}(N)\biggr].\cr}}
It is clear that this may be written as
\eqn\fnonsingsolalt{F_i^{NS}(N,Q^2)=\biggl({\alpha_s(Q_I^2)\over\alpha_s(Q^2)}
\biggr)^{\tilde \gamma_{NS}^{0,l}(N)}
\sum_{n=0}^{\infty}\sum_{m=0}^nF^{NS}_{i,nm}(N,Q_I^2)\alpha^{n-m}_s(Q_I^2)
\alpha_s^m(Q^2)=\sum_{n=0}^{\infty}F^{NS}_{i,n}(N,Q^2),}
and that, once a choice of renormalization scheme and starting scale
have been made, each of the $F^{NS}_{nm}(N)$ must be invariant
quantities under changes of factorization scheme. Any well--ordered 
calculation of the structure function should include all complete terms in 
\fnonsingsolalt\ up to a given
order in $n$ and $m$, and no partial terms. In practice it is
possible to work to a given order in $n$ including all $m\leq n$, i.e. to
expand to a given order in powers of $\alpha_s(Q^2)$ plus powers of
$\alpha_s (Q_I^2)$, if
the $\tilde \gamma$'s and $C_i$'s are known to this order. 

We shall briefly describe how to construct this ordered solution for the
structure functions, working up from zeroth order. Consider first calculating
$F^{NS}_2(N,Q^2)$ by working
to zeroth order in $C^{NS}_2(N,\alpha_s)$, $\tilde
\gamma_{NS}(N,\alpha_s)$ and the starting
distribution (remembering that this is $Q_I^2$--independent). To this order
\eqn\fnonsingsolzero{F^{NS}_{2,0}(N,Q^2)=\sum_{j=1}^{N_f}e^2_j
f^{NS}_{q_j,0}(N)
\biggl({\alpha_s(Q_I^2)\over\alpha_s(Q^2)}\biggr)^{\tilde
\gamma^{0,l}_{NS}(N)}.}
Using the one--loop expression for the running coupling, as is
appropriate for a lowest--order calculation, each of the quantities 
in this expression is factorization scheme
independent and also renormalization scheme independent.
Therefore we have a consistent leading--order (LO) expression.
If we calculate $F_L^{NS}(N,Q^2)$ the zeroth--order
coefficient function is zero. However, the only contribution for 
$n=1$ in \fnonsingsolalt\ comes from
working to first order in $C^{NS}_{L}(N,\alpha_s)$ and to zeroth order
in $\tilde \gamma_{NS}(N,\alpha_s)$ and the starting distribution.
This leads to the
LO expression for $F_{L}^{NS}(N,Q^2)$ of
\eqn\fnonsingsolzerolong{F^{NS}_{L,1}(N,Q^2)=\alpha_s(Q^2)C^{NS}_{L,1,l}(N)
\sum_{j=1}^{N_f}e^2_j f^{NS}_{q_j,0}(N)
\biggl({\alpha_s(Q_I^2)\over\alpha_s(Q^2)}\biggr)^{\tilde
\gamma^{0,l}_{NS}(N)}.}
Again, using the one--loop expression for the running coupling, every
term in this expression is both factorization scheme and
renormalization scheme independent, giving a well--defined
LO expression.

We now consider the first correction to these expressions.
The first--order expression for the renormalization group equation is 
\eqn\firstorderrg{\alpha_s(Q^2) {d \, f^{NS}_{q_j,1}(N,Q^2) \over d
\alpha_s(Q^2)} = -\tilde \gamma_{NS}^{0,l}(N,\alpha,s)f^{NS}_{q_j,1}(N,Q^2)
- \alpha_s(Q^2)\tilde\gamma_{NS}^{1,l}(N)f^{NS}_{q_j,0}(N,Q^2),}
with solution
\eqn\firstordsol{f^{NS}_{q_j,1}(N,Q^2)=\Bigl[(\alpha_s(Q_I^2)-\alpha_s(Q^2))
\tilde \gamma_{NS}^{1,l}(N) f^{NS}_{q_j,0}(N)+\alpha_s(Q_I^2)
f^{NS}_{q_j,1}(N,Q_I^2)\Bigr]
\biggl({\alpha_s(Q_I^2)\over\alpha_s(Q^2)}\biggr)^{\tilde 
\gamma^{0,l}_{NS}(N)}.}
Multiplying $f^{NS}_{q_j,0}$ by $\alpha_s(Q^2)C^{NS}_{2,1,l}(N)$ and adding
to $f^{NS}_{q_j,1}$, we clearly obtain all terms in the expression 
\fnonsingsolalt\ for $F^{NS}_{2}(N,Q^2))$ at $n=1$. Adding this to 
\fnonsingsolzero\ we obtain the factorization--scheme--independent
expression for $F_2^{NS}(N,Q^2)$ up to $n=1$.
{\it We note that this is not the same
as finding the complete solution to the renormalization group equation
including all terms in the anomalous dimension up to
first order in $\tilde \gamma$ and multiplying the solution by the
coefficient function up to first order. This procedure would involve
the exponentiation of the anomalous dimension, and thus would include
incomplete parts of the $F_{nm}^{NS}(N)$'s for $n\geq 1$, and would be
a factorization--scheme--dependent, and hence physically ambiguous quantity.} 

Similarly to $F^{NS}_{2}(N,Q^2)$ we can obtain the NLO
factorization--scheme--independent expression for $F_L^{NS}(N,Q^2)$. The
expression at $n=2$ is obtained by
adding the product of the first--order coefficient function and
$f^{NS}_{q_j,1}(N,Q^2)$ to the product of the second--order coefficient
function and $f^{NS}_{q_j,0}(N,Q^2)$. The NLO
$F_L^{NS}(N,Q^2)$ is the sum of this and \fnonsingsolzerolong.
When working to NLO for either structure function 
we now have expressions which are
renormalization scheme dependent. This scheme dependence compensates
for the renormalization scheme dependence of the  
two--loop coupling constant (which has a 
renormalization--scheme--dependent value for $\Lambda_{QCD}$), and it is this
expression for the coupling that we should use at this level. Doing so
guarantees the renormalization scheme independence of the structure
functions up to corrections of higher order in $\alpha_s$, i.e.
${\cal O}(\alpha_s^2 F^{NS}_{2(L),0(1)})$.  

It is now simple to see how to construct 
structure functions order by order. Defining the 
$n_{\rm th}$ order renormalization group equation by  
\eqn\nthorderrg{\alpha_s(Q^2) {d \, f^{NS}_{q_j,n}(N,Q^2) \over d
\alpha_s(Q^2)} = -\sum_{m=0}^n\tilde \gamma_{NS}^{m,l}(N,\alpha,s)
f^{NS}_{q_j,n-m}(N,Q^2),}
it is easy to prove by induction that the solution contains all terms
in the full solution with a given sum of powers of $\alpha_s(Q^2)$ and
$\alpha_s(Q_I^2)$ multiplying the everpresent
$(\alpha_s(Q_I^2)/\alpha_s(Q^2))^{\tilde\gamma^{0,l}_{NS}(N)}$ factor. Thus,
$F_{i,n}^{NS}(N,Q^2)$ is given by
\eqn\fnonsingnexp{F_{i,n}^{NS}(N,Q^2)=\sum_{j}^{Nf}e^2_j\sum_{m=0}^n 
C^{NS}_{i,m,l}\alpha_s^m(Q^2)f^{NS}_{q_j,n-m}(N,Q^2).}
The $n_{\rm th}$--order
scheme--independent structure function is then the sum
of the $F_{i,n}^{NS}(N,Q^2)$ up to order $n$. 
Including all $F_{2,m}^{NS}(N,Q^2)$ up to order $n$ is
working to $(n+1)_{\rm th}$ nontrivial order, and requires the $(n+1)$--loop
coupling in order to make the expression renormalization scheme
invariant up to higher orders in $\alpha_s$.
Similarly, including all $F_{L,m}^{NS}(N,Q^2)$ up to order $n$ is
working to $n_{\rm th}$ nontrivial order, and requires the $n$--loop
coupling.

This procedure clearly provides factorization scheme independence and 
renormalization scheme independence for this method of expansion.
We can also see how it relates to the discussion of the
factorization--scheme--invariant evolution equations in terms of the
structure functions. In order to do this let us consider the solution
for the non--singlet structure function $F_2^{NS}(N,Q^2)$ again. We
may rewrite our general solution \fnonsingsol\ in the form 
\eqn\fnonsingsol{\eqalign{F^{NS}_2&(N,Q^2)=
\biggl[\Bigl(1 +\sum_{m=1}^{\infty}
C^{NS}_{2,m,l}(N)\alpha_s^m(Q_I^2)\Bigr)\sum_{j=1}^{N_f}e_j^2
\sum_{k=0}^{\infty}\alpha_s^k(Q_I^2)
f^{NS}_{q_j,k}(N, Q_I^2)\biggr]
\biggl({\alpha_s(Q_I^2)\over\alpha_s(Q^2)}
\biggr)^{\tilde \gamma_{NS}^{0,l}(N)}\times\cr
&\hskip -0.2in \exp\biggl[\sum_{n=1}^{\infty}
{(\alpha^n_s(Q_I^2)-\alpha_s^n(Q^2))\over n}\tilde\gamma_{NS}^{n,l}(N)
+\int_{\ln Q_I^2}^{\ln Q^2}{d \over d \ln q^2}
\ln\Bigl(1 +\sum_{m=1}^{\infty}
C^{NS}_{2,m,l}(N)\alpha_s^m(q^2)\Bigr)d\ln q^2\biggr].\cr}}
This way of writing $F^{NS}_2(N,Q^2)$ is particularly useful since it
separates the solution into the value of the structure function
at $Q_I^2$ (the term in square brackets), and the ratio of its value at
a different $Q^2$ with this initial value (the rest of the expression). 
Clearly these two quantities must be separately 
factorization scheme independent. Also, it is clear that this solution
is of exactly the same form as \nonsingfsisol, and we can express it
simply in terms of an input for $F^{NS}_2(N,Q^2)$ at $Q_I^2$ and a
physical anomalous dimension which governs the evolution, i.e. our
input and our prediction. 

The input and 
the evolution will mix with each other if we make a change in the starting 
scale, however. Examining the effects of 
such a change for the full physical quantity gives us information about 
the form of the input. The lowest--order input for the structure function 
is just $F^{NS}_{2,0}(N)=\sum_{j=1}^{N_f}e_j^2f^{NS}_{q_j,0}(N)$.
Making the change of starting scale and consequently of $\alpha_s(Q_I^2)$ 
already considered, the change in the lowest--order evolution is as in 
\zeroevolcha, and this leads to a change in the lowest--order structure 
function which is of higher order, and which can be absorbed by a change in 
the NLO input for the structure function of
\eqn\chainputsf{\Delta F^{NS}_{2,1}(N,Q_I^2)=\delta b_0 \tilde 
\gamma^{0,l}_{NS}(N) F^{NS}_{2,0}(N).}
In terms of parton distributions and coefficient functions 
\eqn\nonsinginone{F^{NS}_{2,1}(N,Q_I^2)=\sum_{j=1}^{N_f}e^2_j
(f_{q_j,1}^{NS}(N,Q_I^2)+C_{2,1,l}^{NS}(N)f_{q_j,0}^{NS}(N)).}
We chose the change in $f_{q_j,1}^{NS}(N,Q_I^2)$ in \chainput\ so that the 
structure function would be independent of $Q_I^2$, and it is clear that 
that is consistent with \chainputsf\ and \nonsinginone. However, we can now 
say more about the form of $f_{q_j,1}^{NS}(N,Q_I^2)$. Because it is 
the leading 
term in the expression for the input which depends on $\alpha_s(Q_I^2)$, 
$F_{2,1}^{NS}(N,Q_I^2)$ must be renormalization scheme independent. However,
$C_{2,1,l}^{NS}(N)$ is renormalization scheme dependent, so 
$f_{q_j,1}^{NS}(N,Q_I^2)$ must also be renormalization scheme dependent in a 
way such as to cancel this. 
Hence, $f_{q_j,1}^{NS}(N,Q_I^2)$ must not only have a
part like $\ln(Q_I^2)\gamma_{NS}^{0,l}(N)f_{q_j,0}^{NS}(N)$ in order to 
maintain $Q_I^2$--independence of the structure function, but also a 
part like $-C_{2,1,l}^{NS}(N)
f_{q_j,0}^{NS}(N)$ in order to maintain renormalization scheme 
independence, i.e.
\eqn\addon{f^{NS}_{q_j,1}(N,Q_I^2)=(\ln(Q_I^2/A_{NS})\gamma^{0,l}_{NS}(N)-
C^{NS}_{2,1,l}(N))f^{NS}_{q_j,0}(N),}
where $A_{NS}$ is some unknown scale parameter. So we see that
\eqn\addoni{F^{NS}_{2,1}(N,Q_I^2)=\ln(Q_I^2/A_{NS})\gamma^{0,l}_{NS}(N)
F^{NS}_{2,0}(N)\equiv \ln(Q_I^2/A_{NS})\Gamma^{0,l}_{NS}(N)
F^{NS}_{2,0}(N).}
It is 
clear that this does not spoil our argument that $f_{q_j,1}^{NS}(N,Q_I^2)$ 
consists of perturbatively calculable quantities multiplying
the fundamentally nonperturbative $f_{q_j,0}^{NS}(N)$, and
that all the higher--order inputs may be chosen to be perturbative functions 
multiplying this nonperturbative input, and therefore that the input for
the structure function is a perturbative power series 
(depending on the physical anomalous dimension) multiplying the single
nonperturbative factor $f_{q_j,0}^{NS}(N)$, which may also be 
interpreted as a fundamentally nonperturbative input for the 
structure function $F_{2,0}^{NS}(N)$. Hence, demanding invariance of our 
expression for the structure function under changes in $Q_I^2$ leads us to
a power series expression for the input, but with only one (for each quark)
really nonperturbative factor for this input.  We also see that if
$Q_I^2=A_{NS}$, the first--order perturbative correction 
to $F_2^{NS}(N,Q_I^2)$ vanishes. Hence, we might expect $A_{NS}$
to be some scale typical of the transition between perturbative and 
nonperturbative physics, i.e $A_{NS} \lsim 1\Gev$.  

Separating the expression for the structure function into a definite
input and evolution part also enables us to view
the loop expansion in an alternative manner. We see that when
expanding out to $n_{\rm th}$ order in the loop expansion we are including
all terms where the order of the input part for the structure function
added to the order of the evolution part of the structure function
is less than or equal to $n$. 
Writing the solution as in \fnonsingsol\ does, however, also illustrate that
demanding factorization scheme invariance does not on its own force us
into the strictly defined loop expansion. 
It is clear that we could, if we wished,
expand the input and evolution term out to different orders in $\alpha_s$, 
still maintaining factorization scheme independence.  However, this is
not a sensible approach for reasons of renormalization scheme dependence. 
If we were to expand out the input and evolution terms to
different orders in $\alpha_s$ we should really use $\alpha_s$ itself
calculated to a different order in each case, surely a
perverse thing to do. It makes far more sense to combine physical quantities 
to a given order. 

\medskip

The solution for the longitudinal structure function is much the same as for 
$F^{NS}_{2,0}(N,Q^2)$. 
The situation for the singlet structure function is Exactly analogous to that
for the nonsinglet, but is complicated by the fact that we now
have coupled evolution equations for the quark and gluon
distributions. This makes it impossible to write a closed form for the
solution to the renormalization group equations in the way we did for
the nonsinglet case in \fnonsingsol, but the 
equations may be solved order by order in the same way, i.e. 
the lowest--order solution is 
\eqn\evolsolsing{\eqalign{ f^{S}_{0}(N,Q^2) &=
f^{S,+}_{0}(N)\biggl({\alpha_s(Q_I^2) \over 
\alpha_s(Q^2)}\biggr)^{\tilde\gamma^{0,l}_+(N)}+
f^{S,-}_{0}(N)\biggl({\alpha_s(Q_I^2) \over 
\alpha_s(Q^2)}\biggr)^{\tilde\gamma^{0,l}_-(N)},\cr
g_{0}(N,Q^2) &=
g^+_{0}(N)\biggl({\alpha_s(Q_I^2) \over 
\alpha_s(Q^2)}\biggr)^{\tilde\gamma^{0,l}_+(N)}+
g^-_{0}(N)\biggl({\alpha_s(Q_I^2) \over 
\alpha_s(Q^2)}\biggr)^{\tilde\gamma^{0,l}_-(N)},\cr}}
where $\tilde \gamma^{0,l}_+(N)$ and $\tilde \gamma^{0,l}_-(N)$ are the
eigenvalues of the zeroth--order matrix for the $\tilde \gamma$'s, and
$f^{S,+}_{0}(N)+f^{S,-}_{0}(N)= f^S_{0}(N)$ and 
$g^+_{0}(N)+g^-_{0}(N)= g_{0}(N)$.  
As with the nonsinglet quark distributions, the lowest--order 
inputs for the partons are $Q_I^2$--independent.
Also, if we expect any powerlike behaviour to come only from
perturbative effects, then these $Q_I^2$--independent
inputs for the quark and gluon are analytic for $N>0$.

We may also discuss the relationship to the solutions using the evolution
equations for the singlet structure functions. 
Unlike the nonsinglet case it is rather
difficult to see how to express the solution in terms of
factorized inputs and evolution parts; the evolution
parts will not be simple exponentials, as in \fnonsingsol. However,
the solution for the structure functions can be
written as a series of terms each of which can be 
factored into a part dependent only on $\alpha_s(Q_I^2)$ and one
depending on both $\alpha_s(Q_I^2)$ and $\alpha_s(Q^2)$, the latter either
vanishing at $Q^2=Q_I^2$, or being equal to $(\alpha_s(Q_I^2)/
\alpha_s(Q^2))^{\tilde\gamma^{0,l}_{+,-}(-1)}$ and thus equal to 
unity at $Q^2=Q_I^2$. For each of these terms the former part may
be interpreted as an input and the latter part may be interpreted as
the evolution. Within the loop expansion we then include all terms where
the order of the input part plus the order of the evolution part sums
to less than or equal to some integer $n$. 

By examining the form of the 
inputs under a change in starting scale, as with the nonsinglet 
structure functions, we find that the only fundamentally nonperturbative 
parts of the inputs are the zeroth--order parts, $F^S_{2,0}(N)$ and 
$F^S_{L,0}(N)$, with all other inputs being in
principle expressed as perturbative functions of the physical
anomalous dimensions
multiplying one of these nonperturbative components. They are given in terms 
of the parton inputs by
\eqn\linkf{F^S_{2,0}(N)=f^S_0(N),} 
and
\eqn\linkg{F^S_{L,0}(N)= C^f_{L,1}(N)f^S_0(N)+C^g_{L,1}(N)g_0(N).}
We choose to think of the expressions for the structure functions as the real
nonperturbative functions, since they are, of course, the physically relevant 
quantities.

The extra complexity of the solution in the singlet case, when compared to
the nonsinglet case, also means it
is far from obvious how to express the physical anomalous dimensions in
terms of the parton anomalous dimensions and coefficient functions
simply by comparing the forms of the solutions in the partonic
language and purely in terms of structure functions. However, the
expressions for the physical anomalous dimensions can
be found at each order using \physanom. 
One could then solve for the structure functions in a manner different
from the loop expansion, e.g. calculate the solution 
to the whole evolution equation for
the structure functions using the physical anomalous dimensions up to a given
order. This would be by definition factorization scheme invariant, but
would only be renormalization scheme invariant up to the same order in
the solution as the order of the anomalous dimensions. The
rest of the solution would contain only a subset of the possible terms
of a given form obtained from the full calculation. This subset will not 
necessarily be a good representation of the full set of terms obtained at
this order, and would therefore
have no real significance, and should be dropped. Hence, the
evolution equations in terms do nothing to alter the
strict ordering of the solution using this method of expansion.     

\bigskip

This whole discussion of the order--by--order--in--$\alpha_s$ 
expansion scheme
is perhaps a little academic since the errors invoked in performing the 
calculations
without paying strict heed to the formally correct procedures 
are rather small. For example, when using this standard method of expansion 
only two different factorization schemes are generally considered, the 
$\overline{\hbox{\rm MS}}$ scheme and the DIS scheme; the latter related
to the former by a change of parton
distributions, so that the singlet quark distribution is equal to the
singlet structure function. 
If a calculation is made in one scheme to a well--defined order, 
and then a transformation to the other
scheme made correctly, precisely the same result will be
obtained. In practice, small differences are 
noticed between calculations
using different schemes within 
the loop expansion to NLO, but these come 
from well--understood sources. One
common source is that the starting distributions in both schemes are
described by a simple functional form, e.g.
\eqn\start{xf(x,Q_I^2) = A(Q_I^2)x^{-\lambda(Q_I^2)}(1-x)^{\eta(Q_I^2)}
(1+\epsilon(Q_I^2) x^{1/2}+\gamma(Q_I^2) x),}
rather than the formally correct expression of a power series in 
$\alpha_s(Q_I^2)$ with essentially perturbative coefficients convoluted with
a $Q_I^2$--independent nonperturbative function of $x$. 
If the starting distribution is of the
form above in one scheme, then in the other
scheme it will not be modelled precisely by a function of the
same form. However, the error is in general small.
Alternatively, if the
calculations are not done in a well--ordered manner, 
then differences between the calculation done in the two schemes, or
between this type of calculation and the correct NLO
calculation, will
be of NNLO. Again this usually results in
only small differences.\foot{For a comparison of calculations done 
at NLO using different methods see \ref\comparison{J. Bl\"umlein 
{\it et al}, Proc. of the 1996 HERA Physics Workshop, eds. G. Ingelman,
R. Klanner, and A. De Roeck, DESY, Hamburg, 1996, Vol. 1, p. 23.}.}
Similarly, small differences would be
obtained by working in two different renormalization schemes.

Nevertheless, there are a couple of points we wish to make here,
concerning the loop expansion. Firstly we note that 
if one considers the input to be consistent with the loop expansion, e.g. 
NLO--order evolution should be accompanied by a NLO input, 
the power of $\lambda(Q_I^2)$ should not correspond to parton 
distributions much 
steeper than flat for the singlet quark or gluon at this order. 
This is because the  first--order--in--$\alpha_s(Q_I^2)$ 
input should be accompanied by no more than a single power 
of $\ln(1/x)$.  
Restricting $\lambda(Q_I^2)$ in this way is rather important 
for the fits to the low $x$ data, and would mean that NLO fits to small--$x$ 
data would be very poor. The only way to avoid this is to let
$\lambda(Q_I^2)$ be an artificial 
free parameter, in which case, if \start\ describes the 
singlet quark density, it must be $\sim 0.2-0.3$ for practically any $Q_I^2$. 
This value is totally unjustified, and the need for this steepness in the 
input for the quark is a clear sign of the limited usefulness of the 
NLO--in--$\alpha_s$ calculation at small $x$. 

There is another reason for being 
concerned about the validity of the loop expansion at small $x$.
The reason for the relative smallness of the differences between inaccurately
performed NLO calculations noted above, even at small $x$, is
that the differences between these calculations do not contain terms
which are any more leading in $\ln (1/x)$ than the NLO
calculation itself. Hence they are genuinely an order of $\alpha_s$ down on
the NLO calculation, with no small--$x$
enhancement. This is also true for calculations done in different
renormalization schemes (or at different renormalization scales).
However, the real NNLO
contribution are higher order in $\alpha_s$ but also contain
terms at higher order in $\ln(1/x)$, and  are potentially
large at small $x$. Hence, the relative
insensitivity of structure functions to changes in
renormalization or factorization scheme for
calculations which are not carefully ordered 
is no guarantee that genuine higher--order
corrections will be small at small $x$ when using the loop
expansion. Indeed we would naively expect them to be large.  

\medskip

In contrast to the insensitivity when using the loop expansion, when
using the leading--$\ln (1/x)$
expansion very large differences between calculations done in a large
number of different factorization schemes have been noted. This is an clear 
sign that the calculations are not being done in a well--ordered
manner, and that the ambiguity introduced by lack of care in the 
calculations is
greater in this method of expansion than the standard loop expansion.   
We will now demonstrate that this is indeed the case.

\subsec{The Leading--$\ln (1/x)$ Expansion.}

It should clearly be possible to define a well--ordered expansion in 
leading powers of $\ln (1/x)$, or
equivalently, in leading powers of $1/N$ in moment space. We will now
demonstrate that this is indeed the case. As in the loop expansion, we
will first work in terms of the traditional parton distribution
functions and coefficient functions, and see how this results in
expressions containing the physical anomalous dimensions. Doing this
enables us to see how large factorization scheme dependence can arise
when calculating less carefully within this expansion scheme. 

First we must set up our notation. In this expansion scheme we write,
\eqn\tildegamgen{\tilde\gamma(N,Q^2) =
\sum_{n=0}^{\infty}\alpha_s^n(Q^2)\sum_{m=1-n}^{\infty}\tilde
\gamma^{nm}\alpha_s^{m}(Q^2) N^{-m} \equiv
\sum_{n=0}^{\infty}\alpha_s^n(Q^2)\tilde\gamma^n(\alpha_s(Q^2)/N).}
In particular, in the $\overline{\hbox{\rm MS}}$ renormalization and
factorization scheme we may write
\eqn\tildegamglu{\tilde \gamma_{gg}(N,\alpha_s(Q^2)) = \sum_{n=0}^{\infty}
\alpha_s^n(Q^2) \tilde\gamma_{gg}^n(\alpha_s(Q^2)/N),}
where the series expansion for $\tilde\gamma^0_{gg}(\alpha_s/N)$ is
known to all orders (all the coefficients being positive). This expression for
$\tilde\gamma^0_{gg}(\alpha_s/N)$ is renormalization scheme independent.
$\tilde\gamma^1_{gg}(\alpha_s/N)$ is of 
course renormalization
scheme dependent, since it must change to absorb part of the effect of the
${\cal O}(\alpha_s^2)$ change in the coupling on
$\tilde\gamma^0_{gg}(\alpha_s/N)$. The renormalization scheme
independence is also true for $\tilde\gamma^0_{gf}(\alpha_s/N)$, which
obeys \gammadef, and also for $\tilde\gamma^0_{ff}(\alpha_s/N)$. 
There is also a
renormalization--scheme--independent relationship between
$\tilde\gamma^1_{ff}(\alpha_s/N)$ and $\tilde\gamma^1_{fg}(\alpha_s/N)$,
which tells us that 
\eqn\gamrel{\tilde\gamma^1_{ff}(\alpha_s/N)={4\over 9}
\biggl(\tilde\gamma^1_{fg}(\alpha_s/N)-{2N_f
\over 6\pi b_0}\biggr),}
where the
second term in the brackets is the one--loop contribution to
$\tilde\gamma^1_{fg}(\alpha_s/N)$. It is also known that neither the 
nonsinglet anomalous
dimension nor the nonsinglet
coefficient function have poles at $N=0$.
Hence, the nonsinglet sector makes very little
contribution to the structure function at small $x$, and as such we
will ignore it for the remainder of this section. 

A general change in factorization scheme may be expressed by writing
an element of the transformation matrix $U$ as 
\eqn\Udef{U_{ab}(N,\alpha_s(Q^2)) = \sum_{n=0}^{\infty} \alpha_s^n(Q^2)
\sum_{m=1-n}^{\infty}
U^{nm}_{ab}\alpha_s^m(Q^2) N^{-m}\equiv
\sum_{n=0}^{\infty}\alpha_s^n(Q^2)U^n_{ab}(\alpha_s(Q^2)/N),}
with condition on the $U^{nm}_{ab}$ such that $U$ obeys
$U_{ab}=\delta_{ab} +{\cal O}(\alpha_s)$. 
A change of factorization scheme with $U^0_{ab}\not= 0$ can 
introduce scheme dependence into the $\tilde
\gamma^0_{ab}(\alpha_s/N)$'s, as we see from \transgamma.
Insisting that $\tilde\gamma^0_{gg}(\alpha_s/N)$,
$\tilde\gamma^0_{gf}(\alpha_s/N)$,
$\tilde\gamma^0_{ff}(\alpha_s/N)$ and $\tilde\gamma^0_{fg}(\alpha_s/N)$
are unaltered by scheme changes leads to the requirement,
\eqn\urequire{\eqalign{&U^0_{fg}(N,\alpha_s)=0, \hskip 1in 
U^0_{ff}(N,\alpha_s)=1,\cr
&U^0_{gg}(\alpha_s/N) = 1
+\sum_{n=1}^{\infty}U^{0,n}_{gg}(\alpha_s(Q^2)/N)^n,
\hskip 0.5in U^0_{gf}(\alpha_s/N)=\fourninths(U^0_{gg}(\alpha_s/N)-1).\cr}}
This requirement also preserves the relationship \gamrel\ between
$\tilde\gamma^1_{ff}(\alpha_s/N)$ and
$\tilde\gamma^1_{fg}(\alpha_s/N)$. For simplicity, and due to
factorization scheme invariance of physical quantities, we will only
consider factorization scheme changes away from $\overline {\hbox{\rm
MS}}$ scheme of the type \urequire. The $U^n_{ab}(\alpha_s/N)$ for $n> 0$
will have no restrictions.

Restricting ourselves to these schemes we may write 
\eqn\defcstwo{C^S_2(N,\alpha_s) = 1 +
\sum_{n=1}^{\infty}\alpha_s^n(Q^2)\sum_{m=1-n}^{\infty}
C^{S}_{2,n,m}\alpha_s^m(Q^2)N^{-m} \equiv
1+\sum_{n=1}^{\infty}\alpha_s^n(Q^2) C^S_{2,n}(\alpha_s/N),}
and all other coefficient functions as
\eqn\defcoeffs{C^a_i(N,\alpha_s) = 
\sum_{n=1}^{\infty}\alpha_s^n(Q^2)\sum_{m=1-n}^{\infty}
C^{a}_{i,n,m}\alpha_s^m(Q^2)N^{-m} \equiv
\sum_{n=1}^{\infty}\alpha_s^n(Q^2) C^a_{i,n}(\alpha_s/N).}
All the $C^a_{i,n}(\alpha_s/N)$ are both renormalization--scheme-- and
factorization--scheme--dependent quantities. Indeed, all of the
$C^{a}_{i,n,m}$ are renormalization scheme and
factorization scheme dependent, except for the $C^a_{L,n,1-n}$, which
come from the one--loop longitudinal coefficient functions which, as
we saw in the previous subsection, are totally scheme independent. 
There are also two
renormalization and factorization scheme (with our restrictions)
independent relationships between coefficient
functions:
\eqn\coeffreltwo{C^S_{2,1}(\alpha_s/N) ={4\over 9}\Bigl(C^g_{2,1}(\alpha_s/N)
-C^g_{2,1,0}\Bigr),}
where the second term in brackets is the one--loop
contribution to $C^g_{2,1}(\alpha_s/N)$, which is itself
renormalization scheme and factorization scheme dependent (being equal
to $(N_f/6\pi)$ in $\overline{\hbox{\rm MS}}$ scheme); 
and 
\eqn\coeffrellong{\Bigl(C^S_{L,1}(\alpha_s/N)-{2\over 3\pi}\Bigr) 
={4\over 9}\Bigl(C^g_{2,1}(\alpha_s/N)-{2N_f\over 6\pi}\Bigr),}
where the second terms in the
brackets are the one--loop contributions to $C^S_{L,1}(\alpha_s/N)$
and $C^g_{L,1}(\alpha_s/N)$, both of which are renormalization and
factorization scheme independent.  

Working in an arbitrary factorization scheme 
(up to the above restrictions) and using the general
expressions for the $\tilde\gamma$'s and coefficient functions, we may
find expressions for the structure functions. The first step towards
this is solving the renormalization group equations for the parton 
distributions. The lowest--order part of the equation is,
\eqn\llevolzero{\alpha^2_s(Q^2) {d\over d \alpha_s(Q^2)} 
\pmatrix{f^S_{0}(N,Q^2)\cr g_0(N,Q^2)\cr} =-
\pmatrix{0 & 0\cr 
\fourninths\tilde \gamma^0_{gg}(\alpha_s/N) 
& \tilde \gamma^0_{gg}(\alpha_s/N)\cr}
\pmatrix{f^S_{0}(N,Q^2)\cr g_0(N,Q^2)\cr}.}
This may easily be solved to give
\eqn\solzero{\eqalign{f^S_0(N,Q^2)&=f^S_0(N,Q_I^2),\cr
g_0(N,Q^2)&= (g_0(N,Q_I^2)+\fourninths f_0^S(N,Q_I^2))\eigplus 
-\fourninths f_0^S(N,Q_I^2).\cr}}
This is analogous to the lowest--order solution within the loop
expansion and contains two factors, one of which must appear in all the
higher terms in the expansion: instead of
$(\alpha_s(Q_I^2)/\alpha_s(Q^2))^{\tilde\gamma^{0,l}_{+,-}(N)}$,
corresponding to the two eigenvalues of $\tilde\gamma^{0,l}(N)$ in the loop
expansion, we have $\eigplus$ and $1$, corresponding to the two
eigenvalues of $\tilde\gamma^0(\alpha_s/N)$ in the leading--$\ln (1/x)$
expansion, i.e. $\tilde\gamma^0_{gg}(\alpha_s/N)$ and $0$. 

The fact that the leading order part of $\tilde \gamma_{gg}$ and 
$\tilde \gamma_{gf}$ is more leading than the leading part of $\tilde 
\gamma_{fg}$ and $\tilde \gamma_{ff}$, in this expansion scheme makes 
obtaining a well--ordered solution rather more complicated than in the 
loop--expansion. (The same is true when using the physical anomalous 
dimensions since $\tilde \Gamma_{2L}$ and $\tilde \Gamma_{22}$ both equal 
$0$.) Hence, before using our solutions for the parton densities to 
construct expressions for the structure functions we will need to
solve higher--order renormalization group equations in order to
determine the general form of the solutions. This is done iteratively 
using evolution equation ordered analogously to \nthorderrg, and in order to 
get useful solutions we must go out to $n = \infty$, keeping only the 
most leading terms each time we iterate. With some work and a 
straightforward inductive proof we obtain
\eqn\lof{\eqalign{f^S(N,Q^2)=& \alpha_s(Q^2)
{\tilde\gamma^1_{fg}(\alpha_s(Q^2)/N)\over
\tilde\gamma^0_{gg}(\alpha_s(Q^2)/N)}(\eigvecplus+\tilde g_0(N,Q_I^2))\times\cr
&\hskip 0.5in \eigpluscor\cr
&+f^S_0(N)\biggl({\alpha_s(Q_I^2)\over
\alpha_s(Q^2)}\biggr)^{-\fourninths{2N_f\over 6\pi b_0}}+
\tilde f^S_1(N,Q_I^2)+\higherorder,\cr}}
and,
\eqn\loglu{\eqalign{g(N,Q^2)=& (\eigvecplus+\tilde g_0(N,Q_I^2))\times\cr
&\hskip 0.2in \eigpluscor\cr
&-\fourninths f^S_0(N)\biggl({\alpha_s(Q_I^2)\over
\alpha_s(Q^2)}\biggr)^{-\fourninths{2N_f\over 6\pi b_0}}+ \higherorder.\cr}}
We may also write the expression for the rate of change of the quark
distribution 
\eqn\loderivf{\eqalign{-\alpha_s^2(Q^2) {d\, f^S(N,Q^2)\over
d\alpha_s(Q^2)}
=& \alpha_s(Q^2)
\tilde\gamma^1_{fg}(\alpha_s(Q^2)/N)(\eigvecplus+\tilde g_0(N,Q_I^2))\times\cr
&\eigpluscor\cr
&-\fourninths \alpha_s(Q^2)
f^S_0(N){2N_f\over 6\pi b_0} \biggl({\alpha_s(Q_I^2)\over
\alpha_s(Q^2)}\biggr)^{-{4\over 9}{2N_f\over 6\pi b_0}}+\higherorder.\cr}}
This last expression will be important since at leading order
$\alpha_s^2(Q^2)(d\,F_2(N,Q^2)/d\alpha_s(Q^2))$ is directly related to 
$d\, F_2(N,Q^2)/d\,\ln Q^2$ which  we will wish to study as well as
$F_2(N,Q^2)$. 

The input $g_0(N,Q_I^2)$ is chosen so that  
the change in $\eigplus$ under a change in $Q_I^2$ can be compensated 
for by a change in
$g_0(N,Q_I^2)$ up to corrections of higher order. Hence, the gluon input
may be written as 
\eqn\inexp{\eqalign{(\eigvecplus+&\tilde g_0(N,Q_I^2))\cr
&\hskip -0.3in =(\eigvecplus)
\biggl(1+\sum_{m=1}^{\infty}\tilde g_{0,m} \biggl({\alpha_s(Q_I^2)\over N}
\biggr)^m\biggr)
\exp[\ln(Q_I^2/A_{gg})\gamma^0_{gg}(\alpha_s(Q_I^2)/N)],\cr}}
where $A_{gg}$ is an unknown scale. 
The series $\sum_{m=1}^{\infty}\tilde g_{0,m}(\alpha_s(Q_I^2)/ N)^m$ 
is at yet undetermined, but is potentially renormalization and 
factorization scheme dependent. It will only be determined when we come to 
construct the structure functions themselves. 
In the same manner we can determine the general form of our input, 
$f_1^S(N,Q_I^2)$, obtaining
\eqn\inexpfc{f^S_1(N,Q_I^2)=(\eigvecplus +\tilde g_0(N,Q_I^2))\alpha_s(Q_I^2)
{\gamma^1_{fg}(\alpha_s(Q_I^2)/N)\over\gamma^0_{gg}(\alpha_s(Q_I^2)/N)}
+\tilde f^S_1(N,Q_I^2),}
where
\eqn\ftildedef{\tilde f^S_1(N,Q_I^2)=N\sum_{n=0}^{\infty}\tilde f^S_{1,m}
\biggl({\alpha_s(Q_I^2)\over N}\biggr)^m(\eigvecplus)}
is potentially
renormalization and factorization scheme dependent, but not yet determined. 

We also consider the form of the $N$--dependence of our inputs. If we take
the point of view that any steep behaviour in the parton distributions
only comes about due to perturbative effects, then we assume that
$f^S_0(N)$ and $g_0(N)$ are both soft, i.e. either flat or even
valence--like when the transform to $x$--space is performed (or at most
going like a finite, small power of $\ln(1/x)$). This
requires that they both be analytic for $N> 0$. As in 
the loop expansion, we may think of $g_0(N)$, and $f^S_0(N)$ as 
fundamentally soft, nonperturbative parts of the input. \foot{Of course, we 
should be trying to find fundamentally nonperturbative inputs for the 
structure functions rather than the partons, as discussed for the 
loop expansion. However, in this expansion scheme $f^S_0(N)$ and $g_0(N)$ are
trivially related to $F_{2,0}(N)$ and $\hat F_{L,0}(N)$ as we will see
soon.} The parts multiplying 
these are then really determined by perturbation theory.  
Since we are meant to be expanding our solution for the structure functions in
powers of both $\alpha_s$ and $N$, it might be argued that we should expand
$f^S_0(N)$ and $g_0(N)$ in powers of $N$. We feel this is not really 
appropriate since it is the perturbative part of the solution for which we 
are able to solve, and thus which we are able to order correctly, 
and in the expressions 
for the structure functions the whole of the nonperturbative 
inputs should multiply the well--ordered perturbative parts of the solution.

\medskip

It is now possible to examine the form of the solutions for the
structure functions. We can construct the leading part of
the full solutions by combining our solutions for the
parton distributions with the zeroth-- and first--order
coefficient functions. This gives  
\eqn\fullfl{\eqalign{F_L(N,Q^2)&= \alpha_s(Q^2)C^g_{L,1}(\alpha_s(Q^2)/N)
(\eigvecplus+\tilde g_0(N,Q_I^2))\times \cr
&\hskip 0.2in \eigpluscor\cr
&+\alpha_s(Q^2) (C^S_{L,1,0}-\fourninths C^g_{L,1,0})f^S_0(N)
\biggl({\alpha_s(Q_I^2)\over
\alpha_s(Q^2)}\biggr)^{-\fourninths{2N_f\over 6\pi b_0}}+\higherorder,\cr}}
where \coeffrellong\ has been used. Also
\eqn\fullftwo{\eqalign{F_2(N,Q^2)&= 
\alpha_s(Q^2)
{\tilde\gamma^1_{fg}(\alpha_s(Q^2)/N)\over
\tilde\gamma^0_{gg}(\alpha_s(Q^2)/N)}+
C^g_{2,1}(\alpha_s(Q^2)/N))
(\eigvecplus +\tilde g_0(N,Q_I^2))\times\cr
&\eigpluscor\cr
&+f^S_0(N)\biggl({\alpha_s(Q_I^2)\over
\alpha_s(Q^2)}\biggr)^{-\fourninths{2N_f\over 6\pi b_0}}+\tilde f^S_1(N,Q_I^2)
+\higherorder,\cr}}
and 
\eqn\fullftwoderiv{\eqalign{-\alpha_s^2(Q^2){d\,F_2(N,Q^2)\over
d\alpha_s(Q^2)}= &(\alpha_s(Q^2)
\tilde\gamma^1_{fg}(\alpha_s(Q^2)/N)+\cr
&\hskip -0.35in \alpha_s(Q^2)
C^g_{2,1}(\alpha_s(Q^2)/N)\tilde\gamma^0_{gg}(\alpha_s(Q^2)/N))
(\eigvecplus+\tilde g_0(N,Q_I^2))\times \cr
&\eigpluscor\cr
&\hskip -0.3in -\alpha_s(Q^2){4\over 9}
{2N_f\over 6\pi b_0}f^S_0(N)\biggl({\alpha_s(Q_I^2)\over
\alpha_s(Q^2)}\biggr)^{-{4\over 9}{2N_f\over 6\pi b_0}}+\higherorder,\cr}}
where, we drop the superscript $S$ for the structure functions for
the rest of this subsection.

Each of these expressions will be factorization
scheme independent since each represents the expression for a physical
quantity up to corrections of a form different from the terms explicitly
appearing and which we have deemed to be higher order in our expansion scheme.
However, they are not in themselves the leading order expressions. In order
to derive these we begin with the 
longitudinal structure function. Making the definition
\eqn\phioneplus{\Phi^+_1(Q^2,Q_I^2)=\!\!\lneigcomiC,}
where $\Phi^+_1(Q^2,Q_I^2)$ is factorization--scheme independent
(as can be checked using the rules \transpart--\transgamma)
and $\Phi^+_1(Q_I^2,Q_I^2)=1$, we may
factorize \fullfl\  completely into input
and evolution parts, i.e. 
\eqn\fullflalt{\eqalign{F_L(N,Q^2)&= \alpha_s(Q_I^2){2N_f\over 6\pi}
(\eigvecplus)\eigplus\times \cr
&\biggl({C^g_{L,1}(\alpha_s(Q_I^2)/N)\over
C^g_{L,1,0}}.\biggl(1+{\tilde g_0(N,Q_I^2)\over \eigvecplus}\biggr)\biggr)
\exp\biggl[\Phi^+_1(Q^2,Q_I^2)-\ln\biggl({\alpha_s(Q_I^2)
\over \alpha_s(Q^2)}\biggr)\biggr]\cr
&+\alpha_s(Q^2) \biggl({18-4N_f\over 27\pi}\biggr)f^S_0(N)
\biggl({\alpha_s(Q_I^2)\over
\alpha_s(Q_I^2)}\biggr)^{-\fourninths{2N_f\over 6\pi b_0}-1} + 
\higherorder.\cr}}
It is now possible to attach direct physical significance to each of the
factorization--scheme--independent pieces appearing in this
expression. 

We first consider the inputs. Going to \fullftwo\ for the 
moment we see that
$f^S_0(N)$ is the only term in $F_2(N,Q_I^2)$ which is zeroth--order 
in our expansion scheme. As such it is the
zeroth order input for $F_2(N,Q^2)$, and we may write 
\eqn\defftwozero{f^S_0(N)=F_{2,0}(N),}
and this is one of our two fundamentally nonperturbative inputs.
Also, from \fullflalt\ we see that
the total $\alpha_s(Q_I^2)$--independent input for $F_L(N,Q_I^2)$ (once
we have divided out a single power of $\alpha_s(Q_I^2)/(2\pi)$)
is
\eqn\defflzero{{2N_f \over 3}(g_0(N)+{2\over N_f} f^S_0(N))=
\hat F_{L,0}(N).}
This is therefore equal to our other fundamentally nonperturbative physical 
input
We also make a similar definition for the part of the input 
for $\hat F_L(N,Q^2)$ which is of the form
$\sum_{n=1}^{\infty}a_n(Q_I^2)(\alpha_s(Q_I)/N)^n$ multiplying 
$(\eigvecplus)$, or more correctly multiplying $(\hat F_{L,0}(N)-((36-8N_f)/27)
F_{2,0}(N))$, to complete our
definition of the input in \fullflalt\ and write
\eqn\defflone{\eqalign{{2N_f\over 3}(g_0(N)+\fourninths f^S_0(N)+
\tilde g_0(N,Q_I^2)) &\biggl({C^g_{L,1}(\alpha_s(Q_I^2)/N)
\over C^g_{L,1,0}}\biggr)\cr
&=\hat F_{L,0}(N)-\biggl({36-8N_f\over 27}\biggr)F_{2,0}(N)
+\tilde {\hat F}_{L,0}(N,Q_I^2).\cr}}
This whole expression must be both factorization scheme and 
renormalization scheme independent, facts which reveal information 
about the form of the gluon 
input. $C^g_{L,1}(N,\alpha_s(Q_I^2))$ is both renormalization and 
factorization--scheme dependent, so the scheme dependence of $\tilde 
g_0(N,Q_I^2)$ must be precisely so as to cancel this out, 
Hence, there is no reason 
for $\tilde {\hat F}_{L,0}(\alpha_s(Q_I^2)/N)$ to depend on the 
leading--order longitudinal gluon coefficient function at all, 
and indeed, a natural 
choice seems to be that $(1+\sum_{m=1}^{\infty}\tilde g_{0,m}
(\alpha_s(Q_I^2)/N)^m)$ is chosen equal to 
$(C^g_{L,1,0}/C^g_{L,1}(N,\alpha_s(Q_I^2)))$ (it is difficult to see
what else it could be chosen equal to), 
and therefore 
\eqn\inexpfl{\eqalign{\hat F_{L,0}(N) -\biggl(&{36-8N_f\over 27}\biggr)
F_{2,0}(N)+\tilde {\hat F}_{L,0}(N,Q_I^2)\cr
&=\biggl(\hat F_{L,0}(N) 
-\biggl({36-8N_f\over 27}\biggr)F_{2,0}(N)\biggr)
\exp[\ln(Q_I^2/A_{LL})\gamma^0_{gg}(\alpha_s(Q_I^2)/N)].\cr}} 
Hence, we have a prediction for the input for the longitudinal structure 
function at small $x$ in terms of the nonperturbative inputs
and some scale $A_{LL}$ (where $A_{LL}=A_{gg}$ from the previous subsection).
As with $A_{NS}$ earlier, $A_{LL}$ is the scale at which the input is equal to 
the nonperturbative input alone, and hence we would expect it to be typical
of the scale where perturbation theory starts to break down. 
$Q_I^2$ is a completely free parameter. We have constructed the solution 
to be formally insensitive to 
$Q_I^2$ at leading order, but there is clearly some
residual $Q_I^2$--dependence. Hence, there 
will also be some optimum $Q_I^2$ to choose as the starting scale. 

We may now examine the terms governing the evolution. $\eigplus$ 
is the scheme--independent factor
governing the small--$x$ growth with $Q^2$, and we make the definition 
\eqn\phizeroplus{\Phi^+_0(Q^2,Q_I^2) =
\int_{\alpha_s(Q^2)}^{\alpha_s(Q_I^2)}{\tilde
\gamma^0_{gg}(\alpha_s(q^2)/N)\over
\alpha^2_s(q^2)}d\alpha_s(q^2).}
Of course, the other evolution factor resulting from the eigenvalues
of the zeroth--order anomalous dimension was simply unity and as such
$\Phi^-_0(Q^2,Q^2_I)$ does exist, but
is implicitly zero. However, there is a
correction to this factor of unity in \fullfl--\fullftwoderiv, and we make
the definition 
\eqn\phioneminus{\Phi^-_1(Q^2,Q_I^2)=-{4\over 9}{2N_f\over
6\pi b_0}\ln\biggl({\alpha_s(Q_I^2)\over \alpha_s(Q^2)}\biggr).}

Having made these factorization--scheme--invariant definitions for the
inputs and evolution we may write the solution for $F_L(N,Q^2)$ as
\eqn\fullflalti{\eqalign{F_L(N,Q^2)&= {\alpha_s(Q_I^2)
\over 2\pi}(\hat F_{L,0}(N)+\tilde {\hat F}_{L,0}(N,Q_I^2)
-\biggl({36-8N_f\over 27}\biggr)F_{2,0}(N))\times\cr
&\exp\biggl[\Phi^+_0(Q^2,Q_I^2)+\Phi^+_1(Q^2,Q_I^2)-\ln\biggl(
{\alpha_s(Q_I^2)\over \alpha_s(Q^2)}\biggr)\biggr]\cr
&+{\alpha_s(Q_I^2) \over 2\pi} \biggl({36-8N_f\over 27}\biggr)F_{2,0}(N)
\exp\biggl[\Phi^-_1(Q^2,Q_I^2)-\biggl(\ln\biggl(
{\alpha_s(Q_I^2)\over \alpha_s(Q^2)}\biggr)\biggr)\biggr]\cr
&+\higherorder.\cr}}
This is still of mixed order, but
it is now relatively obvious how we may separate out the ``leading
part'' from this expression. $[\Phi^+_1(Q^2,Q_I^2)-\ln(
\alpha_s(Q_I^2)/ \alpha_s(Q^2))]$ contains 
the same type of terms as $\Phi^+_0(Q^2,Q_I^2)$, but each is a power of
$N$ higher. Thus $[\Phi^+_1(Q^2,Q_I^2)-\ln(
\alpha_s(Q_I^2)/ \alpha_s(Q^2))]$ is subleading to
$\Phi^+_0(Q^2,Q_I^2)$, and indeed $\Phi^+_1(Q^2,Q_I^2)$ is a 
renormalization--scheme--dependent
quantity, as it must be in order to absorb the change in 
$\Phi^+_0(Q^2,Q_I^2)$ resulting from 
a change in the definition of $\alpha_s(Q^2)$ under a change 
in renormalization scheme when working beyond leading order. 
Hence, we should factor $\exp[\Phi^+_1(Q^2,Q_I^2)-\ln(
\alpha_s(Q_I^2)/ \alpha_s(Q^2))]$ out of the first term. Since
$[\Phi^-_1(Q^2,Q_I^2)-\ln(\alpha_s(Q_I^2)/ \alpha_s(Q^2))]$ is of the 
same form as the zeroth--order--in--$N$ 
part of $\Phi^+_1(Q^2,Q_I^2)$, then it 
should also be factored out of the leading--order expression for
$F_L(N,Q^2)$. Thus, we are left with 
\eqn\lofl{\eqalign{F^0_L(N,Q^2)= &{\alpha_s(Q_I^2)\over 2\pi}
(\hat F_{L,0}(N)+\tilde {\hat F}_{L,0}(N,Q_I^2)-
\biggl({36-8N_f\over 27}\biggr)F_{2,0}(N))
\exp[\Phi^+_0(Q^2,Q_I^2)]\cr
&+{\alpha_s(Q_I^2)\over 2\pi} 
\biggl({36-8N_f\over 27\pi}\biggr)F_{2,0}(N).\cr}}
Using the one--loop running coupling constant, as is appropriate for a
leading--order expression, the whole of \lofl\ is not only manifestly
factorization scheme independent, but also renormalization scheme
independent, as one would hope. Also, $\tilde {\hat F}_{L,0}(N,Q_I^2)$ is 
constructed precisely so as to make the expression unchanged,
at this order, under a change in
starting scale. Hence, within this expansion scheme \lofl\
is genuinely the leading--order expression for $F_L(N,Q^2)$.

We now turn our attention to the more phenomenologically important
case of the structure function $F_2(N,Q^2)$. 
For technical simplicity we begin with the expression for
$\alpha^2_s(Q^2)(d\,F_2(N,Q^2)/d\alpha_s(Q^2))$. 
Using the definitions introduced in
our discussion for the longitudinal structure function we may write
\fullftwoderiv\  as
\eqn\fullftwoderivalt{\eqalign{-\alpha^2_s(Q^2){d \, F_2(N,Q^2)\over
d\alpha_s(Q^2)} =& (\alpha_s(Q^2)\tilde\gamma^1_{fg}(\alpha_s(Q^2)/N)+
\alpha_s(Q^2)C^g_{2,1}(\alpha_s(Q^2)/N)\tilde\gamma^0_{gg}(\alpha_s(Q^2)/N))
\times\cr 
&\hskip -1in {3\over 2N_f}\biggr({C^g_{L,1,0}\over
C^g_{L,1}(\alpha_s(Q^2)/N)}\biggl)(\hat F_{L,0}(N) +\tilde {\hat 
F}_{L,0}(N,Q_I^2)-\biggl({36-8N_f\over 27}\biggr)F_{2,0}(N))\times\cr
& \hskip -1in \exp[\Phi^+_0(Q^2,Q_I^2)+\Phi_1^+(Q_I^2,Q^2)] 
-\alpha_s(Q^2){4\over 9}{2N_f \over
6\pi b_0}F_{2,0}(N)\exp[\Phi_1^-(Q^2,Q_I^2)]\cr
& + \higherorder.\cr}}
The factorization scheme independence of this complete expression  
guarantees that the term $(\tilde\gamma^1_{fg}(\alpha_s(Q^2)/N)+
C^g_{2,1}(\alpha_s(Q^2)/N)\tilde\gamma^0_{gg}(\alpha_s(Q^2)/N)) 
(C^g_{L,1,0}/C^g_{L,1}(\alpha_s(Q^2)/N))$ is a 
factorization--scheme--invariant quantity. Indeed, it was
shown by Catani and Hautmann that  
$(\tilde\gamma^1_{fg}(\alpha_s(Q^2)/N)+
C^g_{2,1}(\alpha_s(Q^2)/N)\tilde\gamma^0_{gg}(\alpha_s(Q^2)/N))$ and
$ C^g_{L,1}(\alpha_s(Q^2)/N)$ could each always be expressed
in terms of the product of a factorization--scheme-- and 
renormalization--scheme--independent factor (which they calculated) 
and a scheme--dependent factor, where the
scheme--dependent part is the same for both \cathaut. Using their results 
it is a trivial matter to find that 
\eqn\gamtwol{\eqalign{(\gamma^1_{fg}&(\alpha_s(Q^2)/N)
+C^g_{2,1}(\alpha_s(Q^2)/N)\gamma^0_{gg}(\alpha_s(Q^2)/N)))
{3\over 2N_f}\biggl({C^g_{L,1,0}\over C^g_{L,1}(\alpha_s(Q^2)/N)}\biggr)\cr
&= {1\over 2\pi}\biggl({3\over
2}\gamma^0_{gg}(\alpha_s(Q^2)/N)
+\sum_{n=0}^{\infty}(\gamma^0_{gg}(\alpha_s(Q^2)/N))^n\biggr)
=\alpha_s(Q^2)\gamma^1_{2L}(\alpha_s(Q^2)/N),\cr}}
which is clearly both factorization scheme and renormalization scheme
independent. 
Thus, our expression for the explicit part of \fullftwoderiv\ is entirely in
terms of factorization--scheme--invariant, and hence physically
meaningful quantities, i.e. 
\eqn\fullftwoderivalt{\eqalign{-\alpha^2_s(Q^2){d \, F_2(N,Q^2)\over
d\alpha_s(Q^2)} =& \alpha_s(Q^2)\tilde\gamma^1_{2L}(\alpha_s(Q^2)/N)
(\hat F_{L,0}(N) +\tilde {\hat F}_{L,0}(N,Q_I^2)-
\biggl({36-8N_f\over 27}\biggr)F_{2,0}(N)) \times\cr
& \exp[\Phi^+_0(Q^2,Q_I^2)+\Phi_1^+(Q_I^2,Q^2)] 
-\alpha_s(Q^2){4\over 9}{2N_f \over
6\pi b_0}F_{2,0}(N)\exp[\Phi_1^-(Q^2,Q_I^2)] \cr
&+ \higherorder.\cr}}

This expression is analogous to that for
$\hat F_L(N,Q^2)$ in \fullflalti, except that it has still 
not been explicitly 
separated into inputs and evolution terms. In order to do this we must 
rewrite \fullftwoderivalt\ as 
\eqn\fullftwoderivalti{\eqalign{-\alpha^2_s(Q^2){d \, F_2(N,Q^2)\over
d\alpha_s(Q^2)} =& \alpha_s(Q_I^2)\tilde\gamma^1_{2L}(\alpha_s(Q_I^2)/N)
(\hat F_{L,0}(N) +\tilde {\hat F}_{L,0}(N,Q_I^2)-
\biggl({36-8N_f\over 27}\biggr)F_{2,0}(N))\times\cr
& \hskip 1.4in \exp[\Phi^+_0(Q^2,Q_I^2)+\Phi_1^+(Q_I^2,Q^2)+
\tilde \Phi^+_1(Q_I^2,Q^2)] \cr
&\hskip -1in -{4\over 9}{2N_f \over
6\pi b_0}\alpha_s(Q_I^2)F_{2,0}(N)\exp\biggl[\Phi_1^-(Q^2,Q_I^2)-\ln\biggl(
{\alpha_s(Q_I^2)\over \alpha_s(Q^2)}\biggr)\biggr] + \higherorder,\cr}}
where $\tilde \Phi^+_1(Q^2,Q_I^2)$ is a series of the same form as 
$\Phi^+_1(Q^2,Q_I^2)$, defined as 
\eqn\deftwidphi{\tilde \Phi^+_1(Q^2,Q_I^2)=
-\int_{\alpha_s(Q^2)}^{\alpha_s(Q_I^2)} {d\over d\, \alpha_s(q^2)}\ln(
\alpha_s(q^2)\tilde \gamma^1_{2,L}(N,\alpha_s(q^2)))d\alpha_s(q^2).}

It is now clear how we obtain the ``leading part'' of this expression, 
i.e. factor out the 
subleading parts of the evolution, $[\Phi^+_1(Q^2,Q_I^2)+\tilde 
\Phi^+_1(Q^2, Q_I^2)]$ (we note that 
$\tilde \Phi^+_1(Q^2,Q_I^2)$ is  a renormalization--scheme--independent 
contribution to
this subleading evolution), and $[\Phi^-_1(Q^2,Q_I^2)-\ln(\alpha_s(Q_I^2)/ 
\alpha_s(Q^2))]$. This leaves us with the leading--order expression
\eqn\loftwoderiv{\eqalign{-\biggl(&\alpha^2_s(Q^2){d \, F_2(N,Q^2)\over
d\alpha_s(Q^2)}\biggr)_0 =\alpha_s(Q_I^2)
\tilde\gamma^1_{2L}(\alpha_s(Q_I^2)/N)\times\cr
&(\hat F_{L,0}(N) +\tilde {\hat F}_{L,0}(N,Q_I^2)-
\biggl({36-8N_f\over 27}\biggr)F_{2,0}(N)) \exp
(\Phi^+_0(Q^2,Q_I^2))
-\alpha_s(Q_I^2){4\over 9}{2N_f \over
6\pi b_0}F_{2,0}(N).\cr}}
Again, this expression is insensitive to changes in starting scale, 
up to higher order, and using the one--loop coupling constant, is
renormalization scheme independent as well as factorization scheme independent.

Finally we consider the expression for $F_2(N,Q^2)$ itself.
It is now a relatively simple matter to write this in terms of
factorization--scheme--independent quantities and in terms of inputs and 
evolution terms. Using the definitions we
have already made and defining
$\hat \Phi^+_1(Q^2,Q_I^2)$ by
\eqn\defhatphi{\hat \Phi^+_1(Q^2,Q_I^2)=
-\int_{\alpha_s(Q^2)}^{\alpha_s(Q_I^2)} {d\over d\, \alpha_s(q^2)}\ln
\biggl({\alpha_s(q^2)\gamma^1_{2,L}(\alpha_s(q^2)/N)\over 
\gamma^0_{gg}(N,\alpha_s(q^2))}\biggr)d\alpha_s(q^2).}
we may factor terms into inputs and evolutions. 
Also, by construction it is guaranteed that
the change of \fullftwo\ at $Q_I^2$ under a change in starting scale
will cancel the change in the evolution in the first term 
under a change in starting scale up to higher orders.
It is also clear that \fullftwo\ at $Q_I^2$ is 
both renormalization scheme and factorization scheme 
independent, as we require, as long as $\tilde f^S_1(N,Q_I^2)$ is scheme
independent. The requirement that if $Q_I^2=A_{LL}$ the input 
reduces to the nonperturbative input $F_{2,0}(N)$ determines 
$\tilde f^S_1(N,Q_I^2)$ uniquely. It must be the scheme--independent quantity
$-\alpha_s(Q_I^2)(\tilde\gamma^1_{2L}(\alpha_s(Q_I^2)/N)/ \tilde
\gamma^0_{gg}(\alpha_s(Q_I^2)/N))
(\hat F_{L,0}(N)-\biggl({36-8N_f\over 27}\biggr)F_{2,0}(N))$. Hence, the 
input for $F_2(N,Q^2)$ is
\eqn\defftwooneinc{F_{2}(N,Q_I^2)=F_{2,0}(N)+\alpha_s(Q_I^2)
{\tilde\gamma^1_{2L}(\alpha_s(Q_I^2)/N)\over \tilde
\gamma^0_{gg}(\alpha_s(Q_I^2)/N)}
\tilde {\hat F}_{L,0}(N,Q_I^2)+\higherorder,}
and the expression for $F_2(N,Q^2)$ is
\eqn\fullftwoaltc{\eqalign{F_2(N,Q^2)=& \alpha_s(Q_I^2)
{\tilde\gamma^1_{2L}(N,\alpha_s(Q_I^2))\over \tilde
\gamma^0_{gg}(N,\alpha_s(Q_I^2))}
(\hat F_{L,0}(N)
+ \tilde {\hat F}_{L,0}(N,Q_I^2) -\biggl({36-8N_f\over 27}\biggr)
F_{2,0}(N))\times\cr
&\exp[\Phi^+_0(Q^2,Q_I^2)+\Phi_1^+(Q_I^2,Q^2)+\hat\Phi^+_1(Q^2,Q_I^2)]
+F_{2,0}(N)\exp[\Phi_1^-(Q^2,Q_I^2)]\cr
&-\alpha_s(Q_I^2)
{\tilde\gamma^1_{2L}(N,\alpha_s(Q_I^2))\over \tilde
\gamma^0_{gg}(N,\alpha_s(Q_I^2))}
(\hat F_{L,0}(N)-\biggl({36-8N_f\over 27}\biggr)
F_{2,0}(N))\cr
&+ \higherorder.\cr}}

Looking at
\fullftwoaltc\ it is clear that there is only one formally LO (in our 
expansion scheme) input multiplied by a LO evolution, and that is 
$F_{2,0}(N)$ multiplying unity. 
But this is obviously completely independent of $\alpha_s$, 
and is rather trivial. At next--to--leading order, or equivalently, at 
leading--$\alpha_s$--dependent order, we include the whole of \fullftwoaltc\
except that we factor $\exp[\Phi^+_1(q^2,Q_I^2)+ \hat
\Phi^+_1(Q^2,Q_I^2)]$, out of the first term. Hence, at 
leading--$\alpha_s$--dependent order we have
\eqn\firstftwo{\eqalign{F_{2,0}&(N,Q^2)= F_{2,0}(N)\exp[\Phi_1^-(Q^2,Q_I^2)]\cr
&+\alpha_s(Q_I^2)
{\gamma^1_{2L}(\alpha_s(Q_I^2)/N)\over \gamma^0_{gg}(\alpha_s(Q_I^2)/N)}
(\hat F_{L,0}(N)
+\tilde {\hat F}_{L,0}(N,Q_I^2) -\biggl({36-8N_f\over 27}\biggr)F_{2,0}(N))
\exp[\Phi^+_0(Q^2,Q_I^2)]\cr
&-\alpha_s(Q_I^2)
{\gamma^1_{2L}(\alpha_s(Q_I^2)/N)\over \gamma^0_{gg}(\alpha_s(Q_I^2)/N)}
(\hat F_{L,0}(N)-\biggl({36-8N_f\over 27}\biggr)F_{2,0}(N)).\cr}}
In this expression there are clearly no terms which mix if we were to make a
change in definition of the coupling $\alpha_s \to \alpha_s + \epsilon
\alpha_s^2$, and hence we can consider it as a leading--order 
expression. If the one--loop coupling is used, it is both
factorization scheme and renormalization scheme independent.

\medskip

We now have the full set of LO expressions in the leading--$\ln(1/x)$ 
expansion scheme. We could obtain the correct scheme--independent expressions
for the structure functions at higher orders within this expansion scheme, 
but we choose to finish at 
leading order. The labour required to obtain higher--order 
expressions becomes progressively greater and we would obtain 
expressions requiring unknown anomalous dimensions and coefficient functions.
Working to NLO we would need to calculate all the NLO evolutions and all
the NLO inputs. This would require all the 
$\tilde \gamma^1_{g,a}$'s, $\tilde \gamma^2_{f,a}$'s and $C^a_{i,2}$'s.
There is optimism that these NLO terms will soon be known
\ref\nlobfkl{L.N. Lipatov and V.S. Fadin, \SJNP \vyp{50}{1989}{712}\semi
V.S. Fadin, R. Fiore and A. Quartarolo, \PR \vyp{D53}{1996}{2729}\semi
V. Del Duca, \PR \vyp{D54}{1996}{989}\semi
G. Camici and M. Ciafaloni, \PL \vyp{B386}{1996}{341}.}, and once this
is so the full NLO scheme--independent expressions should be calculated.

We should make some comments
about our LO scheme--independent expressions for $F_L(N,Q^2)$, 
$F_2(N,Q^2)$ and $(d F_2(N,Q^2)/d\ln (Q^2))$. First we note that all depend on
factorization--scheme--independent combinations of the $\tilde
\gamma^0_{ga}$'s, $\tilde\gamma^1_{fa}$'s 
and $C^a_{i,1}$'s (along with the input
parton distributions, where $g_0(N,Q_I^2)$ is also factorization
scheme dependent). There is, however, 
unlike in the loop expansion no terribly simple prescription
for how one uses the anomalous dimensions and coefficient functions in order 
to arrive at the expressions. One must simply examine the form of the 
solutions and factor out subleading parts. 

By comparing \defftwooneinc\ with \lofl\ we now see that there is a very
direct relationship 
between the inputs for our two structure functions at small $x$, i.e. 
we have a definite prediction, up to additive nonperturbative parts
which are flat at small $x$, for one in terms of the other, as well  
as approximate predictions for the form of each. 
Also comparing with the form 
of \loftwoderiv, these two inputs for the structure functions are directly
related to the slope of $(d\,F_2/d\ln Q^2)$ for small $x$ at $Q_I^2$.  
We do not yet know at what $Q_I^2$ it is most appropriate to choose the 
inputs, but it is a nontrivial requirement that the inputs for the 
three expressions are of the correct form and related in the above manner at 
any $Q_I^2$.    

We should also make some mention of why factorization scheme dependence can 
be very large in this expansion scheme. In order to do this let us consider
\fullfl\ as an example. A representative example of the way in which 
factorization--scheme--dependent calculations are done is to consider 
this expression evaluated with $\tilde\gamma^0_{gg}(\alpha_s(Q^2)/N)$, 
$\tilde\gamma^1_{fg}(\alpha_s(Q^2)/N)$ 
and $C^g_{L,1}(\alpha_s(Q^2)/N)$
known in some particular factorization scheme, 
but $\tilde\gamma^1_{gg}(\alpha_s(Q^2)/N)$, which is 
unknown, either set equal to zero, or guessed by imposing some ansatz such 
as momentum conservation.
As a first comment we consider the input. The input
in terms of parton distributions is multiplied by 
$C^g_{L,1}(\alpha_s(Q_I^2)/N)$, a series which, in general,  
diverges at $N=\lambda(Q_I^2)$. Under a transformation of the type 
\urequire\ this series will be multiplied by 
$(U^0_{gg}(\alpha_s(Q_I^2)/N))^{-1}$. 
A number of scheme transformations that are
considered have the series $U^0_{gg}(\alpha_s(Q_I^2)/N)$ 
also becoming singular at $N=
\lambda(Q_I^2)$ (e.g. \sdis\qzero). This leads to 
powerlike growth of the form $x^{-1-\lambda(Q_I^2)}$ as $x\to 0$, but the 
magnitude of the powerlike behaviour and the manner in which  
in which it is approached depends on the 
the precise behaviour of the coefficients in the series. Hence, changes in
factorization scheme can lead 
to very marked differences in the form of the gluon at small $x$, or if the 
gluon is kept roughly constant (e.g. it is attempted to predict the form
of the input structure function by assuming a form for the input gluon),
to significant changes in the form of 
$F_L(x, Q_I^2)$. Very similar considerations also hold for 
$F_2(x,Q^2)$ because the 
input depends strongly on $C^g_{2,1,0}(\alpha_s(Q_I^2)/N)$, which 
transforms in the same way as $C^g_{L,1,0}(\alpha_s(Q_I^2)/N)$.  

Examining \fullfl\ we can also see how the evolution may be strongly
factorization scheme dependent. As we have already mentioned, the whole
of $\Phi^+_1(Q^2,Q_I^2)$, including $\gamma^1_{gg}(\alpha_s(Q^2)/N)$,
 must be used in order to obtain a factorization--scheme--independent 
expression. The integrand in 
$\Phi^+_1(Q^2,Q_I^2)$ is a series which is a power of $\alpha_s(Q^2)$
down on the series $\gamma^0_{gg}(\alpha_s(Q^2)/N)$. However, in many popular 
factorization schemes the coefficients in the incomplete, or incorrect
series for this integrand are much larger than 
those in $\gamma^0_{gg}(\alpha_s(Q^2)/N)$ (helped by the fact that many of 
the early coefficients in $\gamma^0_{gg}(\alpha_s(Q^2)/N)$ are zero),
e.g. they commonly behave roughly like $(12\ln 2/\pi)^n n^{-3/4}$, whereas the 
coefficients in $\gamma^0_{gg}(\alpha_s(Q^2)/N)$ behave roughly like 
$(12\ln 2/\pi)^n n^{-3/2}$. Hence, the incorrect $\Phi^+_1(Q^2,Q_I^2)$ 
can have a dominant effect on the evolution. Under changes of 
factorization scheme the coefficients in the 
incorrect series can change by amounts similar to 
their own magnitude. Therefore, the evolution
of the structure functions in terms of a given input can appear to have 
a very strong factorization scheme dependence. Once again, this is true for 
the evolution of $F_2(x,Q^2)$ as well as for $F_L(x,Q^2)$: the influence 
of the incorrect $\Phi^+_1(Q^2,Q_I^2)$ can be more important than that of 
$\Phi^+_0(Q^2,Q_I^2)$ and $\gamma_{2L}(N,\alpha_s(Q_I^2))$ combined, 
where for the latter the coefficients in the series are again 
relatively small.

Alternative calculational procedures can lead to expressions which are 
not only similar to 
\fullfl--\fullftwoderiv\ with incorrect or missing $\gamma^1_{gg}
(\alpha_s(Q^2)/N)$, but which have additional factorization--scheme--dependent 
terms. These will be formally of higher order than the terms in 
\fullfl--\fullftwoderiv, but again can have very large coefficients in 
the series expansions. This can lead to even more dramatic effects than 
those outlined above, e.g. \frt.
    
Once $\gamma^1_{gg}(N,\alpha_s(Q^2))$ is known in a given 
scheme and \fullfl--\fullftwoderiv\ can be calculated correctly 
there is no guarantee that the correct $\Phi^+_1(Q^2,Q_I^2)$ is not larger
than $\Phi^+_0(Q^2,Q_I^2)$. If this is the case, $\Phi^+_1(Q^2,Q_I^2)$ 
will then have a large, but at least definite, effect. 
However, because it is a formally NLO correction
to the structure functions, if $\Phi^+_1(Q^2,Q_I^2)$ is introduced
then the full set of NLO expressions, both evolution
and input factors, must be calculated at the
same time: the correct 
calculational method respects renormalization scheme independence as well 
as factorization scheme dependence. 
This requires many more terms than just $\Phi^+_1(Q^2,Q_I^2)$.
Hopefully, the complete NLO
expression, as well as being factorization scheme independent, will also 
cause only fairly small changes to the LO expressions. 

Having obtained our full set of leading--order expressions, we can
also examine the difference between $F_2(N,Q^2)$ and
$(d F_2(N,Q^2)/d \ln Q^2)$ in this expansion scheme. 
There is a distinction between $F_2(N,Q^2)$ and 
$(d\,F_2/d\,\ln Q^2)$ even in the usual loop expansion. Differentiating a 
fixed--order expression for $F_2(N,Q^2)$ and using the $\beta$--function
evaluated to the appropriate order in $\alpha_s$ results
in the fixed--order expression for $(d\,F_2/d\,\ln Q^2)$ plus terms of higher
order in $\alpha_s(Q^2)$, which depend on the $\beta$--function. 
The size of these extra terms is of the same order as
the renormalization scheme uncertainty, and the distinction 
between the fixed--order expressions for $F_2(N,Q^2)$ and 
$(d\,F_2/d\,\ln Q^2)$ is of similar magnitude to the 
distinction between renormalization schemes.   
The distinction when using the small--$x$ expansion appears
more graphically, and is not only dependent on terms in the
$\beta$-function. If we differentiate  
\firstftwo\ with respect to $\alpha_s(Q^2)$ we obtain 
\eqn\derfirstftwo{\eqalign{-\alpha^2_s(Q^2){d \, F_{2,0}(N,Q^2)\over
d\alpha_s(Q^2)} =&\alpha_s(Q_I^2)
\tilde\gamma^1_{2L}(\alpha_s(Q_I^2)/N)\biggl({\gamma^0_{gg}(\alpha_s(Q^2)/N)
\over \gamma^0_{gg}(\alpha_s(Q_I^2)/N)}\biggr)
\times\cr
&(\hat F_{L,0}(N) +\tilde{\hat F}_{L,0}(N,Q_I^2)-
\biggl({36-8N_f\over 27}\biggr)F_{2,0}(N)) \exp
[\Phi^+_0(Q^2,Q_I^2)]\cr
&-\alpha_s(Q^2){4\over 9}{2N_f \over
6\pi b_0}F_{2,0}(N)\exp[\Phi^-_1(Q^2,Q_I^2)].\cr}}
This is clearly not exactly the same as \loftwoderiv, the 
difference being due to the additional terms  
$(\gamma^0_{gg}(\alpha_s(Q^2)/N)/ \gamma^0_{gg}(\alpha_s(Q_I^2)/N))$ 
in the first term, 
and the factors $(\alpha_s(Q^2)/\alpha_s(Q_I^2))$ and  
 $\exp[\Phi^-_1(Q^2,Q_I^2)]$ in the second term. All of these
factors are unity at the boundary of the evolution, and the two 
expressions are therefore identical in this limit, i.e the inputs are 
the same. Therefore, it is the evolution terms which are different when 
comparing \loftwoderiv\ and \derfirstftwo. Writing 
\eqn\addonii{\biggl({\gamma^0_{gg}(\alpha_s(Q^2)/N)\over 
\gamma^0_{gg}(\alpha_s(Q_I^2)/N)}\biggr)=\exp\biggl[
-\int_{\alpha_s(Q^2)}^{\alpha_s(Q_I^2)} {d\over d\, \alpha_s(q^2)}\ln
(\gamma^0_{gg}(\alpha_s(q^2)/N))d\alpha_s(q^2)\biggr],}
\eqn\addoniii{\alpha_s(Q^2)=\alpha_s(Q_I^2)\exp\biggl[-\ln\biggl(
{\alpha_s(Q_I^2)\over \alpha_s(Q^2)}\biggr)\biggr]}
we see that the terms present in \derfirstftwo\ but absent in \loftwoderiv\ 
are NLO evolution terms. Thus, as in the loop 
expansion, the difference between the fixed--order expression for $(d\,F_2/d
\ln Q^2)$ and the $\ln Q^2$ derivative of the fixed--order $F_2$ consists
of terms of higher order. However, in the leading--$\ln (1/x)$ expansion 
this difference exists between even LO expressions, and 
even at leading order we have to decide 
which of the two expressions to
use (although there is very little difference between
the choices in practice). 
It may be
argued that in certain senses $(d F_2(N,Q^2)/d \ln Q^2)$ 
is more natural because it is
a ``real'' perturbative quantity, beginning at first order in
$\alpha_s$, as does $F_L(N,Q^2)$. 
We will discuss and confirm this choice in \S 4.4. 

We should also make some comment about longitudinal momentum
conservation. The first moment of the parton distributions is
interpreted as the fraction of momentum carried by that type of
parton, and it is usually required that $f^S(1,Q^2)+g(1,Q^2)=1$. 
For this to be true for all $Q^2$ then, of course,
$(d\,(f^S(1,Q^2)+g(1,Q^2))/d\ln Q^2)=0$, and from the renormalization
group equations this is true if 
\eqn\momcons{\gamma_{ff}(1,\alpha_s(Q^2))+
\gamma_{gf}(1,\alpha_s(Q^2))=0, \hskip 0.5in \gamma_{fg}(1,\alpha_s(Q^2))+
\gamma_{gg}(1,\alpha_s(Q^2))=0.}  
When expanding the anomalous dimensions order by
order in $\alpha_s(Q^2)$, it is easy to specify
that \momcons\ be true for the anomalous dimensions at each order, and
to define a wide variety of factorization schemes which maintain this. 
This guarantees that the fraction
of momentum carried by the $n_{\rm th}$--order parton distributions is
conserved at each $n$. Sometimes, all the momentum is designated 
to be carried by the zeroth order part of the solution, 
but this need not be the case.
Most often the input is implicitly assumed to be the all--orders 
input. In this case it is 
true that it must carry all the momentum, but this method destroys the 
strict ordering of the solution.

The situation is not as simple for the
leading--$\ln (1/x)$ expansion. As can be seen from the matrix in
\llevolzero, the leading--order $\gamma$ contains entirely positive
entries for $N=1$, and is clearly not consistent with
momentum conservation: $f^S_0(Q^2)$ carries a
constant amount of momentum while that carried by $g_0(Q^2)$ is
constantly increasing with $Q^2$. In a general factorization scheme 
there is no reason that working to a finite higher
order will restore the relationship \momcons.
Two general methods have been proposed to restore momentum conservation 
\ehw, \bfresum, \ref\bfmomcon{R.D. Ball and 
S. Forte, \PL \vyp{B359}{1995}{362}.}.Both methods guarantee
momentum conservation only if one truncates the series for the
$\gamma$'s at the $\gamma^1$'s and solves the whole
renormalization group equation using this truncated $\gamma$.
Hence, both
prescriptions destroy any sense of ordering the solution correctly. 
We advocate that the most 
sensible approach is to obtain a well--ordered solution 
for the structure functions, i.e. regard a correct treatment
of the physical quantities as of paramount importance. 
Thus, we simply take the hint offered us
by the zeroth--order anomalous dimension, and accept the fact that
momentum is not conserved order by order in this method of expansion.   
How badly it appears to be violated, however, will depend very much on which
factorization scheme we claim to work in, and we can choose the violation to 
be negligible. We will discuss this in more detail in \S 4.5.

\medskip

Finally, we can also discuss the relationship between our scheme--independent
solutions and the ones which would be obtained using Catani's physical
anomalous dimensions. One can solve for the LO structure functions 
using these effective anomalous dimensions in exactly the same way as we 
solved for the parton distributions in the previous subsection. In the same 
way that we have the relationships between the LO anomalous
dimensions, \gammadef\ and \gamrel, we have relationships between the 
effective anomalous dimensions  as seen in \physanomval.
Using these relationships it is 
straightforward to obtain the analogous
expressions to \lof, \loglu\ and \loderivf:
\eqn\plof{\eqalign{F_2(N,Q^2)=& \alpha_s(Q^2)
{\Gamma^1_{2L}(\alpha_s(Q^2)/N)\over
\Gamma^0_{LL}(\alpha_s(Q^2)/N)}(\hat F_{L,0}(N)+\tilde 
{\hat F}_{L,0}(N,Q_I^2)
-\biggl({36-8N_f \over 27}\biggr)F_{2,0}(N))\times\cr
& \hskip -0.2in \eigpluscorp\cr
&\hskip -0.2in -\tilde F_{2,1}(N,Q_I^2)+F_{2,0}(N)\biggl({\alpha_s(Q_I^2)\over
\alpha_s(Q^2)}\biggr)^{-\fourninths{2N_f\over 6\pi b_0}}+
\higherorder,\cr}}
\eqn\ploglu{\eqalign{\hat F_L(N,Q^2)=& (\hat F_{L,0}(N)+
\tilde{\hat F}_{L,0}(N,Q_I^2)
-\biggl({36-8N_f \over 27}\biggr)F_{2,0}(N))\times\cr
&\hskip -0.2in \eigpluscorp\cr
&+ \biggl({36-8N_f \over 27}\biggr)F_{2,0}(N)\biggl({\alpha_s(Q_I^2)\over
\alpha_s(Q^2)}\biggr)^{-\fourninths{2N_f\over 6\pi b_0}}+ \higherorder,\cr}}
and 
\eqn\loderivfp{\eqalign{-\alpha_s^2(Q^2) {d\, F_2(N,Q^2)\over
d\alpha_s(Q^2)}
=& \alpha_s(Q^2)
\tilde\Gamma^1_{2L}(\alpha_s(Q^2)/N)(\hat F_{L,0}(N)+
\tilde {\hat F}_{L,0}(N,Q_I^2)
-\biggl({36-8N_f \over 27}\biggr)F_{2,0}(N))\times\cr
&\hskip -1in \eigpluscorp\cr
&-\fourninths \alpha_s(Q^2) 
F_{2,0}(N){2N_f\over 6\pi b_0} \biggl({\alpha_s(Q_I^2)\over
\alpha_s(Q^2)}\biggr)^{-{4\over 9}{2N_f\over 6\pi b_0}}+\higherorder,\cr}}
where $\tilde {\hat F}_{L,0}(N,Q_I^2)$ is a function of $\Gamma^0_{LL}$ rather
than $\gamma^0_{gg}$.

We see that, once we make the identifications 
\eqn\identi{\gamma^0_{gg}(\alpha_s(Q^2)/N)=\Gamma^0_{LL}
(\alpha_s(Q^2)/N),\hskip 0.8in \gamma^1_{2L}(\alpha_s(Q^2)/N)=
\Gamma^1_{2L}(\alpha_s(Q^2)/N),}
and
\eqn\identii{\eqalign{\tilde \gamma^1_{gg}(\alpha_s(Q^2)/N)+\fourninths \tilde 
\gamma^1_{fg}(\alpha_s(Q^2)/N)-&{d\over d\,\alpha_s(Q^2)}\biggl(\ln
\biggl({C^g_{L,1}(\alpha_s(Q^2)/N)\over C^g_{L,1,0}}\biggr)\biggr)\cr
&=\tilde\Gamma^1_{LL}(\alpha_s(Q^2)/N)-
\biggl({36-8N_f\over 27}\biggr)\tilde\Gamma^1_{2L}(\alpha_s(Q^2)/N),\cr}}
\ploglu\ is identical to \fullflalti, 
\loderivfp\ is identical to \fullftwoderivalt\ and  \plof\ is identical to
\fullftwoaltc. Of course, the identifications \identi--\identii\ are 
exactly what we obtain from the definitions of the physical anomalous 
dimensions in \S 3. 

Thus, we are able to reach these expressions 
for the structure functions somewhat more directly by 
using the physical anomalous dimensions, and do not have to worry about 
problems with factorization scheme dependence (though we do have 
to calculate the physical anomalous dimensions in terms of known 
coefficient functions and anomalous dimensions of course). 
Once we have obtained these expressions using the 
physical anomalous dimensions we may then separate each of the terms into 
input parts and evolution parts (where more of this, {\it but certainly not 
all}, has already 
been done automatically when using the physical anomalous dimensions) and
keep the most leading parts, obtaining once again the LO expressions
\lofl, \loftwoderiv\ and \firstftwo.
So, using the physical anomalous dimensions leads us
in a rather more direct manner to the correct leading--order expressions. 
If we were to work to higher orders, the amount of simplification obtained by
using the physical anomalous dimensions rather than working in terms of 
parton densities would increase significantly.

However, we stress that one will always
automatically obtain factorization--scheme--independent answers by 
working to well--defined orders in physical quantities even when working 
in terms of partons. Also, we stress very strongly that 
{\it even if one uses the physical anomalous dimensions, 
care is still needed to obtain expressions which are consistent with 
renormalization scheme dependence}, and there is no simple prescription for 
obtaining the correct results even in terms of the physical anomalous 
dimensions. If we were to solve the evolution 
equations for structure functions using  $\tilde \Gamma^0_{LL}$,
$\tilde \Gamma^0_{L2}$, $\tilde \Gamma^1_{2L}$ and $\tilde \Gamma^1_{22}$
we would obtain the explicit parts of \plof--\loderivfp\ (with 
$\tilde \Gamma^1_{LL}=0$) plus corrections of higher order. Terms must still
be factored out of these expressions in order to obtain 
the true leading order structure functions. 
We will discuss the phenomenological consequences of this in \S 6.

\bigskip

In this section, we have derived well--ordered, and consequently, 
factorization--scheme--independent
expressions for structure functions in the leading--$\ln (1/x)$
expansion (which should be useful at small $x$) 
up to the order which is useful at present. This expansion does, however,
sacrifice any attempt to describe the structure functions at large $x$. 
We would hope there is some expansion scheme
which will be useful at all values of $x$. In the next section we will
show that there is indeed an expansion scheme which satisfies this 
criterion, and argue that it is the only really correct expansion scheme.

\subsec{The Renormalization--Scheme--Consistent Expansion.}

In order to devise an expansion scheme which is useful at both large
and small $x$ we would {\it a priori} expect that we would need to
use anomalous dimensions and coefficient
functions at low orders in both $\alpha_s$ and the leading--$\ln (1/x)$
expansion. There have already been various methods along these lines,  
however, they have all been scheme dependent.
We demand complete consistency of our 
expressions for physical quantities with renormalization scheme
invariance (which then guarantees factorization scheme 
independence). Consequently, our approach will be different 
from those used previously, and the results and conclusions will also be 
somewhat different. 

To begin, let us consider what we have meant by 
``consistency with renormalization scheme dependence'' so far in this paper.
In both the loop expansion and the leading--$\ln (1/x)$ 
expansion we demanded that once we had chosen a particular renormalization
scheme and chosen to work to a particular order in this renormalization
scheme then we would include all terms in our expressions for the 
structure function which were of greater magnitude than the uncertainty
due to the freedom of choice of renormalization scheme (i.e. the uncertainty 
in the definition of the coupling constant), and no others. In each case the 
leading--order term consisted of the lowest--order inputs multiplying the 
lowest--order evolution terms. If working with the $n$--loop coupling
constant the
uncertainty in its definition is of order $\alpha_s^{n+1}$. 
Thus, the uncertainty of the 
input or evolution when working to $n_{\rm th}$--order is the change in the 
leading--order input or evolution if $\alpha_s \to \alpha_s(1+\epsilon
\alpha_s^n)$. Hence, the uncertainty in the whole structure function
is of the order of the change of the leading--order part under such 
a change in the coupling. Therefore the $n_{\rm th}$--order 
renormalization--scheme--independent expression includes all complete terms
smaller than this change. 

This definition does give us a well--defined
way of building up an ordered solution to the structure functions, 
but relies upon the definition of a given expansion scheme.
It leaves an ambiguity about how we define the leading--order 
expressions and in how we define the order of terms compared to this 
leading--order term. Our two examples, the loop expansion, where the size
of a term is determined simply by its order in $\alpha_s$, and the 
leading--$\ln(1/x)$ expansion, where $\ln(1/x)$ is put on an equal footing 
to $\alpha_s$, are just two examples of expansion schemes. 
Both have potential problems: in the former one does not 
worry about the large--$\ln(1/x)$ terms which can cause enhancement 
at small $x$ of terms which are higher order in $\alpha_s$, 
and in the latter one does not worry about the fact that 
at large $x$, especially as $Q^2$ increases, it is the terms of 
lowest order in $\alpha_s$ that dominant. Hence, both should have limited 
regions of validity.

The shortcomings of these two expansion schemes 
come about because, even though any given order contains no terms which are 
inconsistent with working to the same given order in a particular
renormalization scheme, in neither case does it include every one of the terms 
which are consistent to working to a given order in the  
renormalization scheme. In each expansion scheme some of the terms 
appearing at what we call higher orders are not really
subleading in $\alpha_s$ to any terms which have already appeared. Thus,
despite the fact that for a given expansion method these terms are 
defined to be the same order as 
uncertainties due to the choice of renormalization scheme, they are not terms 
which can actually be generated by a change in renormalization 
scheme.\foot{Similarly, they cannot be generated by a change in 
renormalization scale.} 

In order to demonstrate this point more clearly we consider a simple toy 
model. Let us imagine some hypothetical physical quantity which can be
expressed in the form
\eqn\hypothet{H(N,\alpha_s(Q^2))=\sum_{m=1}^{\infty}\alpha_s(Q^2)
\sum_{n=-m}^{\infty}a_{mn}N^n\equiv \sum_{i=0}^{\infty}\alpha^i_s(Q^2)
\sum_{j=1-i}^{\infty}b_{ij}\biggl({\alpha_s(Q^2)\over N}\biggr)^j,}
where the expansion in powers of $N$ about $N=0$ is convergent for all $N$. 
The first way of writing $H(N,\alpha_s(Q^2))$ as a power series corresponds to
the loop expansion, where we work order by order in $m$, out to $m=k$,  
and use the $k$--loop
coupling. The second corresponds to the leading--$\ln(1/x)$ expansion 
where we work order by order in $i$, out to $i=l$,
and use the $(l+1)$--loop coupling.
Let us, for a moment, consider the LO expression in the loop expansion,
$\alpha_s(Q^2) \sum_{n=-1}^{\infty}a_{1n}N^n$.
The coupling is uncertain by ${\cal O}(\alpha^2_s(Q^2))$ and hence the 
uncertainty of the leading--order expression (i.e. the change due to 
a change of the coupling) is $\sim\alpha_s^2(Q^2)\sum_{n=-1}^{\infty}
b_{1n}N^n$. There is no change with powers of $N$ less than $-1$,
and hence any such term is not really subleading.
Similarly, the uncertainty of the leading--order expression in the leading
$\ln (1/x)$ expansion contains no terms at first order in $\alpha_s$ (or
with positive powers of $N$), and such terms are not really 
subleading either. The full set of terms contained within the combination
of both leading--order expressions is genuinely leading order, and 
is therefore renormalization scheme independent by definition. 

Perhaps the best way in which to write our expression for 
$H(N,\alpha_s(Q^2))$ in order to appreciate these points is 
\eqn\hypotheti{H(N,\alpha_s(Q^2))=\sum_{m=-1}^{\infty}N^m
\sum_{n=1}^{\infty}c_{mn}\alpha_s^n(Q^2)+ \sum_{m=2}^{\infty}N^{-m}
\sum_{n=m}^{\infty}c_{mn}\alpha^n_s(Q^2),}
i.e. as an infinite number of power series in $\alpha_s(Q^2)$, one 
for each power on $N$. Each of these series in $\alpha_s(Q^2)$ is 
independent of the others, and the lowest order in $\alpha_s(Q^2)$ of each 
is therefore renormalization scheme independent and part of
the complete LO expression for $H(N,\alpha_s(Q^2))$.  
The full LO expression for $H(N,\alpha_s(Q^2))$ is therefore
\eqn\hypothetlo{\eqalign{H_0(N,\alpha_s(Q^2))&=\sum_{m=-1}^{\infty}N^m
c_{m1}\alpha_s(Q^2)+ \sum_{m=2}^{\infty}c_{mm}N^{-m}
\alpha^{m}_s(Q^2)\cr
&\equiv \alpha_s(Q^2)
\sum_{n=-1}^{\infty}a_{0n}N^n+ 
\sum_{j=2}^{\infty}b_{0j}\biggl({\alpha_s(Q^2)\over N}\biggr)^j.}}
Hence, the combined set of terms considered LO in the two expansion schemes comprise the full set of renormalization scheme
invariant, and thus truly leading--order, terms.
By considering $H(N,\alpha_s(Q^2))$ written in the form \hypotheti, 
and considering
how the coefficients in the expression must change in order to make the 
whole expression invariant under a redefinition of the coupling constant,
$\alpha_s(Q^2)\to \alpha_s(Q^2)+{\cal O}(\alpha^m_s(Q^2))$, we see that the 
$n_{\rm th}$--order expression for $H(N,\alpha_s(Q^2))$, which should be 
used with the $n$--loop coupling constant, consists of the sum of the first 
$n$ terms in each of the power series in $\alpha_s(Q^2)$. Thus, the full 
$n_{\rm th}$--order expression consists of the $n_{\rm th}$--order 
expression in the loop expansion plus additional terms with inverse powers 
of $N$ greater than $n$.

Similar arguments have already been applied to the anomalous dimensions 
and coefficient functions, for example  \cathaut\ and particularly \bfresum,
though in a somewhat weaker form.
Here we take a strong, inflexible viewpoint and insist that the complete 
renormalization--scheme--consistent expressions (with no artificial 
suppression of leading--$\ln (1/x)$ terms \bfresum), must be used. 
Furthermore, and very importantly, the 
expressions used must be those for the physical structure functions, not for 
the factorization--scheme-- and renormalization--scheme--dependent 
coefficient functions and anomalous dimensions.

When considering the real structure functions the situation is  
technically a great deal more  complicated than our toy model, 
but the principle is exactly the same. This can be seen by examining the 
the LO expressions for the structure functions in the 
two expansion schemes already considered. There is some overlap between 
the two, 
but each contains an infinite number of terms not present in the other. 
Considering the change of each 
leading--order expression under a change of coupling of  
${\cal O}(\alpha_s^2)$ we see that, as with our 
toy model, the change in the LO 
structure functions in the loop expansion contains no terms in the 
LO expressions in the leading--$\ln (1/x)$ expansion, and
{\it vice versa} i.e., none of the terms contained within each of the LO
expressions are generated by uncertainties at higher order
in the other. Therefore, they  
should really all be regarded as genuinely LO,
and be included in the full ``leading order'' expressions for the 
structure functions which use the one--loop coupling constant. 
Since these expressions
contain all the parts of the one--loop expressions, and also contain
leading--$\ln(1/x)$ terms as well, they should be able to 
describe the data over the full range of parameter space
(except very low $Q^2$, of course), as we would like 
from our correct LO expressions. 
However, there are two main complications when considering structure
functions in 
comparison to our simple toy model. One is that the structure functions 
are combinations of evolution parts and input parts, rather 
than one simple power series in $\alpha_s(Q^2)$. The other
is that in general the physical anomalous dimensions, out of   
which the perturbative parts are constructed, are nonanalytic 
functions which cannot be expressed as power series about a particular
value $N_0$ for all $N$. 
They have singularities at $N=0$, and also at
negative powers of $N$ (as well as possible $\alpha_s(Q^2)$--dependent 
nonanalyticities due to 
resummation effects, e.g. the branch point in $\Gamma^0_{LL}
(N,\alpha_s(Q^2))$ at $N=\lambda(Q^2)$). We will deal with this second 
complication first. 

Let us consider the perturbative parts of the expressions for the 
structure functions. The singularities at negative integer values of $N$
mean that we cannot write any physically meaningful quantity as just a power
series about $N=0$ (or about $(N+1)$ for the nonsinglet case).
Any such power series expansion will have a radius of 
convergence of unity, and a series 
expansion which applies over the whole range of $N$ does not exist.
In order to overcome this problem let us consider making the 
inverse transformation of some 
physically relevant perturbative function $A(N,\alpha_s(Q^2))$ to $x$--space.
The inverse of the Mellin transformation \melltranssf\ is 
\eqn\inverstran{{\cal A}(x,\alpha_s(Q^2))={1\over 2\pi i}
\int_{c-i\infty}^{c+i\infty} \exp[\xi N]A(N,\alpha_s(Q^2))dN.}
where the line of integration is to the right of all nonanalyticities, and 
$\xi =\ln(1/x)$.   
Since $A(N,\alpha_s(Q^2))$ has, in general, singularities for all 
nonpositive integers, this whole integral may be evaluated by performing an 
infinite series of integrals, each with a contour centred on a given 
singularity, and not extending as far as unity from this singularity, i.e.
not reaching any of the other singularities. Within each of these contours 
the function $A(N,\alpha_s(Q^2))$ may be expanded as a power series about 
the singularity, i.e. we may write
\eqn\inverstranps{{\cal A}(x,\alpha_s(Q^2))={1\over 2\pi i}\sum_{n=0}^{\infty}
x^{n}\int_{c_n} \exp[\xi (N+n)]A_n((N+n),\alpha_s(Q^2))d(N+n),}
where $A_{n}((N+n),\alpha_s(Q^2))$ denotes $A(N,\alpha_s(Q^2))$ expanded as 
a power series about $N=-n$. The integrals will produce functions of $\xi$,
which do not sum to integer powers of $x$, and each of the integrals 
in \inverstranps\ will be independent and physically relevant in its own 
right. Hence, we must consider the 
complete moment--space expression as an infinite number of expressions of the
form \hypotheti, each one having power series expansions in terms of $(N+n)$,
where $n=0
\to \infty$. The expression for each $n$ is then related to the part of 
the $x$--space expression behaving $\sim x^{n}$. 
Of course, in practice, unless we 
want to examine the details of the 
structure function for $x$ very close to $1$, we can ignore all $n$ 
greater than a finite, relatively small constant. 

Thus, when calculating the expressions for the perturbative part of the 
singlet structure functions, we are only 
concerned about LO terms beyond lowest order in $\alpha_s$ 
for the specific case of $n=0$. For $n>0$ we take the 
whole LO expression to be the one--loop expression. The 
terms we ignore by making this necessary decision are those which are 
LO in $\ln (1/x)$ at first order in $x$. Although these terms 
grow like $\alpha_s^m\ln^{2m-1}(1/x)$, there is no evidence that 
their coefficients are any
larger than those for the zeroth--order--in--$x$ logarithms. Since the 
resummed terms at zeroth order in $x$ only begin to make an impact as $x$ 
falls to $\sim 0.1$ (as we will see), and only become dominant 
for $x$ much smaller than this,
the effect of terms like  $x\alpha_s^m\ln^{2m-1}(1/x)$ should be 
very small in comparison. Indeed, the effect of those terms of
the form $x\alpha_s^m\ln^{2m-1}(1/x)$ which are actually known, 
i.e. $m=2$, are indeed negligible. In a similar manner, we only consider 
the one--loop expressions for the nonsinglet structure functions in 
practice: the other LO parts of the expressions again lead to small--$x$ 
enhancement of the form $x\alpha_s^m\ln^{2m-1}(1/x)$, which is very small
compared to the singlet small--$x$ enhancement, and there is only 
detailed data at very small $x$ for the total structure function.

\medskip
    
We now return to our first problem, i.e. the fact that the 
structure functions are expressed in terms of both inputs and evolution parts.
We discuss the case of the nonsinglet structure 
functions as an example first.      

\subsec{The Renormalization--Scheme--Consistent (RSC)
Nonsinglet Structure Functions.}

We consider a nonsinglet longitudinal structure function. For the nonsinglet 
structure functions the physical anomalous dimensions contain no singularities
at $N=0$, so the leading--$\ln (1/x)$ behaviour comes from singularities at 
$N=-1$. Expanding about $N=-1$, the full LO physical 
anomalous dimension can be written in the form,
\eqn\lonslpan{\eqalign{\Gamma^{NS}_{L,0}(N+1, \alpha_s(Q^2))=&
\alpha_s(Q^2)\biggl[
\sum_{m=-1}^{\infty}a_m (N+1)^m \cr
&\hskip 0.3in +\sum_{m=1}^{\infty}b_m
(N+1)^{-1}\biggl({\alpha_s(Q^2)
\over (N+1)^2}\biggr)^m+\sum_{m=1}^{\infty}c_m\biggl({\alpha_s(Q^2)
\over (N+1)^2}\biggr)^m\biggr].\cr}}
The first sum is just $\Gamma_{L,0,l}^{NS}(N+1)$, the one--loop 
anomalous dimension expanded in powers of $(N+1)$. The second sum 
contains the leading singularities in $(N+1)$ for all other orders
in $\alpha_s(Q^2)$.
The final sum is included because, despite the obvious fact that it is a 
power of $(N+1)^{-1}$ down on the second sum, 
a series of this form cannot be created from the second sum by
a change in the definition of the coupling of ${\cal O}(\alpha^2_s(Q^2))$.
Therefore, the third sum is not subleading in $\alpha_s(Q^2)$ to 
the second sum, and 
must be renormalization scheme independent. Integrating  
\lonslpan\ between $Q_I^2$ and $Q^2$, and including the overall 
power of $\alpha_s(Q^2)$ for the 
longitudinal structure function we obtain the leading--order evolution 
\eqn\loevolnslsf{\biggl({\alpha_s(Q_I^2)\over \alpha_s(Q^2)}
\biggr)^{\tilde\Gamma_{L,0,l}^{NS}(N+1)-1}
\exp\biggl[
\int_{\alpha_s(Q^2)}^{\alpha_s(Q_I^2)}
\biggl(\sum_{m=1}^{\infty} \biggl({1\over N+1}b_m+c_m\biggr)
\biggl({\alpha_s(q^2)
\over (N+1)^2}\biggr)^m\biggr){d\alpha_s(q^2)
\over b_0\alpha^2_s(q^2)}\biggr].}
In order 
to construct the full LO input we must consider how the evolution 
term \loevolnslsf\ changes under a change in starting scale,
and therefore how the input must change in order to 
compensate for this. With a little work 
we can see that the full LORSC input is    
\eqn\loinputnslsf{\eqalign{\alpha_s&(Q_I^2)F^{NS}_{2,0}(N)\biggl(
C^{NS}_{L,1,l}(N+1) + C^{NS}_{L,1,l}(N=-1)
\biggl(\ln(Q_I^2/A^{NS}_L)\alpha_s(Q_I^2)\times\cr
&\biggl({1\over N+1}
\sum_{m=0}^{\infty} b_m\biggl({\alpha_s(Q_I^2)
\over (N+1)^2}\biggr)^m
+ \sum_{m=0}^{\infty}\bigl(c_m
+\half \sum_{n=0}^{m}b_nb_{m-n}\ln(Q_I^2/A^{NS}_L)
\bigr)\biggl({\alpha_s(Q_I^2)
\over (N+1)^2}\biggr)^m\biggr)\biggr)\biggr),\cr}}
where $b_0\equiv a_{-1}$ and $c_0=0$.
The first term is just the lowest--order input in the loop expansion, 
while the second includes all the leading--$\ln (1/x)$ terms. 
 
Hence, the full LORSC expression
for the nonsinglet longitudinal structure function, expanded about 
$N=-1$, is 
\eqn\rscflns{\eqalign{F^{NS}_{L,RSC,0}&(N,Q^2)=
\alpha_s(Q_I^2)F^{NS}_{2,0}(N)\biggl(C^{NS}_{L,1,l}(N+1) +
C^{NS}_{L,1,l}(N=-1)
\biggl(\ln(Q_I^2/A^{NS}_L)\alpha_s(Q_I^2)\times\cr
&\hskip -0.2in \biggl({1\over N+1}
\sum_{m=0}^{\infty} b_m\biggl({\alpha_s(Q_I^2)
\over (N+1)^2}\biggr)^m
+ \sum_{m=0}^{\infty}\bigl(c_m+
\half \sum_{n=0}^{m}b_nb_{m-n}\ln(Q_I^2/A^{NS}_L)\bigr)
\biggl({\alpha_s(Q_I^2)
\over (N+1)^2}\biggr)^m\biggr)\biggr)\biggr)\times\cr
&\biggl({\alpha_s(Q_I^2)\over \alpha_s(Q^2)}
\biggr)^{\tilde\Gamma_{L,0,l}^{NS}(N+1)-1}\exp\biggl[
\int_{\alpha_s(Q^2)}^{\alpha_s(Q_I^2)}
\biggl(\sum_{m=1}^{\infty} \biggl({1\over N+1}b_m+c_m\biggr)
\biggl({\alpha_s(q^2)
\over (N+1)^2}\biggr)^m\biggr){d\alpha_s(q^2)\over 
b_0\alpha^2_s(q^2)}\biggr].\cr}}
In practice, we will only use the one-loop expression for the 
nonsinglet structure functions for all the singularities in the anomalous 
dimensions. This is because of the phenomenological 
reasons given at the end of the last subsection,
and also because of lack of knowledge of the full
physical anomalous dimensions. 
The terms $\sim \alpha_s^m(N+1)^{2m-1}$ in the anomalous dimension are
all known \ref\nsresum{R. Kirschner and L.N. Lipatov, \NP  
\vyp{B213}{1983}{122}.}, however, there is little knowledge yet of the 
terms of the sort $\sim \alpha_s^m(N+1)^{2m-2}$. These 
are an intrinsic part of the LORSC
expression for the nonsinglet structure function, and should be 
calculated and included in order to give a true indication of the effect
of leading--$\ln (1/x)$ terms. Hence, we believe that calculations of the 
nonsinglet \brv\ref\nsresumphen{B.I. Ermolaev, S.I. Manaenkov and M.G. Ryskin,
\ZP\vyp{C69}{1996}{259} \semi J. Bl\"umlein and A. Vogt, 
{\it Acta Phys. Polon.} \vyp{B27}{1996}{1309}; \PL \vyp{B370}{1996}{149}.}  
(and polarized \ref\resumpol{J. Bartels, B.I. Ermolaev and M.G. Ryskin, 
{\tt hep-ph/9603204},
preprint DESY--96--025, March 1996; \ZP \vyp{C70}{1996}{273}\semi
J. Bl\"umlein and A. Vogt, \PL \vyp{B386}{1996}{350};
{\tt hep-ph/9610203}, Proc. of the 1996 HERA Physics Workshop, 
eds. G. Ingelman,
R. Klanner, and A. De Roeck, DESY, Hamburg, 1996, Vol. 2, p. 799;
Proc. of the International Conference SPIN'96, Amsterdam, September 1996,
eds. K. de Jager, P. Mulders {\it et al}, (World Scientific, Singapore, 1996),
in print.}) structure functions which claim to 
include leading--$\ln(1/x)$ corrections are presently incomplete.

\medskip

We now consider the nonsinglet structure function $F^{NS}_2(N,Q^2)$. This 
leads us back to our previous question of whether we should use the
full RSC expression for $F_2(N,Q^2)$ or that
for $(d\, F_2(N,Q^2)/d\,\ln Q^2)$. In order to illustrate the
difference between the two we consider the simpler nonsinglet case, and 
also pretend for the moment 
that there is no small--$x$ enhancement
at higher orders in $\alpha_s$. Hence, the LO term
in the evolution is just $\biggl({\alpha_s(Q_I^2)\over \alpha_s(Q^2)}
\biggr)^{\tilde\Gamma_{2,0,l}^{NS}(N+1)}$.
The input may be written as a power series in $\alpha_s(Q_I^2)$, 
as we saw in \S 4.1, and is of the form 
\eqn\nsinputftwo{F^{NS}_{2,0}(N,Q_I^2)=F^{NS}_{2,0}(N)[1+\alpha_s(Q_I^2)
\ln(Q_I^2/A^{NS}_2)\Gamma^{2,0,l}_{NS}(N+1)+ \hbox{\rm higher order in 
$\alpha_s(Q_I^2)$} ].}
Therefore, as well as the lowest order 
$\alpha_s$--, and hence $Q_I^2$--dependent part of the input there is also 
the ``sub--lowest--order'', $Q_I^2$--independent part. These two terms are 
clearly of different order, but
under a change in renormalization scheme, both remain unchanged and both
should therefore appear in the LO definition of the structure function.
This mixing of orders seems rather unsatisfactory, and comes about 
because for 
$F^{NS}_2(N,Q^2)$ the structure function still exists at zeroth order in 
$\alpha_s$ and hence, it is not a
perturbative quantity in quite the sense way as $F_L^{NS}(N,Q^2)$
or $(d\, F_2^{NS}(N,Q^2)/d\ln Q^2)$, both of which vanish with $\alpha_s$.
So in our simplified model the LORSC expression for $F_2^{NS}(N,Q^2)$, 
obtained by combining the LO input and evolution, is
\eqn\loexpnssfpret{F^{NS}_{2,RSC,0}(N,Q^2)=F^{NS}_{2,0}(N)[1+\alpha_s(Q_I^2)
\ln(Q_I^2/A^{NS}_2)\Gamma^{2,0,l}_{NS}(N+1)]
\biggl({\alpha_s(Q_I^2)\over \alpha_s(Q^2)}
\biggr)^{\tilde\Gamma_{2,0,l}^{NS}(N+1)}.}
This consists of two parts which are clearly of different magnitude, and 
it is clear that the same effect will be 
seen for the singlet structure function.

If we instead consider $(d\,F_2^{NS}(N,Q^2)/d\,\ln Q^2)$ 
the full expression is 
\eqn\ftwoderivnsexample{\eqalign{{d\,F_2^{NS}(N,Q^2)\over d\,\ln Q^2}  
 =&\Gamma_{2,NS}(N+1,\alpha_s(Q_I^2))F_2^{NS}(N,Q_I^2)\times\cr
&\exp\biggl[\int_{\ln Q_I^2}^{\ln Q^2}
\biggl(\Gamma_{2,NS}(N+1,\alpha_s(q^2))
-{d\over d\,\ln q^2}\ln(\Gamma_{2,NS}(N+1,\alpha_s(q^2)))\biggr)
d\ln q^2\biggr].\cr}}
Hence, the input may be written as
\eqn\nsinputderivftwo{\eqalign{\biggl({d\,F^{NS}_{2,0}(N,Q^2)\over d\,\ln Q^2}
\biggr)_{Q_I^2}=F^{NS}_{2,0}(N)\bigl[
\alpha_s(Q_I^2)\Gamma^{2,0,l}(N+1)
+\alpha^2_s(Q_I^2)&
\bigl(\ln(Q_I^2/A^{NS}_2)(\Gamma^{2,0,l}_{NS}(N+1))^2\cr
&+\Gamma^{2,1,l}(N+1)\bigr)+{\cal O} (\alpha^3_s(Q_I^2)) \bigr].\cr}}
The ${\cal O}(\alpha^2_s(Q_I^2))$ piece is 
renormalization scheme dependent, 
so this time we have a LORSC input which is of a 
given order in $\alpha_s(Q_I^2)$. The full LO expression for 
$(dF_2(N,Q^2)/d\ln Q^2)$ is then 
\eqn\nsderivftwoex{\biggl({d\,F^{NS}_{2,0}(N,Q^2)\over d\,\ln Q^2}
\biggr)_{0}=F^{NS}_{2,0}(N)
\alpha_s(Q_I^2)\Gamma^{NS}_{2,0,l}(N+1)\biggl({\alpha_s(Q_I^2)\over 
\alpha_s(Q^2)}\biggr)^{\tilde\Gamma_{2,0,l}^{NS}(N+1)-1}.}
This is rather more satisfactory than the 
renormalization--scheme--independent expression for $F_2^{NS}(N,Q^2)$ itself
\loexpnssfpret, and hence we choose $(dF_2(N,Q^2)/d\ln Q^2)$ 
to be the perturbative
quantity we calculate, in both the nonsinglet and singlet case. 

If we wish to calculate the structure function $F_2(N,Q^2)$ itself to a 
given order we will do this by integrating the given order expression for 
$(dF_2(N,Q^2)/d\ln Q^2)$ between $Q_I^2$ and $Q^2$, 
and adding it to $F_2(N,Q_I^2)$
evaluated to the same order. For example, in our simplified nonsinglet model 
we would integrate \nsderivftwoex\ and add this to the explicitly 
written part of \nsinputftwo. This results in the effective LO expression 
\eqn\loexpnssfpretalt{F^{NS}_{2,RSC,0}(N,Q^2)=F^{NS}_{2,0}(N)\biggl[
\biggl({\alpha_s(Q_I^2)\over \alpha_s(Q^2)}
\biggr)^{\tilde\Gamma_{2,0,l}^{NS}(N+1)} +\alpha_s(Q_I^2)
\ln(Q_I^2/A^{NS}_2)\Gamma^{2,0,l}_{NS}(N+1)\biggr].}

Of course, this whole discussion of $F^{NS}_{2}(N,Q^2)$ has been rather 
simplified by the assumption that the higher--order--in--$\alpha_s$ 
terms in the
physical anomalous dimension do not contain higher singularities in $(N+1)$. 
Recognizing that they do, we obtain 
\eqn\rscftwons{\eqalign{\biggl(&{d\,F^{NS}_{2,RSC,0}(N,Q^2)\over d\,\ln Q^2}
\biggr)_0=\alpha_s(Q_I^2)F^{NS}_{2,0}(N)
\biggl(\Gamma^{NS}_{2,0,l}(N+1) +
{1\over N+1}\times\cr
&\biggl(\sum_{m=1}^{\infty}\bigl( \tilde b_m
+ (N+1)\tilde c_m\bigr)\biggl({\alpha_s(Q_I^2)
\over (N+1)^2}\biggr)^m+{\tilde a_{-1}\alpha_s(Q_I^2)\over (N+1)}
\ln(Q_I^2/A^{NS}_2) \sum_{m=0}^{\infty}
\tilde b_m\biggl({\alpha_s(Q_I^2)\over (N+1)^2}\biggr)^m
\biggr)\times\cr
&\biggl({\alpha_s(Q_I^2)\over \alpha_s(Q^2)}
\biggr)^{\tilde\Gamma_{2,0,l}^{NS}(N+1)-1}
\exp\biggl[
\int_{\alpha_s(Q^2)}^{\alpha_s(Q_I^2)}
\biggl(\sum_{m=1}^{\infty} \biggl({1\over N+1}\tilde b_m+\tilde c_m\biggr)
\biggl({\alpha_s(q^2)\over (N+1)^2}\biggr)^m\biggr){d\alpha_s(q^2)\over 
b_0\alpha^2_s(q^2)}\biggr]\times\cr
&\biggl({1+\sum_{m=1}^{\infty} \tilde b_m(\alpha^{m+1}_s(Q^2)/ (N+1)^{2m})
\over 1+\sum_{m=1}^{\infty} \tilde b_m(\alpha^{m+1}_s(Q_I^2)/ (N+1)^{2m})}
\biggr)\biggr)\cr}}
and
\eqn\loinputnstwosf{\eqalign{F^{NS}_{2,RSC,0}&(N,Q_I^2)=F^{NS}_{2,0}(N)\biggl(
1 +\ln(Q_I^2/A^{NS}_2)\alpha_s(Q_I^2)\biggl(\Gamma^{NS}_{2,0,l}(N+1)+\cr
&{1\over N+1}
\sum_{m=1}^{\infty} \tilde b_m\biggl({\alpha_s(Q_I^2)
\over (N+1)^2}\biggr)^m
+ \sum_{m=0}^{\infty}\bigl(\tilde c_m+
\half\sum_{n=0}^{m}\tilde b_n\tilde b_{m-n}\ln(Q_I^2/A^{NS}_2)
\bigr)\biggl({\alpha_s(Q_I^2)
\over (N+1)^2}\biggr)^m\biggr)\biggr),\cr}}
where for $F^{NS}_{2}(N,Q^2)$ the coefficients in the series in 
$(N+1)^{-1}$are not necessarily 
the same as for $F^{NS}_{L}(N,Q^2)$, hence the slightly 
different notation. However, $a_{-1}=\tilde a_{-1}$ and
$b_m =\tilde b_m$ (probably \nsresumphen). The $c_m$ are not equal to the 
$\tilde c_m$ though, so there is no guarantee that 
$F^{NS}_{2,RSC,0}(x,Q^2)$ and $F^{NS}_{L,RSC,0}(x,Q^2)$ will behave in the 
same way in the small--$x$ limit. 

\medskip

When all the $b_m$ and $c_m$ are known they can be used to 
present an argument 
for the form of the small--$x$ behaviour of nonsinglet structure functions. 
Until this happens 
our discussion of the LORSC calculation of the nonsinglet structure
functions is rather academic. 
However, it has enabled us to discuss many of the issues in a simpler 
framework than if we had gone directly to the singlet structure functions.
We will discuss these singlet structure functions next.

\subsec{The Renormalization--Scheme--Consistent Singlet Structure Functions.}

When calculating the singlet structure functions
we cannot just construct the complete LO
evolution and input and combine these to obtain the LO 
expression. 
Each of the component parts of the LO expressions for $F_2(N,Q^2)$ and 
$F_L(N,Q^2)$ (we omit the superscript $S$ in this section) must
consist of LO input parts and evolution parts, but 
it is not obvious what these are. In order to find the full LORSC 
expressions for the singlet structure functions we will have to work in 
steps. We will consider only the full LO expression with the perturbative 
factors expanded about the particular value of $N=0$, 
and the simplest way to proceed is to solve
the evolution equations in terms of physical anomalous 
dimensions and structure functions. We have already proved that in the loop
expansion the LO expressions only depend on the one--loop physical 
anomalous dimensions, and in the leading--$\ln (1/x)$ expansion the  
LO expressions depend only on \physanomval. Hence,  
it is only the combination of these anomalous dimensions which is 
considered in our solution. 

We cannot simply write the physical anomalous dimension matrix
\eqn\inclmatrix{\pmatrix{\alpha_s(Q^2)\Gamma^{0,l}_{LL}(N)+
\Gamma^{\tilde 0}_{LL}(\alpha_s(Q^2)/N)& \alpha_s(Q^2) 
\Gamma^{0,l}_{L2}(N)+\Gamma^{\tilde 0}_{L2}(\alpha_s(Q^2)/N)\cr
\alpha_s(Q^2)\Gamma^{0,l}_{2L}(N)+\alpha_s(Q^2)
\Gamma^{\tilde 1}_{2L}(\alpha_s(Q^2)/N)&\alpha_s(Q^2)\Gamma^{0,l}_{22}(N) +
\alpha_s(Q^2)\Gamma^{\tilde 1}_{22}(\alpha_s(Q^2)/N)\cr}}
(where $\Gamma^{\tilde 0}_{LL}(\alpha_s(Q^2)/N)=
\Gamma^{0}_{LL}(\alpha_s(Q^2)/N)$ with the one--loop component subtracted 
out, etc.), and solve the renormalization group equations because 
the full solution contains terms 
which are not properly of leading order. We must choose some way of 
solving for the structure functions systematically which enables us to 
extract the true LO behaviour in as simple a manner as possible. 
In order to do this we take account of the fact that the one--loop 
solutions for $F_L(N,Q^2)$ and $(dF_2(N,Q^2)/d\ln Q^2)$ must be part of the 
complete LORSC solutions. Hence, we split our anomalous dimension 
matrix up into the form 
\eqn\inclmatrixsplit{\alpha_s(Q^2)\pmatrix{\Gamma^{0,l}_{LL}(N)+
& \Gamma^{0,l}_{L2}(N)\cr
\Gamma^{0,l}_{2L}(N)& 
\Gamma^{0,l}_{22}(N)\cr}
+\pmatrix{\Gamma^{\tilde 0}_{LL}(\alpha_s(Q^2)/N)& 
\Gamma^{\tilde 0}_{L2}(\alpha_s(Q^2)/N)\cr
\alpha_s(Q^2)\Gamma^{\tilde 1}_{2L}(\alpha_s(Q^2)/N)&
\alpha_s(Q^2)\Gamma^{\tilde 1}_{22}(\alpha_s(Q^2)/N)\cr},}
and solve by treating the second matrix as a perturbation to the first.
Doing this we
obtain the one--loop solutions as the lowest--order solutions and can
systematically calculate corrections to this, extracting the parts of these 
``corrections'' which are leading order. 

So, first let us consider the solution to the ${\cal O}(\alpha_s)$
renormalization group equation
with boundary conditions $\hat F^{0,l}_L(N,Q_I^2)=\hat F_L(N)$ and
$F^{0,l}_2(N,Q_I^2)=F_2(N)$. We may write the solution for the longitudinal
structure function as
\eqn\fullsolii{\hat F^{0,l}_L(N,Q^2)=\hat F^{0,l,+}_L(N)\biggl({\alpha_s(Q_I^2)
\over \alpha_s(Q^2)}\biggr)^{\tilde \Gamma^{0,l,+}(N)}+\hat F^{0,l,-}_L(N)
\biggl({\alpha_s(Q_I^2)
\over \alpha_s(Q^2)}\biggr)^{\tilde \Gamma^{0,l,-}(N)},}
where $\tilde \Gamma^{0,l,+,-}(N)$ are the two eigenvalues of the 
zeroth--order 
physical anomalous dimension matrix (which are the same as the eigenvalues 
of the zeroth--order parton anomalous dimension matrix),
and in practice
\eqn\fullsolc{\hat F^{0,l,+}_L(N)=\hat F_L(N)-
{36-8N_f\over 27}F_2(N)+{\cal O}(N),\hskip 0.2in \hat F^{0,l,-}_L(N)={36-8N_f
\over 27} F_2(N)+{\cal O}(N).} 
Having chosen to 
write the lowest--order 
solution for the longitudinal structure function in this way we may then 
write the lowest--order solution for $F_2(N,Q^2)$ as
\eqn\fullsoliii{\hat F^{0,l}_2(N,Q^2)=e^+(N)\hat F^{0,l,+}_L(N)
\biggl({\alpha_s(Q_I^2)
\over \alpha_s(Q^2)}\biggr)^{\tilde\Gamma^{0,l,+}(N)}+e^-(N)\hat F^{0,l,-}_L(N)
\biggl({\alpha_s(Q_I^2)
\over \alpha_s(Q^2)}\biggr)^{\tilde \Gamma^{0,l,-}(N)},}
where 
\eqn\fullsola{\eqalign{e^+(N)= &\biggl({\Gamma^{0,l,+}(N)-
\Gamma^{0,l}_{LL}(N)\over
\Gamma^{0,l}_{L2}(N)}\biggr)=N/6+{\cal O}
(N^2),\cr 
e^-(N)= &\biggl({\Gamma^{0,l,-}(N)-\Gamma^{0,l}_{LL}
(N)\over \Gamma^{0,l}_{L2}(N)}\biggr)=\biggl({27\over 36-8N_f}\biggr)
+{\cal O}(N),\cr}}

The first corrections to the one--loop solution may be obtained by solving 
the equations,
\eqn\fullsoliv{\eqalign{{d\over d\, \ln Q^2}\pmatrix{\hat F^{c1}_L(N,Q^2)\cr 
F^{c1}_2(N,Q^2)\cr}&=\alpha_s(Q^2)\pmatrix{\Gamma^{0,l}_{LL}(N)
&\Gamma^{0,l}_{L2}(N)\cr
\Gamma^{0,l}_{2L}(N)& 
\Gamma^{0,l}_{22}(N)\cr}\pmatrix{\hat F^{c1}_L(N,Q^2)\cr 
F^{c1}_2(N,Q^2)\cr}\cr
&\hskip -0.6in+\pmatrix{\Gamma^{\tilde 0}_{LL}(\alpha_s(Q^2)/N)& 
-{(36-8N_f)\over 27}\Gamma^{\tilde 0}_{LL}(\alpha_s(Q^2)/N)\cr
\alpha_s(Q^2)\Gamma^{\tilde 1}_{2L}(\alpha_s(Q^2)/N)&
-{(36-8N_f)\over 27}\alpha_s(Q^2)\Gamma^{\tilde 1}_{2L}(\alpha_s(Q^2)/N)\cr}
\pmatrix{\hat F^{0,l}_L(N,Q^2)\cr F^{0,l}_2(N,Q^2)\cr},\cr}}
where we have used the relationships between the physical anomalous 
dimensions in order to simplify the second matrix. 
We proceed as follows. First we define the vectors
\eqn\fullsolv{\underline e^+(N)=\pmatrix{1\cr e^+(N)\cr}, \hskip 0.5in
\underline e^-(N)=\pmatrix{1\cr e^-(N)\cr},\hskip 0.5in \underline 
F^{c1}(N,Q^2)=
\pmatrix{F^{c1}_L(N,Q^2)\cr F^{c1}_2(N,Q^2)\cr},}
and write 
\eqn\neeqni{\underline F^{c1}(N,Q^2)= \underline e^+(N)F^{c1,+}(N,Q^2)+
\underline e^-(N)F^{c1,-}(N,Q^2).}
We also define projection operators $\underline p^+(N)$ and 
$\underline p^-(N)$ ($\underline p^+(N)\cdot\underline e^+(N) =1$,
$\underline p^+(N)\cdot\underline e^-(N) =0$ etc.) which in practice are
\eqn\fulsolvii{\underline p^+(N)=\pmatrix{1\cr{8N_f-36\over 27}\cr}
+ {\cal O}(N), \hskip 1in \underline p^-(N)=\pmatrix{0\cr{27 \over 8N_f-36}\cr}
+ {\cal O}(N).}
Multiplying \fullsoliv\ by $\underline p^+(N)$ or $p^-(N)$ leads to  
straightforward first--order differential equations for $F^{c1,+}(N,Q^2)$
or $F^{c1,-}(N,Q^2)$. 
Writing the zeroth--order solution $\underline F^{0,l}(N,Q^2)$, 
in the form 
\eqn\neweqnii{\underline F^{0,l}(N,Q^2)=
\underline e^+(N)F^{0,l,+}(N,Q^2)+\underline e^-(N)F^{0,l,-}(N,Q^2),}
these equation can now be solved by using the power series expansions of 
$\underline e^{+(-)}(N)$ and $\underline p^{+(-)}(N)$ in terms of $N$. 
Only a small part of the overall solution contributes at LO.

Solving for the complete leading--order part of $F^+(N,Q^2)$ 
inductively in this 
manner, and making it formally insensitive to changes of $Q_I^2$ we get
\eqn\fullsolxvii{\eqalign{F^+_{RSC,0}(N,Q^2)=&\biggl[
\biggl(\hat F_L(N)-\biggl({36-8N_f\over 27}\biggr)F_2(N)\biggr)
(\exp[\ln(Q_I^2/A_{LL})
\Gamma^0_{LL}(\alpha_s(Q_I^2)/N)]-1)\cr
+\hat F^{0,l,+}_L(N)\biggr]
&\biggl({\alpha_s(Q_I^2)\over
\alpha_s(Q^2)}\biggr)^{\tilde \Gamma^{0,l,+}(N)}
\exp\biggl[\int_{\alpha_s(Q^2)}^{\alpha_s(Q_I^2)}
{\tilde \Gamma^{\tilde 0}_{LL}(\alpha_s(q^2)/N)\over \alpha^2_s(q^2)}
d\alpha_s(q^2)\biggr].\cr}}
We can solve for the corrections which are proportional to $\underline 
e^-(N)$ in exactly the same manner, using $\underline p^-(N)$ instead of 
$\underline p^+(N)$.
Inserting the whole of $F^+_{RSC,0}(N,Q^2)$ into the right--hand side
the solution is relatively simple if we wish to keep only
the LO parts. 
Adding to the solution at one--loop, the whole of the LO part of 
$F^-(N,Q^2)$ is
\eqn\fullsolxxv{\eqalign{F^{-}_{RSC,0}(N,Q^2)&=
\biggl(F^{0,l,-}_L(N) -\biggl({36-8N_f\over 27}\biggr)
\biggl(\hat F_L(N)-
\biggl({36-8N_f\over 27}\biggr)F_2(N)\biggr)\times\cr
&\biggl({\alpha_s(Q_I^2)
\Gamma^{\tilde 1}_{2L}(\alpha_s(Q_I^2)/N)
-{N\over 6}\Gamma^{\tilde 0}_{LL}(\alpha_s(Q_I^2)/N)\over 
\Gamma^{0}_{LL}(\alpha_s(Q_I^2)/N)}\biggr)\biggr)
\biggl({\alpha_s(Q_I^2)\over
\alpha_s(Q^2)}\biggr)^{\tilde \Gamma^{0,l,-}(N)}+\cr
&\hskip -0.7in\biggl({36-8N_f\over 27}\biggr)
\biggl({\alpha_s(Q_I^2)
\Gamma^{\tilde 1}_{2L}(\alpha_s(Q_I^2)/N)
-{N\over 6}\Gamma^{\tilde 0}_{LL}(\alpha_s(Q_I^2)/N)\over 
\Gamma^{0}_{LL}(\alpha_s(Q_I^2)/N)}\biggr)
\biggl(\hat F_L(N)-
\biggl({36-8N_f\over 27}\biggr)F_2(N)\biggr)\times\cr
&\hskip -0.5in \exp[\ln(Q_I^2/A_{LL})
\Gamma^0_{LL}(\alpha_s(Q_I^2)/N)]\biggl({\alpha_s(Q_I^2)\over
\alpha_s(Q^2)}\biggr)^{\tilde \Gamma^{0,l,+}(N)}\exp\biggl[
\int_{\alpha_s(Q^2)}^{\alpha_s(Q_I^2)}
{\tilde \Gamma^{\tilde 0}_{LL}(\alpha_s(q^2)/N)\over \alpha^2_s(q^2)}
d\alpha_s(q^2)\biggr]\cr
&+\hbox{higher order},\cr}}
where, in order to make the expression invariant under changes in starting 
scale, and also ensure that for $Q_I^2$ =$A_{LL}$ we have $F^{-}(N,Q_I^2)=
\hat F_L^{0,l,-}(N)$, we make the choice
\eqn\deffmininput{\eqalign{F^{cfull,-}(N,Q_I^2)=\biggl({36-8N_f\over 27}\biggr)
&\biggl({\alpha_s(Q_I^2)
\Gamma^{\tilde 1}_{2L}(\alpha_s(Q_I^2)/N)
-{N\over 6}\Gamma^{\tilde 0}_{LL}(\alpha_s(Q_I^2)/N)\over 
\Gamma^{0}_{LL}(\alpha_s(Q_I^2)/N)}\biggr)\times\cr
&\hskip -0.6in \biggl(\hat F_L(N)-
\biggl({36-8N_f\over 27}\biggr)F_2(N)\biggr)(\exp[\ln(Q_I^2/A_{LL})
\Gamma^0_{LL}(\alpha_s(Q_I^2)/N)]-1).\cr}}

We must now use the leading parts of $F^+(N,Q^2)$ and $F^{-}(N,Q^2)$
in order to obtain LORSC expressions for the structure functions. The way 
in which we have set up the calculation 
makes this very straightforward for the longitudinal structure function: we  
multiply $F^+_{RSC,0}(N,Q^2)$ and $F^{-}_{RSC,0}(N,Q^2)$ by $(\alpha_s(Q^2)/
2\pi)$, where the only part of $F^{-}_{RSC,0}(N,Q^2)$
which contributes to the LORSC expression for $F_L(N,Q^2)$ is 
the one--loop part,
\eqn\fullsolfl{\eqalign{F_{L,RSC,0}(N,Q^2)&={\alpha_s(Q_I^2)\over 2\pi}
\Biggl[\biggl({\alpha_s(Q_I^2)\over
\alpha_s(Q^2)}\biggr)^{\tilde\Gamma^{0,l,+}(N)-1}
\exp\biggl[\int_{\alpha_s(Q^2)}^{\alpha_s(Q_I^2)}
{\tilde \Gamma^{\tilde 0}_{LL}(\alpha_s(q^2)/N)\over \alpha_s(q^2)}
d\alpha_s(q^2)\biggr]\times\cr
&\hskip -0.3in\biggl(F^{0,l,+}_L(N)+
\biggl(\hat F_L(N)-\biggl({36-8N_f\over 27}\biggr)F_2(N)\biggr)
(\exp[\ln(Q_I^2/A_{LL})\Gamma^0_{LL}(\alpha_s(Q_I^2)/N)]-1)\biggr)\cr
&\hskip -0.3in +F^{0,l,-}_L(N)\biggl({\alpha_s(Q_I^2)\over
\alpha_s(Q^2)}\biggr)^{\tilde \Gamma^{0,l,-}(N)-1}\Biggr].\cr}}

We also wish to find the LORSC expressions for 
$(d\,F_2(N,Q^2)/d\,\ln Q^2)$ and for the input $F_2(N,Q_I^2)$.
We consider the former first. Using the form of $e^+(N)$ and $e^-(N)$ in 
\fullsola\ it is clear that, besides for the one--loop contributions, 
the LORSC expression   
will come from $(27/(36-8N_f))\times(d\,F^-_2(N,Q^2)/d\,\ln Q^2)$
and from $N/6 \times(d\,F^+_2(N,Q^2)/d\,\ln Q^2)$. Explicitly we obtain 
\eqn\fullsolfderiv{\eqalign{\biggl({d\,F_{2}(N,Q^2)\over d\,\ln Q^2}
\biggr)_{RSC,0}&
=\alpha_s(Q_I^2)\Biggl[e^-(N)\Gamma^{0,l,-}(N)
\hat F^{0,l,-}_L(N)\biggl({\alpha_s(Q_I^2)\over
\alpha_s(Q^2)}\biggr)^{\tilde \Gamma^{0,l,-}(N)-1}\cr
&+\biggl(e^+(N)\Gamma^{0,l,+}(N)\hat F^{0,l,+}_L(N)- \Gamma^{1,0}_{2,L}(N)
\biggl(\hat F_L(N)-\biggl({36-8N_f\over 27}\biggr)F_2(N)\biggr)\cr
&\hskip-0.6in+\Gamma^{1}_{2L}(\alpha_s(Q_I^2)/N)
\biggl(\hat F_L(N)-\biggl({36-8N_f\over 27}\biggr)F_2(N)\biggr)
\exp[\ln(Q_I^2/A_{LL})
\Gamma^0_{LL}(\alpha_s(Q_I^2)/N)]\biggr)\times\cr
&\exp\biggl[
\int_{\alpha_s(Q^2)}^{\alpha_s(Q_I^2)}
{\tilde \Gamma^{\tilde 0}_{LL}(\alpha_s(q^2)/N)\over \alpha^2_s(q^2)}
d\alpha_s(q^2)\biggr]\biggl({\alpha_s(Q_I^2)\over
\alpha_s(Q^2)}\biggr)^{\tilde \Gamma^{0,l,+}(N)-1}\Biggr],\cr}}
where ${N\over 6}(\Gamma^0_{LL}(\alpha_s(Q^2)/N)
-\Gamma^{\tilde 0}_{LL}(\alpha_s(Q^2)/N))=\alpha_s(Q^2)(\Gamma^1_{2L}
(\alpha_s(Q^2)/N)-\Gamma^{\tilde 1}_{2L}(\alpha_s(Q^2)/N))$ 
has been used and $\Gamma^{1,0}_{2,L}(N)\biggl(
\hat F_L(N)-\biggl({36-8N_f\over27}\biggr)F_2(N)\biggr)$ is the part of the 
input which is common to both the one--loop and the leading--$\ln (1/x)$ 
input, and is subtracted from the former in order to avoid double counting. 
Similarly the input for $F_2(N,Q^2)$ is given by the relatively simple form
\eqn\fullsolfin{\eqalign{F_{2,RSC,0}(N,Q_I^2) &=F_2(N)\cr 
&\hskip -0.8in + \alpha_s(Q_I^2)
{\Gamma^1_{2L}(\alpha_s(Q_I^2)/N)\over\Gamma^0_{LL}(\alpha_s(Q_I^2)/N)}
\biggl(\hat F_L(N)-{(36-8N_f)\over 27}
F_2(N)\biggr)(\exp[\ln(Q_I^2/A_{LL})\Gamma^0_{LL}(\alpha_s(Q_I^2)/N)]-1)\cr
&\hskip -0.8in+\ln(Q_I^2/A_{LL})
\alpha_s(Q_I^2)\biggl(e^+(N)\Gamma^{0,l,+}(N)\hat 
F_L^{0,l,+}(N,Q_I^2) + e^-(N)\Gamma^{0,l,-}(N)\hat F_L^{0,l,-}(N,Q_I^2) \cr
&\hskip 1.2in -\Gamma^{1,0}_{2,L}(N)\biggl(
\hat F_L(N)-\biggl({36-8N_f\over27}\biggr)F_2(N)\biggr)\biggr).}}
The third term is the renormalization--scheme--invariant 
order--$\alpha_s(Q_I^2)$ input, which must compensate for changes in the 
one--loop evolution under changes in $Q_I^2$. Again we explicitly
extract a term $\propto\ln(Q_I^2/A)\Gamma^{1,0}_{2,L}(N)\biggl(
\hat F_L(N)-\biggl({36-8N_f\over27}\biggr)F_2(N)\biggr)$ in order to avoid
double counting. The fact that this term can be thought 
of as appearing from two different sources leads us to choose both our 
unknown scale constants in the input equal to the same value $A_{LL}$. 
We note that our choice of inputs not only ensures $Q_I^2$--invariance 
up to higher orders, but are also such that our expressions for
$\hat F_{L,RSC,0}(N,Q_I^2)$ and $F_{2,RSC,0}(N,Q_I^2)$ reduce to the 
nonperturbative inputs if $Q_I^2=A_{LL}$.
Having obtained our LORSC expressions for $(d\,F_2(N,Q^2)/d\,\ln Q^2)$ and 
$F_2(N,Q_I^2)$, then as already argued, in order to obtain our expression 
for $F_2(N,Q^2)$ we integrate $(d\,F_{2}(N,Q^2)/d\,\ln Q^2)_{RSC,0}$ from 
$Q_I^2$ to $Q^2$ and add to the input $F_{2,RSC,0}(N,Q_I^2)$.

\medskip

Thus, we have our complete leading--order
renormalization--scheme--consistent expressions for the structure functions. 
These are significantly different from both the one--loop expressions 
and the leading--$\ln(1/x)$ expressions, although they do reduce to them 
in the appropriate limits. Indeed, once 
we include the ${\cal O}(\alpha_s(Q_I^2))$ inputs for $F_2(N,Q^2)$ in 
the definition of the leading--order inputs in the loop--expansion
(as we should), each of 
the terms in the inputs and evolution terms in \fullsolfl, \fullsolfderiv\ 
and \fullsolfin\ contains a part which appears in both the LO
expression in the loop expansion and the LO expression in the leading
$\ln(1/x)$ expansion. Our full LORSC expressions are obtained rather 
more easily than above by simply letting each of the input and evolution terms 
become the combination of the terms in the two expansion schemes. In a sense 
this result is obvious, but it is necessary to verify this by deriving the
expressions as above.

Let us comment on the form of our final LORSC expressions.
We note that as for the leading--$\ln (1/x)$ expansion, 
the LORSC expansion still leads to 
predictions for all the small--$x$ inputs: predictions of each in terms
of the nonperturbative inputs (which we imagine should be quite flat) and the 
nonperturbative scale $A_{LL}$, and also stronger predictions for the 
relationships between the inputs (although the scale $Q_I^2$ at which they
should be imposed is not determined). 
 
Finally, we notice that each of the terms appearing in our expressions is
manifestly renormalization scheme invariant, and it is clear that 
no terms are subleading in $\alpha_s$ to any other terms, either in the
input or in the evolution. If we had simply solved the renormalization group
equations using the whole of the anomalous dimension matrix \inclmatrix\
then we would have obtained many terms which do not appear in our full 
leading--order expressions \fullsolfl, \fullsolfderiv\ and \fullsolfin. 
These would still be 
renormalization scheme independent, since our input anomalous dimensions 
are renormalization scheme independent. However, 
they would be of the same form as terms which are
renormalization scheme dependent (e.g. we saw 
in 4.2 that the subleading--in--$\ln (1/x)$ evolution 
$\Phi^+_1(Q^2,Q_I^2)$ has a manifestly 
renormalization--scheme--independent part depending on $\Gamma^1_{2L}
(\alpha_s(Q^2)/N)$ as well as a renormalization--scheme--dependent part 
depending on $\Gamma^1_{LL}(\alpha_s(Q^2)/N)$). These terms should 
be dropped, and \fullsolfl, \fullsolfderiv\ and \fullsolfin\ are the 
correct expressions for the structure functions to be used 
with the one--loop coupling constant. 

\newsec{$x$--Space Solutions.}

We shall now discuss how we use the expressions \fullsolfl, \fullsolfderiv\ 
and \fullsolfin\ in order to obtain our expressions for 
the $x$--space structure 
functions and ultimately compare with data. The data on 
${\cal F}_2(x,Q^2)$ exist
over a range of $Q^2$ from $\sim 0.2\Gev^2$ to $5000 \Gev^2$, though 
in practice we will impose a lower cut on the data of $Q^2=2\Gev^2$, 
except for the HERA data where we choose $Q^2=1.5\Gev^2$ simply in order not
to lose some of the very low $x$ data.  The threshold of heavy quark
production is $W^2 = 4m_H^2$, where
$W$ is the invariant mass of the hadronic system created from the struck
proton (or neutron) and in the limit of zero proton (and/or neutron) mass is
given by $W^2=Q^2(x^{-1}-1)$. $m_H$ is the mass of the heavy quark. 
Thus, we clearly work in the limit where the up, down and strange
quarks are effectively massless. However, we cross the $b$--quark
threshold of $W^2 \approx 20 \Gev^2$, and at the lower end of our range are 
in the region of the $c$--quark threshold of $W^2\approx 2\Gev^2$.
Hence, we will need expressions for the structure functions which cross 
quark thresholds.

The correct treatment of these heavy quark thresholds is less well established
than the treatment of effectively massless quarks, and is certainly more
complicated. 
Here we use the prescription for treating heavy quarks outlined in 
\ref\coltung{J.C. Collins and W.K. Tung, \NP \vyp{B278}{1986}{934}.}. This
involves treating all the quarks as massless, but only allowing the
heavy quarks to become active above the simple threshold $Q^2=m_H^2$. Hence,
the value of $N_f$ appearing explicitly in any expressions
changes discontinuously at this threshold. The running
coupling constant, is defined to be continuous at the thresholds. It is 
determined by the relationship 
\eqn\couplink{\alpha_{s,n}(Q^2)=\alpha_{s,n+1}(Q^2)\biggl(1+
{\alpha_{s,n+1}(Q^2)\over 6\pi}\ln(m^2_{n+1}/Q^2)\biggr),}
where the central $\alpha_s(Q^2)$ with $N_f=4$ is defined by 
\eqn\couplinkfour{\alpha_{s,4}(Q^2)={12\pi \over (33-2\cdot(N_f=4))
\ln(Q^2/\Lambda^2_{QCD,4})}.}
This complete prescription for treatment of
the heavy quarks is consistent with the decoupling theorem, as it is 
guaranteed to provide the correct expressions far above or below any 
threshold: the increase in the coupling below a 
threshold compensating for the absence of virtual heavy quarks in calculations 
below this threshold. It is clearly rather 
unsatisfactory in the region of
the threshold, with heavy quark structure functions having an abrupt
threshold in $Q^2$ rather than a smooth threshold 
in $W^2$. Work to rectify
this is in progress, and will certainly involve the use of the heavy 
quark coefficient functions at leading order in $\ln (1/x)$ already 
calculated \kti\cat. 

\medskip

We can now discuss the form 
of the $x$--space solutions for the structure functions. 
It is the perturbative part of the moment--space structure functions for 
which we can produce well--ordered, RSC expressions and 
the nonperturbative parts of the expressions for these structure functions, 
$F_2(N)$ and $\hat F_L(N)$, will be nonanalytic, complicated functions of $N$. 
Our complete moment--space expression for a general structure 
function  (or derivative of a structure function) will be
\eqn\compstrN{ F_i(N,Q^2) = P_{i,2}(N, \alpha_s)F_2(N) + P_{i,L}(N,\alpha_s) 
\hat F_L(N),}
where the $P_{ij}(N,\alpha_s)$'s are the calculable perturbative components  
of the complete expressions which can be expanded as power series in 
both $\alpha_s$ and $N$, i.e. examples of the physically relevant 
perturbative functions we discussed in \S 4.3.
We obtain the $x$--space structure 
functions as follows. We factor out 
the perturbative part of each of these expressions which is proportional to 
either $\hat F_L(N)$ or $F_2(N)$, and take the inverse Mellin transformation
of this perturbative function
by integrating around a contour encircling $N=0$, 
obtaining the leading--in--$\ln(1/x)$, at 
lowest--order--in--$x$ parts of the 
perturbative ${\cal P}(x,\alpha_s)$'s as well as all the 
lowest--order--in--$\alpha_s$ parts at lowest--order--in--$x$.
In order to obtain the leading--order--in--$\ln(1/x)$--and--$\alpha_s$,
at lowest--order--in--$x$, part of the perturbatively calculated 
structure functions we then convolute with the whole of the nonperturbative
$\hat F_L(x)$ and $F_2(x)$ which are obtained by a complete inverse Mellin
transformation of $\hat F_L(N)$ and $F_2(N)$. 
In principle, the full LORSC $x$--space structure functions are calculated by
repeating this procedure for the LORSC moment--space structure functions 
where the perturbative parts are expanded about $N$ equal to all 
negative integers, but in practice one only really needs
include the one--loop, or leading--in--$\alpha_s$ part of the full 
LORSC perturbative parts for all 
negative integers.

In practice the calculation of the structure functions is performed in a 
rather different manner. The structure functions are calculated using a 
modification of the computer program that is used by Martin, 
Roberts and Stirling in 
their global fits to structure function data. This works in terms of parton
densities directly in $x$--space. Input parton densities are 
specified at some scale $Q_0^2$ and the $Q^2$ evolution is calculated 
on a grid in $x$ and $Q^2$. This evolution is obtained by
integrating up the renormalization group equations involving the 
complete parton distributions and full specified splitting functions. 
The structure functions are calculated by numerically 
performing the convolutions of the resulting $Q^2$--dependent 
parton distributions with coefficient functions. 
Thus, in fact, we obtain the scheme--independent LORSC 
structure functions by working in terms of parton densities 
and choosing coefficient functions and splitting functions (i.e. an effective
factorization scheme) which will reproduce the correct expressions
as closely as possible. 

We will briefly describe our choice of effective splitting functions
as follows. At first
order in $\alpha_s$ we choose the normal parton splitting functions. We then 
add corrections to these in order to reproduce our desired results. 
Denoting these corrections by $\Delta \gamma_{ab}(N,\alpha_s(Q^2))$,
for a given parton to parton splitting function, and expressing these 
in moment space and in terms of previously discussed quantities 
for simplicity we get for ${\cal F}_L(x,Q^2)$ 
\eqn\anomcorrl{\eqalign{ \Delta \gamma_{fg}(N,\alpha_s(Q^2))&=
\Delta \gamma_{ff}(N,\alpha_s(Q^2))=0\cr
\Delta \gamma_{gg}(N,\alpha_s(Q^2)) &= 
\ninefourths\Delta \gamma_{gf}(N,\alpha_s(Q^2)) = 
\Gamma^{\tilde 0}_{LL}(\alpha_s(Q^2)/N).\cr}}
This form is not too difficult to understand by 
looking at  e.g. \fullsolfl: the only leading--$\ln (1/x)$
enhancement of the one-loop expression comes from the evolution in the 
longitudinal sector which is directly related to the 
enhanced gluon evolution and hence corrections to the gluon anomalous 
dimensions.
For ${\cal F}_2(x,Q^2)$ the corrections are a little more involved: 
\eqn\anomcorrtwo{\eqalign{ \Delta \gamma_{fg}(N,\alpha_s(Q^2))&=
\alpha_s(Q^2)\hat C^g_{2,1,l}(N) \Gamma^{\tilde 1}_{2L}(\alpha_s(Q_I^2)/N)\cr
\Delta \gamma_{ff}(N,\alpha_s(Q^2))&=\alpha_s(Q^2)\Biggl[\biggl(
{8N_f\over 27}-{4\over 3}\biggr)+{2\over N_f}\hat C^g_{2,1,l}(N)\Biggr]
\Gamma^{\tilde 1}_{2L}(\alpha_s(Q_I^2)/N)\cr
\Delta \gamma_{gg}(N,\alpha_s(Q^2)) &= 
\Gamma^{\tilde 0}_{LL}(\alpha_s(Q^2)/N)-\fourninths \Delta 
\gamma_{fg}(N,\alpha_s(Q^2))\cr
\Delta \gamma_{gf}(N,\alpha_s(Q^2)) &= 
\fourninths \Gamma^{\tilde 0}_{LL}(\alpha_s(Q^2)/N)-\fourninths
\Delta \gamma_{ff}(N,\alpha_s(Q^2)).\cr}}
The corrections to the quark anomalous dimensions are functions of $Q_I^2$ 
rather than $Q^2$, except for the single power of $\alpha_s(Q^2)$, because 
$\Gamma^1_{2L}(\alpha_s/N)$ appears only in the input terms in 
\fullsolfderiv. In the small--$x$, or small--$N$ limit we have the 
simplifications that $\Delta \gamma_{fg}(N,\alpha_s(Q^2))\to
\alpha_s(Q^2)(2N_f/3)\Gamma^{\tilde 1}_{2L}(\alpha_s(Q_I^2)/N)$ and 
$\Delta \gamma_{ff}(N,\alpha_s(Q^2))\to \fourninths \Delta 
\gamma_{fg}(N,\alpha_s(Q^2))$.
Thus, in this limit these corrections take on the standard form of 
$\alpha_s(Q^2)\gamma^1_{fg}(\alpha_s(Q_I^2)/N)$ and 
$\alpha_s(Q^2)\gamma^1_{ff}(\alpha_s(Q_I^2)/N)$ 
minus their one--loop components. The prefactors multiplying these terms, 
which depend on the one--loop longitudinal coefficient functions, 
ensure that the leading--$\ln(1/x)$ enhancement of the rate of growth of 
${\cal F}_2(x,Q^2)$ is directly coupled to the longitudinal structure, 
not to the gluon. This delays the 
enhancement to slightly smaller values of $x$, and reduces it a little. The 
contributions to the gluon anomalous dimensions beyond LO in $\ln (1/x)$ are
present to counter the exponentiation of the corrections to the 
quark anomalous dimensions.

We perform checks to see if the particular
choice above does indeed lead to an accurate representation of the correct 
expressions. In order to do this we use the types of techniques 
outlined in \frt\ to transform the exact expressions \fullsolfl, 
\fullsolfderiv\ and \fullsolfin\ to 
$x$--space obtaining analytic solutions for the structure functions. These 
expressions are very complicated, 
but are almost exact for very low values
of $x$, and are compared, with the 
calculations performed using the computer program in $x$-space. For 
$x \lsim 0.1$ our choice of the anomalous dimensions leads to
agreement with the analytic expressions to an accuracy much better than the
errors on the data in any appropriate range of parameter space. 
At higher $x$ setting to
zero all terms in the MRS program other than those coming from one loop, 
the expressions for the structure functions for $x > 0.1$, are 
very close to the one--loop expressions alone, 
which agrees with calculations taking the 
inverse Mellin transformation of the LORSC moment--space expressions. 
Therefore, the correct choice of effective
splitting functions leads to 
the MRS program producing a very accurate approximation to our correct 
LORSC expressions for the structure functions over the complete range of 
parameter space.

Now that it has been determined that our choice of splitting functions is 
correct, one can vary the input parameters in the standard way to
obtain the best fit. In practice the inputs are of the standard MRS form
\start, 
in terms of partons, but these are constrained by demanding that the resulting
structure functions at $Q_I^2$ are compatible with  
forms in \fullsolfl, \fullsolfderiv\ and \fullsolfin (for some $A_{LL}$) 
where the nonperturbative 
inputs are flat at small $x$, i.e. they must be described well by
a function of the form 
\eqn\softinput{{\cal F}_i(x) =F_i(1-x)^{\eta_i}(1+\epsilon_i x^{0.5}+
\gamma_i x).} 

We note that with our choice of splitting functions and 
definitions of parton densities then momentum is not conserved by the 
evolution: in the best fit to ${\cal F}_2(x,Q^2)$ discussed in the next 
section the total momentum carried by the partons at $Q^2 =2\Gev^2$
is $87\%$ and at $Q^2=5000\Gev^2$ it is about $94\%$. Hence, the amount of 
momentum violation is at the level of a few percent. 
We have already defended this 
violation of momentum conservation in subsection 4.3. We now also point 
out that starting with our definition of the partons we could define 
new parton densities and splitting functions
by defining non--zero $C^{f(g)}_{2,1}(\alpha_s/N)$
and $C^{f(g)}_{L,1}(\alpha_s/N)$ beyond one--loop, and use the 
transformation rules in \S 2 to keep the structure functions unchanged.
If these coefficient functions had negative coefficients then, compared to 
our prescription above, the low $Q^2$ parton distributions would need to 
be larger, and would hence carry more momentum. As $Q^2$ increased the 
effect of these new coefficient functions would decrease, and the extra
amount of momentum carried by the new parton distributions compared to the 
original ones would decrease. Thus, the effect of such a redefinition of 
parton distributions would be to increase the amount of momentum carried by 
the partons at low $Q^2$ and also to 
slow the growth of momentum with $Q^2$ (or even 
to turn it into a fall). With a judicious redefinition of coefficient
functions of this sort (with perhaps some dependence on $\alpha_s(Q_I^2)$ 
as well as $\alpha_s(Q^2)$) it should clearly be possible to find an 
effective factorization scheme where the momentum violation will be 
extremely small, and one may think of this as a ``physical scheme''. We 
have not seriously investigated this redefinition of parton densities in any
quantitative manner since it will not affect any physical quantities.

\medskip

Having specified precisely how we will perform our calculations of 
structure functions we can make some comment on the general form 
the structure functions have to take. 
The expressions for ${\cal F}_L(x,Q^2)$ and ${\cal F}_2(x,Q^2)$
depend on $\Gamma^0_{LL}(\alpha_s/N)$ and $\Gamma^1_{2L}(\alpha_s/N)$. 
Both of these series have coefficients which behave like 
$n^{-3/2}(12\ln 2)/\pi$ for very large $n$. This leads to a cut in 
the $N$--plane
at $N=\bar \alpha_s 4\ln 2$ in both cases, and to the structure functions 
having asymptotic behaviour ${\cal F}(x,Q^2) \sim (\ln x)^{-3/2}
x^{-4\ln 2 \bar \alpha_s}$ as $x \to 0$, where $\alpha_s=\alpha_s(Q_I^2)$ 
and $Q^2\geq Q_I^2$.
However, it was 
convincingly demonstrated in \bfresum\ (and discussed from a different 
point of 
view in \frt) that one only need keep a finite number of terms in the 
leading--$\ln (1/x)$ series if working at finite $\ln (1/x)$. In practice,
if one works down to $x\approx 10^{-5}$ and the nonperturbative inputs behave 
roughly like $(1-x)^5$, then keeping terms up to $10_{\rm th}$ order in 
the series for 
$\Gamma^0_{LL}(\alpha_s/N)$ and $\Gamma^1_{2L}(\alpha_s/N)$ is more than 
sufficient. 
The series up to this order have the explicit form
\eqn\serieslong{\Gamma^0_{LL}(x)=
x +2.40x^4+2.07x^6+
17.3x^7
+2.01x^8+39.8x^9+
168.5x^{10} +\cdots}
and
\eqn\seriestwo{2\pi\Gamma^1_{2L}(x)=1+
2.5x+x^2+x^3+
7.01x^4+
5.81x^5
+13.4x^6+58.1x^7
+64.7x^8+
196.8x^9
+650x^{10}+\cdots.}
The low--order terms in both these series have coefficients 
which generally grow far less quickly than the asymptotic relationship $a_{n+1}
=4\ln 2 a_n$; some fall, and in the case of $\Gamma^0_{LL}(\alpha_s/N)$
are even zero. Thus, in the range of parameter space we are considering we
will have rather less steep behaviour than the asymptotic limit of 
${\cal F}(x,Q^2) \sim (\ln x)^{-3/2}x^{-4\ln 2 \bar \alpha_s}$. Using the 
techniques in \frt\ one can derive analytic expressions for the structure 
functions in the small--$x$ limit. Choosing $Q_I^2= 25\Gev^2$ and 
making a guess that $A_{LL}=0.8\Gev^2$ we obtain $\ln(Q_I^2/A_{LL})\approx
3.4$. We also choose nonperturbative inputs which behave like
$F_i(1-x)^5$ for $i=L,2$. As argued in \frt, for values of $x\lsim 0.01$ 
it is a good approximation to replace the $(1-x)^5$ behaviour with the 
behaviour $\Theta(0.1-x)$.\foot{The argument of the $\Theta$--function 
depends on the power of $(1-x)$. Higher powers would require the step to 
occur at lower $x$ and {\it vice versa}.} Doing this we can derive simple 
expressions for the form of the inputs for $x\lsim 0.01$ which will be 
accurate up to a few percent error. Using the series \serieslong\ and 
\seriestwo\ and the expressions \fullsolfl\ and \fullsolfin\ we can take the 
inverse Mellin transformations, obtaining
\eqn\anflinput{\eqalign{{\cal F}_L(x,Q_I^2) &\approx 
{\alpha_s(Q_I^2)/2\pi}\biggl[\biggl(F_L+{4\over 27}F_2\biggr)
\biggl(1+3.4\bar\alpha_s(Q_I^2)\xi +
5.8{(\bar\alpha_s(Q_I^2)\xi)^2\over 2!}+6.6{(\bar\alpha_s(Q_I^2)\xi)^3
\over 3!}\cr&+13.7{(\bar\alpha_s(Q_I^2)\xi)^4\over 4!}+31.5{
(\bar\alpha_s(Q_I^2)\xi)^5\over 5!}
+56{(\bar\alpha_s(Q_I^2)\xi)^6\over 6!}+136{(\bar\alpha_s(Q_I^2)\xi)^7
\over 7!}\biggr)-{4\over 27}F_2\biggr],\cr}}
and 
\eqn\anftwoinput{\eqalign{{\cal F}_2(x,Q_I^2) &\approx 3.4
{\alpha_s(Q_I^2)/2\pi}\biggl(F_L+{4\over 27}F_2\biggr)\biggl(1+
4.2\bar\alpha_s(Q_I^2)\xi +
7.1{(\bar\alpha_s(Q_I^2)\xi)^2\over 2!}+9.1{(\bar\alpha_s(Q_I^2)\xi)^3\over 
3!}\cr&+19.7{(\bar\alpha_s(Q_I^2)\xi)^4\over 4!}+44.1{(\bar\alpha_s(Q_I^2)
\xi)^5\over 5!}+84{(\bar\alpha_s(Q_I^2)\xi)^6\over 6!}+206{
(\bar\alpha_s(Q_I^2)\xi)^7\over 7!}\biggr)+F_2,}}
where $\xi=\ln(0.1/x)$, and $\alpha_s(Q_I^2)\approx 0.2$ if $\Lambda_{QCD,4}
\approx 100\Mev$. Putting in values of $F_L=2.5$ and $F_2=1$, choices which, as
with the $(1-x)^5$ behaviour, are roughly compatible with the high $x$ data,
we have a rough estimate of the form of the input structure functions at 
$Q_I^2=25\Gev^2$. The coefficients in the series in 
$\alpha_s(Q_I^2)\xi$ multiplying $(F_L+4/27 F_2)$ grow a little more quickly 
for ${\cal F}_2(x,Q_I^2)$ than for ${\cal F}_L(x,Q_I^2)$. However, 
in the former case this contribution which rises with falling $x$ is 
accompanied by the flat $F_2$, whereas in the latter it is accompanied only
by the nearly insignificant $-4\alpha_s(Q_I^2)/(54\pi)F_2$. Thus,    
for $x\approx 10^{-4}$ ${\cal F}_L(x,25)$
behaves approximately like $0.1x^{-0.3}$ but 
${\cal F}_2(x,25)$ is slightly flatter,
behaving approximately like $0.6 x^{-0.28}$. 
These powers of $x$ are clearly somewhat 
less steep than the asymptotic $x^{-0.5}$.
Comparison of the estimate of ${\cal F}_2(x,25)$ with the data in \hone\ 
and \zeus\ shows a very reasonable qualitative agreement. Of course, 
there is no data for ${\cal F}_L(x,25)$
for values of $x$ anything like this low. Being rather more general, we find
that for any $Q_I^2$ between $10\Gev^2$ and $100\Gev^2$, and with any 
sensible choice of $A_{LL}$ (e.g $0.2\Gev^2
\geq A_{LL} \geq 2\Gev^2$), then both $F_2(x,Q_I^2)$ and $F_L(x,Q_I^2)$
behave roughly like $x^{-0.3}$ for $0.01 \geq x \geq 0.00001$. Hence, this type
of behaviour can be taken as a prediction of the theory. 

One can also use the same 
techniques to make an estimate of the structure functions at values of $Q^2$
away from $Q_I^2$. With the particular 
inputs above they do lead to good qualitative agreement with the data 
below $x=0.01$, and we have every reason to feel encouraged by our results.
However, the 
real test of our approach will be a complete global fit to the available data
for ${\cal F}_2(x,Q^2)$ using the rather more accurate calculations,
especially at large $x$, 
of the MRS program. We will therefore discuss these detailed fits next. 

\newsec{Fits to the Data and Predictions.}

Once we have the general LORSC expression then by combining the
singlet and nonsinglet components and varying all the free parameters 
($A_{LL}$, the soft inputs for ${\cal F}^S_L(x,Q^2)$ and 
${\cal F}^S_2(x,Q^2)$ and the 
soft nonsinglet inputs), we obtain the  
best fit for the available ${\cal F}_2$ structure function data using a 
particular starting scale $Q_I^2$. 
Choosing the renormalization scale to be $Q^2$,
the one--loop value for $\Lambda_{N_f=4}$ is fixed at $100 \hbox{\rm MeV}$,
thus giving $\alpha_s(M_Z^2)=0.115$.\foot{Of course, since we are using a 
genuinely leading--order expression, any change in renormalization scale 
is exactly countered by a change in $\Lambda_{N_f=4}$. 
However, it is encouraging
that making the simple choice $\mu_R^2=Q^2$ leads to a value of 
$\alpha_s(M_Z^2)$ which is nicely compatible with the usual value.}
This precise value is not determined by a best fit, but a value near
to this is certainly favoured. 

There are some further details we should mention. Firstly,
when obtaining a fit using the above approach, the values of 
${\cal F}_2(x,Q^2)$ used are 
not precisely those published in \hone\ and \zeus. This is because it 
is not ${\cal F}_2(x,Q^2)$ that is measured directly at HERA, but 
the differential cross--section. This is related to the structure functions as 
follows
\eqn\crosstosf{{d^2\sigma\over d x d Q^2}={2\pi\alpha^2\over Q^4 x}
\biggl(2-2y+{y^2\over 1+R}\biggr){\cal F}_2(x,Q^2),}
where $y=Q^2/xs\approx Q^2/90000x$ at HERA 
and $R={\cal F}_L(x,Q^2)/({\cal F}_2(x,Q^2)-{\cal F}_L(x,Q^2))$.
Over most of the range of parameter space $y$ is very small, 
${\cal F}_L(x,Q^2)$ is likely to be small, and the measurement is 
essentially directly that of ${\cal F}_2(x,Q^2)$. However, at high $y$
the value of ${\cal F}_2(x,Q^2)$ 
must be extracted using some prescription for  
${\cal F}_L(x,Q^2)$. Both H1 and ZEUS, roughly
speaking, obtain their values of ${\cal F}_L(x,Q^2)$ from predictions coming 
from NLO calculations of ${\cal F}_2(x,Q^2)$.
Since the approach in this paper leads in practice to a somewhat lower 
prediction of ${\cal F}_L(x,Q^2)$, the values of ${\cal F}_2(x,Q^2)$ 
used in the fit must be altered to take account of this. 
Thus, the ${\cal F}_2(x,Q^2)$ values are a little (at most about $6\%$) 
lower for the largest values of $y$ than in those in \hone\ and \zeus.

The value of $m_c^2$ is 
chosen to be equal to $4\Gev^2$ in order to obtain a reasonable description 
of the data on the charm structure function coming from EMC 
\ref\emccharm{EMC collaboration: J.J. Aubert {\it et al}, \NP 
\vyp{B213}{1983}{31}.} and from measurements at HERA 
\ref\heracharm{H1 collaboration: C. Adloff et. al., \ZP 
\vyp{C72}{1996}{593}.}. The quality of the fit is shown in 
\fig\loxcharm{The description of the EMC and preliminary H1 data for 
${\cal F}^c_2(x,Q^2)$ using the LORSC fit.}. It is of a fair quality, 
with the predicted ${\cal F}^c_2(x,Q^2)$ 
perhaps being a little large in general at large $x$, and a little small 
at small $x$, a result which is qualitatively
consistent with the fact that we have used a 
threshold at $Q^2=m_c^2$, rather than at $W^2=Q^2(x^{-1}-1)=4m_c^2$. 
Of course, $m_c^2=4\Gev^2$ 
is a little high compared with the values obtained from reliable 
determinations. We will comment on this value later.
The strange quark is treated as being massless in our calculations, which is
presumably a good approximation for the values of $Q^2$ considered. Also,
we insist that the strange contribution to the structure function 
is $0.2$ of the singlet structure function minus the valence
contributions (i.e. the ``sea structure function'') at $Q^2=m_c^2$. This 
ensures compatibility with the data on neutrino--induced deep inelastic 
di--muon production obtained by the CCFR collaboration \ref\strange{CCFR 
collaboration: A.O. Bazarko {\it et al}, \ZP \vyp{C65}{1995}{189}.}. 

Finally we consider the form of the gluon at large $x$. Within our effective
factorization scheme we would expect the gluon distribution to be quite
similar to that in the $\overline{\hbox{\rm MS}}$ scheme at 
NLO--in--$\alpha_s$ for very large $x$.
Thus, we demand that our gluon is qualitatively similar to that obtained 
from the WA70 prompt photon data \ref\gluon{WA70 collaboration: M. Bonesini
{\it et al}, \ZP \vyp{C38}{1988}{371}.} at $x\geq 0.3$, i.e. 
roughly of the form $2.5(1-x)^6$ at 
$Q^2=20\Gev^2$ for values of $x$ this large.
Encouragingly, this is the type of large $x$ behaviour that the best fit 
chooses for the gluon, and no strong constraint is needed.      

\medskip

The fit is performed for a wide variety of data: the H1 \hone\ and Zeus 
\zeus\ data on ${\cal F}^{ep}_2(x,Q^2)$ with $0.000032\geq x \geq 
0.32$ and $1.5\Gev^2 \geq Q^2\geq 5000 \Gev^2$; The BCDMS data 
\ref\bcdms{BCDMS collaboration: A.C. Benvenuti {\it et al}, \PL 
\vyp{B223}{1989}{485}.} on 
${\cal F}^{\mu p}_2(x,Q^2)$ with $0.07 \geq x\geq 0.75$ and 
$ 7.5 \Gev^2 \geq Q^2 
\geq 230 \Gev^2$; the new NMC data \ref\nmc{NMC 
collaboration, {\tt hep-ph/9610231}, \NP B, in print.} on 
${\cal F}^{\mu p}_2(x,Q^2)$ and ${\cal F}^{\mu d}_2(x,Q^2)$ with 
$0.008\geq x\geq 0.5$ and 
$2.5\Gev^2 \geq Q^2 \geq 65 \Gev^2$; NMC data on the ratio of 
${\cal F}^{\mu n}_2(x,Q^2)$ to ${\cal F}^{\mu p}_2(x,Q^2)$ 
\ref\nmcrat{NMC collaboration: 
M. Arneodo {\it et al}, \PL \vyp{B364}{1995}{107}.} with $0.015 \geq
x \geq 0.7$ and $5.5 \Gev^2 \geq Q^2 \geq 160 \Gev^2$, CCFR data 
\ref\ccfr{CCFR collaboration: P.Z. Quintas {\it et al}, \PRL 
\vyp{71}{1993}{1307}.} on ${\cal F}^{\nu N}_2(x,Q^2)$ and 
${\cal F}^{\nu N}_3(x,Q^2)$
with $0.125\geq x\geq0.65$ and $5\Gev^2 \geq Q^2\geq 501.2\Gev^2$; and the 
E665 \ref\esixsixfive{E665 collaboration: M.R. Adams {\it et al}, 
\PR\vyp{D54}{1996}{3006}.} data on ${\cal F}^{\mu p}_2(x,Q^2)$
with $0.0037 \geq x\geq 0.387$
and $2.05\Gev^2 \geq 64.3\Gev^2$. One can see 
that the full range of data used in the fit covers an extremely 
wide range of both $x$ 
and $Q^2$ and thus provides a very stringent test of any approach used to 
describe it.\foot{We have not included ZEUS data for $Q^2\geq 2000\Gev^2$.}
     
The result of the best fit to this data using the LORSC expressions with 
$Q_I^2$ chosen to be equal to $40\Gev^2$, $m_c^2=4\Gev^2$
and $m_b^2=20\Gev^2$ is shown in table 1. 
Also the result of the fit to the 
small--$x$ data is shown in \fig\data{The curves correspond to
the value of the proton structure function ${\cal F}_2(x,Q^2)$ obtained 
from the 
leading--order, renormalization--scheme--consistent (LORSC) calculation at 12
values of $x$ appropriate for the most recent HERA data. For clarity of 
display we add $0.5(12-i)$ to the value of ${\cal F}_2(x,Q^2)$ each time 
the value 
of $x$ is decreased, where $i=1\to12$. The data are assigned to the $x$
value which is closest to the experimental $x$ bin (for more details see
the similar figure displaying the two--loop fits
in \mrsiii). E665 data are also shown on the curves with 
the five largest $x$ values. The H1 and ZEUS data are normalized by $1.00$ 
and $1.015$ respectively in order to produce the best fit.}.   
As one can see there is a very 
good quality fit to the whole selection of data, and thus over the whole 
range of $x$ and $Q^2$. Overall, the fit gives a $\chi^2$ of $1105$ for
$1099$ data points.
The fit shown is for the particular starting scale $Q_I^2=40\Gev^2$,
but the quality of the fit is extremely insensitive to changes in this scale, 
as we expect from the method of construction of the solutions. The fit is
essentially unchanged over the range $20-80 \Gev^2$, 
and we choose $40\Gev^2$ as the geometric mean. 
When $Q_I^2$ drops below $20\Gev^2$ the fit immediately gets markedly worse,
due to the bottom quark threshold.

There are 18 free parameters used in the fit: $\Lambda_{QCD,4}$
(which we choose to describe as a parameter since, although it is fixed at 
$100 \hbox{\rm MeV}$, and small variations would no doubt improve the fit 
slightly, it can certainly not vary too much); four parameters for each of the 
valence quark contributions, which are described by functions of the form
\start\ where the normalization is set by the number of valence quarks; 
four parameters for the nonperturbative inputs for the two singlet 
structure functions, which are of the form \softinput; and the unknown 
scale $A_{LL}$, where we allow $A_{LL}$ to be a free parameter for each 
$Q_I^2$. We do not consider $Q_I^2$ as a free parameter, since it can take 
a very wide range of values.
The parameter $A_{LL}$, which we have argued should be a scale typical of soft 
physics, turns out to be $0.55\Gev^2$ for the fit starting at $Q_I^2=40
\Gev^2$. This decreases a little as $Q_I^2$ increases and {\it vice versa}.
For $Q_I^2=40\Gev^2$ the soft inputs for the fit are roughly
\eqn\softinputs{\hat F^S_L(x)\approx 3(1-x)^5(1+0.1x^{0.5}-0.2x)
, \hskip 0.5in F^S_2(x)\approx
(1-x)^{3.4}(1-0.65x^{0.5}+4.5x).}
These nonperturbative 
inputs lead to complete inputs of $\hat F^S_L(x,Q_I^2)\approx 3x^{-0.33}$
and $F^S_2(x,Q_I^2)\approx 0.65 x^{-0.3}$ for $0.01 \geq x\geq 0.0001$, 
with the effective $\lambda$ increasing in both cases for even smaller $x$. 
Instead of forcing the nonperturbative inputs to be flat as $x\to 0$ we
could allow an asymptotic behaviour $x^{-\lambda}$, where 
$\lambda \lsim 0.08$. This leads to an
equally good fit.

\medskip

Thus, the LORSC fit seems to be a success. However, in order to gauge its true 
quality it is helpful to have some points of comparison, 
and we will discuss some alternative fits. The first we mention is 
that obtained by simply solving the evolution equations for structure 
functions using the full leading order physical anomalous dimensions 
\inclmatrix, rather than using the more careful procedure described above. 
In this case if we drop the restriction on the relationship between the 
inputs and let them each have arbitrary small $x$ form (the arguments 
relating the inputs do not really follow if we simply solve using the full
physical anomalous dimensions) then the fit is nearly as good as the 
LORSC fit for $Q^2\geq 4\Gev^2$. However, it gets markedly worse for
$Q^2\sim 3\Gev^2$ and is extremely poor below this. If we do try to impose
the relationship between the inputs (i.e. retain any explanatory power for
the form of these inputs), then the fit at small $x$ is much worse than the 
LORSC fit, the value of $(d\,{\cal F}_2(x,Q^2)/d\, \ln Q^2)$ 
being rather too steep.
Thus, phenomenologically the naive use of the LO physical anomalous
dimensions is clearly inferior to the full LORSC calculation. 

We also compare to more conventional approaches.
As in \mylet, the most recent MRS fits R$_1$ and R$_2$ are shown. These
are obtained using the standard two--loop method, where R$_1$ allows 
$\Lambda^{N_f=4}_{{\overline{\rm MS}}}$ to be free (giving 
$\Lambda^{N_f=4}_{{\overline{\rm MS}}}=241{\hbox{\rm MeV}}$) and 
R$_2$ fixes $\Lambda^{N_f=4}_{{\overline{\rm MS}}}=344{\hbox{\rm MeV}}$
to force a better fit to the HERA data. The MRS fits 
are useful for purposes of comparison for a number of reasons:
the treatment of the errors in
this paper is identical to that in the MRS fits, so consistency in this 
respect is guaranteed, any systematic differences due to differences
in computer programs is guaranteed to be absent, and the cuts in $Q^2$ 
for each data set are chosen to be the same in this paper as for the 
MRS fits.   
The number of free parameters in the NLO--in--$\alpha_s$ 
fit is the same as in the LORSC 
fit. There is one more parameter in the NLO--in--$\alpha_s$ 
fit due to the powerlike forms
of the input gluon and singlet quark at small $x$ being independent\foot{We 
claim that the allowed powerlike form of the inputs at small $x$ is against 
the spirit of a well--ordered perturbative expansion, 
as already discussed in \S 4.1. However, enforcing this rule would
mean that the quality of the fit to the small--$x$ data was very poor. Thus, we
allow the more usual, unjustified choice for the small--$x$ form of the 
inputs. In fact, at $Q_I^2=1\Gev^2$ the gluon is strongly valence--like, while
the sea--quark distribution $\sim x^{-1-0.15}$at small $x$.} whereas the 
small--$x$ shape of the inputs for both structure functions is completely
determined by the one parameter $A_{LL}$ in the LORSC fit.
However, there is one less parameter in the
NLO fit due to the normalization of the gluon being determined by 
momentum conservation. Essentially, the LORSC fit is a little less 
constrained at large $x$ than the NLO fit, but rather more predetermined at
small $x$.    

Comparing the LORSC fit and the MRS fits, it is clear that
the LORSC scheme--independent fit is much better for the HERA data (even
when compared to R$_2$), much better for the BCDMS data (even when compared
to R$_1$) and similar in standard for the rest of the data. The overall fit is 
$\sim 200$ better for the whole data set; clearly a lot better. 
However, this direct comparison with the MRS fits is rather unfair:  
the MRS fits use the SLAC data \ref\slac{L.W. Whitlow {\it et al},
\PL \vyp{B282}{1992}{475}\semi L.W. Whitlow {\it et al}, preprint SLAC--357
(1990).} on ${\cal F}^{ep}_2(x,Q^2)$ which is not 
included in the LORSC fit due to much greater sensitivity of this 
data to potential higher twist effects than any of the other data sets, and 
also the fact that the quark and gluon 
distributions are input at $1\Gev^2$ in the MRS fits is not helpful to their
quality. If the best fit is obtained for the HERA data using the 
NLO calculation with massless quarks with inputs at $Q_I^2\approx 4\Gev^2$, 
as in \steep, and evolution performed downwards in $Q^2$ then the gluon 
becomes negative at very small $x$ before $Q^2=1\Gev^2$. 
Hence, starting at larger $Q^2$ where the gluon 
distribution is happy to be positive everywhere, produces a better fit.
Thus, in order to make a more meaningful comparison of the LORSC fit with 
a NLO--in--$\alpha_s$ fit we have ourselves performed a NLO--in--$\alpha_s$
fit, called NLO$_1$, allowing 
the normalizations to vary in the same way as in the LORSC fit, 
with exactly the same treatment of 
quark thresholds as the LORSC fit, with the input parton distributions
chosen at $Q_I^2=m_c^2$ and with the fit to exactly the same 
data.\foot{Once again we allow the small--$x$ form of the 
inputs to be unjustified powerlike behaviours. At $Q_I^2=2.75\Gev^2$ the 
gluon is quite flat while the sea--quark distribution $\sim x^{-1-0.22}$.}
As in the MRS fits the values of ${\cal F}_2(x,Q^2)$ at small $x$ are those 
quoted in \zeus\ and \hone. The value of $m_c^2$ is chosen to provide a 
good description of the data on the charm structure function. The value 
needed is $m_c^2=2.75\Gev^2$, and 
the fit to the charm structure function data is shown in 
\fig\nloxcharm{The description of the EMC and preliminary H1 data for 
${\cal F}^c_2(x,Q^2)$ using the NLO$_1$ fit.}.
\foot{As in the LORSC fit 
the choice of $m_c^2$ which gives a good description of the charm data is 
also the choice which gives the best global fit.} 
The value of $\Lambda^{N_f=4}_{{\overline{\rm MS}}}$ 
determined by the fit is $299 \hbox{\rm MeV}$. 
The quality of the fit is shown in table 1. 

The higher starting scale for evolution has produced a better fit to the HERA 
data than the MRS fits. However, the best fit comes from allowing 
the normalization
of the H1 data to be at the lower limit allowed by the error on the 
normalization\foot{The fit to the H1 data continues to improve slightly for
a normalization going down to $0.96$.}, and is still not as
good as the LORSC fit to the HERA data. Not including the SLAC data 
leads to a much better fit to the 
BCDMS data than the MRS fits, but again this is clearly not as good as the 
LORSC fit. The fit to the NMC data is much the same for the LORSC fit as for 
the NLO$_1$ fit, and the NLO$_1$ fit is a little better for the CCFR data. 
The overall quality of the NLO$_1$ fit is $1169$ for the $1099$ points, and 
thus is $64$ worse than the LORSC fit. Therefore, the LORSC fit is still 
clearly better than the NLO$_1$ fit, but not nearly as convincingly as 
appeared to be the 
case when compared to the MRS fits. Nevertheless, it is encouraging that, 
while the overall fits to the relatively high $x$ data, i.e. the BCDMS, 
NMC, E665 and CCFR data are similar in the NLO$_1$ fit and the LORSC fit, 
it is the fit to the small--$x$ HERA data that is 
definitely better for the LORSC
fit, as we would expect. This can be seen even more clearly if we  
examine the quality of the fit by separating the data into two sets: one 
where $x< 0.1$ and one where $x\geq 0.1$. This is shown in table 2, and 
demonstrates that the LORSC fit is superior at small $x$, while not
quite as good as the NLO$_1$ fit at large $x$. 

This qualitative result is exactly what we would expect. 
The importance of the leading--$\ln(1/x)$ terms in the LORSC calculation 
can be quantitatively judged
by how they affect the fit. If, after obtaining the best LORSC fit, all terms
other than those in the one--loop expressions are set to zero, the quality 
of the fit is unchanged above $x=0.3$, begins to 
alter slightly below this, and is clearly worse by the time we reach 
$x=0.1$. Thus, the leading--$\ln(1/x)$ terms are important by this 
value of $x$. However, most of this effect 
is due to the terms at ${\cal O}(\alpha_s^2)$, so the NLO
expression at fixed order in $\alpha_s$ should be insensitive to 
higher--order--in--$\alpha_s$ leading--$\ln(1/x)$ terms down to $x$ somewhat 
lower than $0.1$. Therefore, above $x=0.1$ the NLO fit should in principle 
be better than the LORSC fit since it contains terms at NLO in $\alpha_s$
which are important at large $x$. However, the NLO fit should be 
considerably worse at small $x$ since it does not contain many important
leading--$\ln (1/x)$ terms. This is qualitatively in agreement with the 
comparison of the NLO$_1$ fit and the LORSC fit. 

However, the above comparison is 
somewhat incorrect because in the process of obtaining the best fit for all
the data the NLO$_1$ fit may choose some parameters, particularly   
$\Lambda^{N_f=4}_{{\overline{\rm MS}}}$, such that they mimic the effects 
of the leading--$\ln(1/x)$ terms, and a decent fit for the small
$x$ data is obtained to the detriment of the fit to the large $x$ data. 
In order to check this hypothesis we have also performed a 
NLO--in--$\alpha_s$ fit with 
$\Lambda^{N_f=4}_{{\overline{\rm MS}}}$ fixed at $250 \hbox{\rm MeV}$, 
which we will denote by NLO$_2$. The results of this fit in terms of the 
different data sets is shown in table 1. The fit clearly improves compared 
to the NLO$_1$ fit for the BCDMS and CCFR data, and gives the best overall fit
for the high $x$ data sets. It worsens somewhat for the NMC data, and also
for the HERA data, and overall is slightly worse than the NLO$_1$ fit, 
having a $\chi^2$ of $1184$ for the $1099$ data points. Nevertheless, it is
perhaps a truer representation of a real NLO--in--$\alpha_s$ 
fit than NLO$_1$ since it gives a better fit to high $x$ data, but not such a 
good fit to the small--$x$ data  which presumably require the 
leading--$\ln(1/x)$ terms. Whether one believes this argument or not, 
it is certainly
clear that the LORSC fit does provide a better description of the data than 
any standard NLO--in--$\alpha_s$ fit.  

This leads us to the question of the determination of 
$\alpha_s(M_Z^2)$ using global fits to structure function data. The complete 
RSC expression for structure functions only exists at leading order, and
this leaves the renormalization scale 
undetermined. Hence, a determination of $\alpha_s(M_Z^2)$ does not really 
take place. It is not yet possible to extend the RSC
calculation beyond the leading order due to lack of knowledge of 
NLO--in--$\ln (1/x)$ terms, but hopefully these will shortly become available 
\nlobfkl, and when they do the NLO versions of \fullsolfl--\fullsolfin\ 
can be derived and
put to use, and the NLO coupling constant determined. 
Until these full renormalization--scheme--consistent 
NLO expressions become available, we believe that it is incorrect to use the
present
NLO--in--$\alpha_s$ fits to small--$x$ structure function data in order to 
determine the NLO coupling constant.
However, as already argued, the fixed--order--in--$\alpha_s$ expressions 
should be accurate 
for CCFR, BCDMS and NMC data (except perhaps at the lowest $x$ values), 
which after all are still much more 
precise than HERA data, and fits to these data alone will provide the 
best determination of the NLO $\alpha_s(Q^2)$. 

\medskip

We have given what we hope are convincing arguments for our 
advocated approach for calculating structure functions. 
Not only do we claim theoretical correctness, and 
limited predictive power, but we also have good quality, very 
comprehensive fits to data on ${\cal F}_2(x,Q^2)$.
However, we are well aware that only further experimental tests can prove us 
right (or wrong). Hence, we now discuss our predictions for 
${\cal F}_L(x,Q^2)$. 
   
So far we have only probed ${\cal F}_L(x,Q^2)$ indirectly, 
i.e. it is simply related to the $\ln Q^2$ derivative of ${\cal F}_2(x,Q^2)$
(as well as to the input ${\cal F}_2(x,Q_I^2)$). 
Having tied down the nonperturbative inputs and $A_{LL}$ and $Q_I^2$ from
our fit to ${\cal F}_2(x,Q^2)$, we have a prediction for 
${\cal F}_L(x,Q^2)$. The result of this prediction for the fit with 
$Q_I^2=40\Gev^2$ is shown in \fig\fl{Comparison of predictions for 
${\cal F}_L(x,Q^2)$ using the full leading--order,
renormalization--scheme--consistent (LORSC) fit and the NLO$_1$ 
fit. For both sets of curves ${\cal F}_L(x,Q^2)$ 
increases with increasing $Q^2$ at the lowest $x$ values.}, 
where it is compared to the prediction using the NLO--in--$\alpha_s$ approach, 
in particular the NLO$_1$ fit. As one can see, it is smaller
than the NLO$_1$ ${\cal F}_L(x,Q^2)$, but becomes steeper at very small 
$x$. The NLO$_2$ fit gives a very similar form of ${\cal F}_L(x,Q^2)$ to the 
NLO$_1$ fit. 
The LORSC prediction for ${\cal F}_L(x,Q^2)$ is weakly dependent on the 
value of $Q_I^2$ chosen: the value at $Q^2=5\Gev^2$ and $x=10^{-4}$
varies by $\pm 10\%$ within our range of 
$Q_I^2$ (increasing with $Q_I^2$), and by
less than this for higher $x$ and $Q^2$.

Hence, measurements of ${\cal F}_L(x,Q^2)$ at $x<10^{-2}$
would be a good discriminant between fixed--order--in--$\alpha_s$ 
calculations and those involving leading--$\ln (1/x)$ terms. 
However, the ``determination'' of ${\cal F}_L(x,Q^2)$
already performed by H1 \ref\fldet{H1 collaboration, \PL 
\vyp{B393}{1997}{452}.} is really 
only a consistency check for a particular NLO--in--$\alpha_s$ 
fit, and is by no means a true measurement 
of ${\cal F}_L(x,Q^2)$. In essence, all it proves is that the 
measurements of the cross--section are consistent with a particular 
NLO--in--$\alpha_s$
fit to ${\cal F}_2(x,Q^2)$ when the relationship between the cross--section
and the value of ${\cal F}_2(x,Q^2)$ is determined assuming 
the correctness of the NLO--in--$\alpha_s$ expressions for both structure 
functions. This is a perfectly correct procedure, and should be adopted for
any fit to ${\cal F}_2(x,Q^2)$ data, as it has been for the LORSC fit, 
but in itself says nothing about the validity of a different 
approach, or about the actual value of ${\cal F}_L(x,Q^2)$.
Hence, real measurements of ${\cal F}_L(x,Q^2)$ 
are needed in order to find the real values of ${\cal F}_L(x,Q^2)$.
From \fl\ it is clear that such measurements at HERA would be an important 
(and probably essential) way of determining the validity of the 
approach in this paper, and the genuine importance of leading--$\ln(1/x)$ 
terms in structure functions.

Another potentially important discriminant between different methods of 
calculating structure functions is the measurement of the charm structure 
function. Both the LORSC fit and the NLO$_1$ fit can provide good fits to 
the currently available charm data, as already seen, and the value of $m_c^2$ 
in the NLO$_1$ fit is rather more satisfactory than that in the LORSC 
fit. However, these calculations involve incorrect treatments of the 
charm quark threshold. A more correct treatment of this threshold at NLO in 
$\alpha_s$ \ref\grs{M. Gl\"uck, E. Reya and M. Stratmann, 
\NP \vyp{B422}{1994}{37}.} shows that the value of $m_c^2$ must be
considerably reduced in
order to produce the same sort of value of the charm structure function as 
the approach used in this paper. Indeed, as seen in \ref\ballroeck{R.D. Ball
and A. De Roeck, {\tt hep-ph/9609309},
Proc. of the International Workshop on Deep 
Inelastic Scattering, Rome, April, 1996, in print.}, 
NLO--in--$\alpha_s$ 
calculations with $m_c^2=2.25\Gev^2$ undershoot the small--$x$ 
data. A correct treatment of the charm quark threshold within the framework 
advocated in this paper has not been completely worked out (although, as 
already mentioned, work is in progress), and has certainly not been tested.
However, a 
comparison of the theoretical calculations with the ever--improving data on 
the charm structure function seems potentially to be a very useful way 
of discriminating between different methods of calculation.

In principle there are many other quantities which could be calculated 
within the LORSC framework and compared to those calculated using the 
NLO--in--$\alpha_s$ approach (or any other method) and to experimental data. 
Particularly obvious examples are the distribution of the transverse
energy flow in the final state in lepton--hadron scattering and the 
the cross--section for forward jet production, for both of which there exists 
some experimental data which does not seem to be terribly well described by 
the order--by--order--in--$\alpha_s$ approach, e.g. 
\ref\transverse{H1 collaboration, paper pa02-073, 
submitted to ICHEP 1996, (Warsaw, July, 1996).} and 
\ref\fjets{H1 collaboration, paper pa03-049, 
submitted to ICHEP 1996, (Warsaw, July, 1996).}. We have not 
performed a LORSC calculation of any
such quantities, and it is very difficult to estimate the 
results, other than guess that they will probably lie somewhere 
between the fixed--order--in--$\alpha_s$ predictions and those obtained using 
BFKL physics naively. Such calculations are
obviously a priority, and work will begin soon. Only by comparing our
theoretical predictions with a wide variety of experimental data can we 
determine unambiguously which theoretical approach is correct. 

\newsec{Conclusion and Summary.}

In this paper we have derived expressions for the structure functions 
${\cal F}_2(x,Q^2)$ and ${\cal F}_L(x,Q^2)$
in a theoretically correct, well--ordered manner, within the framework of the 
renormalization group and collinear factorization. We have first done this 
for the particular expansions schemes which order the expressions strictly in
orders of $\alpha_s$, i.e. the standard loop expansion,
or in terms of the leading powers of $\ln (1/x)$ for given power of $\alpha_s$,
i.e. the leading--$\ln (1/x)$ expansion. In both cases we have demonstrated 
that a correct calculation in terms of structure function inputs and evolution
automatically leads to factorization--scheme--invariant 
results which may be expressed in terms of physical 
anomalous dimensions. Thus, these physical 
anomalous dimensions are fundamental 
pieces in the correct expressions for the structure functions.
However, we also demonstrate that in the case of the 
leading--$\ln (1/x)$ expansion the correct expressions are 
more difficult to obtain than in the loop expansion, and in both cases
a correct calculation requires rather more care 
than just the use of these physical anomalous dimensions. 

We have then argued that both the above expansion schemes are restrictive, and
lead to only part of the correct solution at any given order. We have 
shown that the only calculational method which is truly consistent with 
working to a given order in $\alpha_s$ within a given renormalization scheme 
is the renormalization--scheme--consistent expansion, in which one 
works to a given order in both $\alpha_s$, and in $\ln(1/x)$ for given power
of $\alpha_s$, for physical quantities, i.e. the inputs and evolution of 
the structure functions. Doing this one obtains the conventional, 
order--by--order--in--$\alpha_s$ renormalization group results in the high
$x$ limit, and smoothly incorporates the leading $\ln(1/x)$ corrections. 
We have then derived the leading--order, renormalization--scheme--consistent 
(and hence renormalization scheme invariant) expressions for 
the structure functions ${\cal F}_2(x,Q^2)$ and ${\cal F}_L(x,Q^2)$. 
In this derivation we have made use of the physical anomalous 
dimensions (which is not necessary, but is extremely convenient),
but needed to do more work than simply include them in 
their leading form and solve the evolution equations. Indeed,
this naive method of inclusion leads to 
a comparison with data which is far worse than that obtained from the correct
LORSC calculation.  

As part of our overall approach we have also taken an unusual 
view of the starting scale $Q_I^2$. Rather than trying to guess some 
form for the input at some particular $Q_I^2$, or simply allowing the 
inputs at some arbitrary $Q_I^2$ to take any form they wish within a given 
parameterization, we have demanded firstly that our expressions should be 
formally insensitive to the choice of the input scale, and secondly that
any deviation from the flat Regge--type behaviour of the structure functions
must come from perturbative effects. Thus, our inputs take the form of  
nonperturbative functions, which are flat at small $x$, convoluted with 
functions of the physical anomalous dimensions which are evaluated at $Q_I^2$,
and determined by the requirement of insensitivity to the 
value of $Q_I^2$. This leads to our inputs being determined entirely in 
terms of the flat nonperturbative inputs and one arbitrary scale $A_{LL}$ 
which roughly indicates the scale where perturbative physics should break 
down, i.e. $A_{LL}\lsim 1\Gev^2$. This gives us a great deal more predictive 
power than more usual approaches. We have some idea of the form of 
individual structure functions at small $x$: 
for a range of sensible choices of 
$Q_I^2$ ($10\Gev^2 - 100\Gev^2$) and $A_{LL}$ we obtain 
${\cal F}_2(x,Q_I^2)$ roughly $\propto x^{-(0.25-0.33)}$
for $0.01\geq x\geq 0.00001$, which is clearly in good 
qualitative agreement with the data. However, we also have a much stronger 
prediction for the relationship between the small--$x$ inputs. So as well 
as some predictive (or at the very 
least, explanatory) power for individual structure functions, 
once we choose the input for one, in practice ${\cal F}_2(x,Q^2)$, we have 
determined, up to a small amount of freedom, the small--$x$ inputs for
$d\,{\cal F}_2(x,Q^2)/d\ln Q^2$ and ${\cal F}_L(x,Q^2)$.

Not only are the features of the LORSC calculation compelling, but they also 
work rather well in practice (for a range of $Q_I^2$ from $20\Gev^2 -
80\Gev^2$). The LORSC expressions, including this 
constraint on the small--$x$ inputs, lead to very good fits to the data,
the $\chi^2$ for the LORSC fit to 1099 data on ${\cal F}_2(x,Q^2)$ ranging from
from $0.75\geq x\geq 0.000032$ and $1.5\Gev^2 \leq Q^2\leq
5000\Gev^2$ is better by at least 60 than any NLO--in--$\alpha_s$ fit, 
a very significant improvement. This is  even though the NLO--in--$\alpha_s$ 
fit is allowed arbitrary, unjustified 
powerlike behaviour at small $x$, and the small--$x$ inputs for 
${\cal F}_2(x,Q^2)$ and $d{\cal F}_2(x,Q^2)/d\ln Q^2$ are largely
independent. In fact, all of this superiority comes from the 
fit to the data with $x<0.1$. Thus, we find the naively expected, but much 
refuted, result that the leading $\ln(1/x)$ terms are both important and 
helpful in fits to structure functions when included properly.
As a caveat, it is certainly true that
the calculations in this paper must be improved to take account of massive
quark thresholds in a better manner (as is also the case with 
NLO--in--$\alpha_s$ fits), and work towards this end is in progress.
Nevertheless, with the present treatment we feel
that the quality of the fit and the degree of explanatory (if not predictive)
power, not to mention the theoretical correctness, give strong
justification for the LORSC expressions. 

We do, however,  recognize that the quality of the fit alone does not 
necessarily convince one that this approach has to be correct. In 
order to obtain verification we must exploit the factorization theory fully
and compare with more and different 
experimental data. Hence, we have presented a LORSC prediction for 
${\cal F}_L(x,Q^2)$, comparing it to that obtained using the 
NLO--in--$\alpha_s$ approach. Hopefully there will be true measurements of 
${\cal F}_L(x,Q^2)$ at HERA some time in the future with which we can compare 
these predictions. We stress the 
importance of such measurements to the understanding of the physics 
which really underlies hadron interactions. In the near future we will also 
have predictions of the charm contribution to the structure function 
within the 
framework of a correct treatment of the massive quark, and comparison with
the ever--improving data on the charm structure function should also be a 
good discriminant between different theoretical approaches. Calculations and 
measurements of other, less inclusive quantities, such as forward jets, are 
another clear goal.

Finally, our calculation is at present only at leading 
order due to the lack of knowledge of 
next--to--leading--order--in--$\ln (1/x)$ 
coefficient functions and anomalous dimensions, or
equivalently of physical anomalous dimensions. This means that 
the NLO--in--$\alpha_s$ approach is still in principle superior to our 
approach at high $x$, where the leading--$\ln(1/x)$ terms at third 
order in $\alpha_s$ and beyond are not important but the 
${\cal O} (\alpha_s^2)$ terms are. Indeed, 
in practice the NLO--in--$\alpha_s$ fits, 
are slightly better than the LORSC fit for data at
$x\geq 0.1$, and we might expect any predictions coming from the 
NLO--in--$\alpha_s$ approach to be more accurate than those coming from 
the LORSC approach down to values of $x$ somewhere in the region of $0.05$. 
The lack of the NLORSC expressions also means that a true determination of 
$\alpha_s(M_Z^2)$ from a global 
fit to structure function data is not yet possible, but that the best 
determination from fits to structure function data should 
at present come from using the NLO--in--$\alpha_s$
approach, using only 
large $x$ data. For a really fair comparison between the 
renormalization--scheme--consistent method and the conventional
order--by--order--in--$\alpha_s$ approach we really need the full 
NLORSC calculation.
Hopefully the required NLO in $\ln(1/x)$ quantities will soon become 
available, and with some work we will be able to 
make such a comparison. 
We would expect that when using the NLORSC expressions
the fit to the large $x$ data would become at least as good as for the 
NLO--in--$\alpha_s$ approach, and that the fit to the small--$x$ data 
would be of at least the same quality as for the LORSC fit. 
We wait expectantly to discover if this is indeed the case.  

\bigskip

{\bf Acknowledgements.}
\medskip
I would like to thank R.G. Roberts for continual help during the period of 
this work and for the use of the MRS fit program. I would also like to thank 
Mandy Cooper--Sarkar, Robin Devenish, Jeff Forshaw and Werner Vogelsang  
for helpful discussions.

\vfill 
\eject

\noindent {\bf Table 1}\hfil\break
\noindent Comparison of quality of fits using the full 
leading--order (including 
leading--$\ln (1/x)$ terms) renormalization--scheme--consistent expression, 
LORSC, and the two--loop fits MRSR$_1$, MRSR$_2$, NLO$_1$ and NLO$_2$.
For the LORSC fit the H1 data chooses a
normalization  of $1.00$, the ZEUS data of $1.015$, and the BCDMS data
of $0.975$. The CCFR data is fixed at a normalization of $0.95$, and the 
rest is fixed at $1.00$. 
Similarly, for the NLO$_1$ fit the H1 data is fixed at a 
normalization of $0.985$, the ZEUS chooses a normalization of $0.99$, 
and the BCDMS data of $0.975$.
Again the CCFR data is fixed at a normalization of $0.95$, 
and the rest fixed at $1.00$.
Also, for the NLO$_2$ fit the H1 data is fixed at a 
normalization of $0.985$, the ZEUS chooses a normalization of $0.985$, 
and the BCDMS data of $0.97$.
Again the CCFR data is fixed at a normalization of $0.95$, 
and the rest fixed at $1.00$.  
In the R$_{1}$ and R$_2$ fits the BCDMS data has a fixed normalization of
$0.98$, the CCFR data of $0.95$ and the rest of $1.00$. 
 
\medskip

\hfil\vtop{{\offinterlineskip
\halign{ \strut\tabskip=0.6pc
\vrule#&  #\hfil&  \vrule#&  \hfil#& \vrule#& \hfil#&
\vrule#& \hfil#& \vrule#& \hfil#&
\vrule#& \hfil#& \vrule#& \hfil#& \vrule#\tabskip=0pt\cr
\noalign{\hrule}
& Experiment && data   && \omit &\omit& &\omit&
$\chi^2$&\omit& \omit &\omit&\omit&\cr
&\omit       && points && LORSC &\omit& NLO$_1$ &\omit& NLO$_2$ &\omit& R$_1$ 
&\omit& R$_2$ &\cr
\noalign{\hrule}
& H1 ${\cal F}^{ep}_2$ && 193 && 123 && 145 && 145 && 158 && 149 &\cr
& ZEUS ${\cal F}^{ep}_2$ && 204 && 253 && 281 && 296 && 326 && 308 &\cr
\noalign{\hrule}
& BCDMS ${\cal F}^{\mu p}_2$ && 174 && 181 && 218 && 192 && 265 && 320 &\cr
& NMC ${\cal F}^{\mu p}_2$ && 129 && 122 && 131 && 148 && 163 && 135 &\cr
& NMC ${\cal F}^{\mu d}_2$ && 129 && 114 && 107 && 125 && 134 && 99 &\cr
& NMC ${\cal F}^{\mu n}_2/{\cal F}^{\mu p}_2$ && 85 && 142 && 137 && 138 
&& 136 && 132 &\cr
& E665 ${\cal F}^{\mu p}_2$ && 53 && 63 && 63 && 63 && 62 && 63 &\cr
\noalign{\hrule}
& CCFR ${\cal F}^{\nu N}_2$ && 66 && 59 && 48 && 40 && 41 && 56 &\cr
& CCFR ${\cal F}^{\nu N}_2$ && 66 && 48 && 39 && 36 && 51 && 47 &\cr
\noalign{\hrule}}}}\hfil

\bigskip

\noindent {\bf Table 2}\hfil\break
\noindent Comparison of quality of fits using the full leading--order 
(including 
leading--$\ln (1/x)$ terms) renormalization--scheme--consistent expression, 
LORSC, and the two--loop fits NLO$_1$ and NLO$_2$. The fits are identical to 
above, but the data are 
presented in terms of whether $x$ is less than $0.1$ or not.
 
\medskip

\hfil\vtop{{\offinterlineskip
\halign{ \strut\tabskip=0.6pc
\vrule#&  #\hfil&  \vrule#&  \hfil#& \vrule#& \hfil#&
\vrule#& \hfil#& \vrule#& \hfil#& \vrule#\tabskip=0pt\cr
\noalign{\hrule}
& \omit && data   && \omit &\omit&$\chi^2$& \omit&\omit&\cr
&\omit       && points && LORSC &\omit& NLO$_1$& \omit &NLO$_2$&\cr
\noalign{\hrule}
& $x\geq 0.1$&& 551 && 622 && 615 && 595 &\cr
& $x<0.1$ && 548 && 483 && 554 && 589 &\cr
\noalign{\hrule}
& total && 1099 && 1105 && 1169 && 1184 &\cr
\noalign{\hrule}}}}\hfil

\footatend\vfill\supereject\immediate\closeout\rfile\writestoppt
\baselineskip=14pt\centerline{{\bf References}}\bigskip{\frenchspacing%
\parindent=20pt\escapechar=` \input refs.tmp\vfill\eject}\nonfrenchspacing

\vfill\eject\immediate\closeout\ffile{\parindent40pt
\baselineskip14pt\centerline{{\bf Figure Captions}}\nobreak\medskip
\escapechar=` \input figs.tmp\vfill\eject}

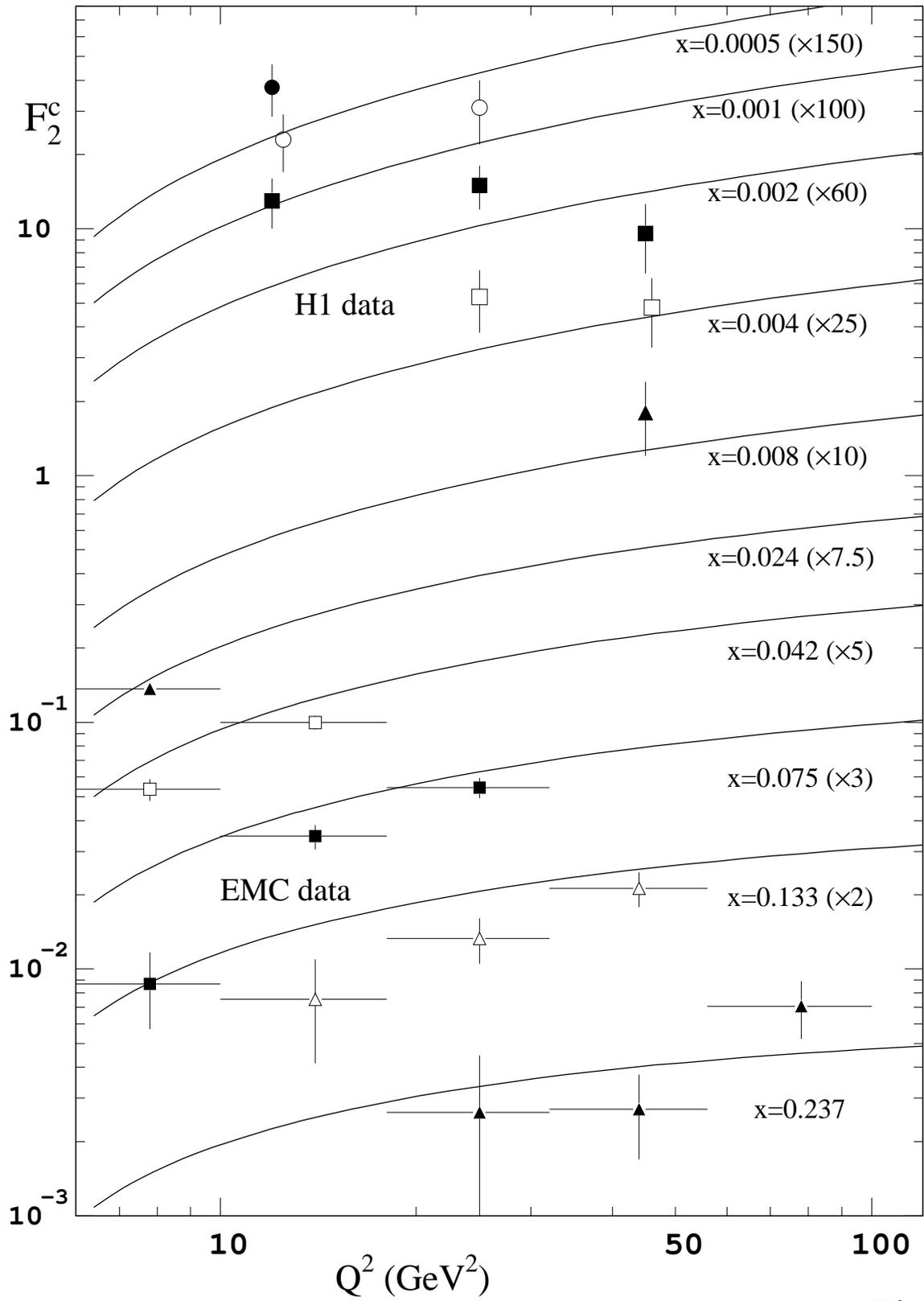
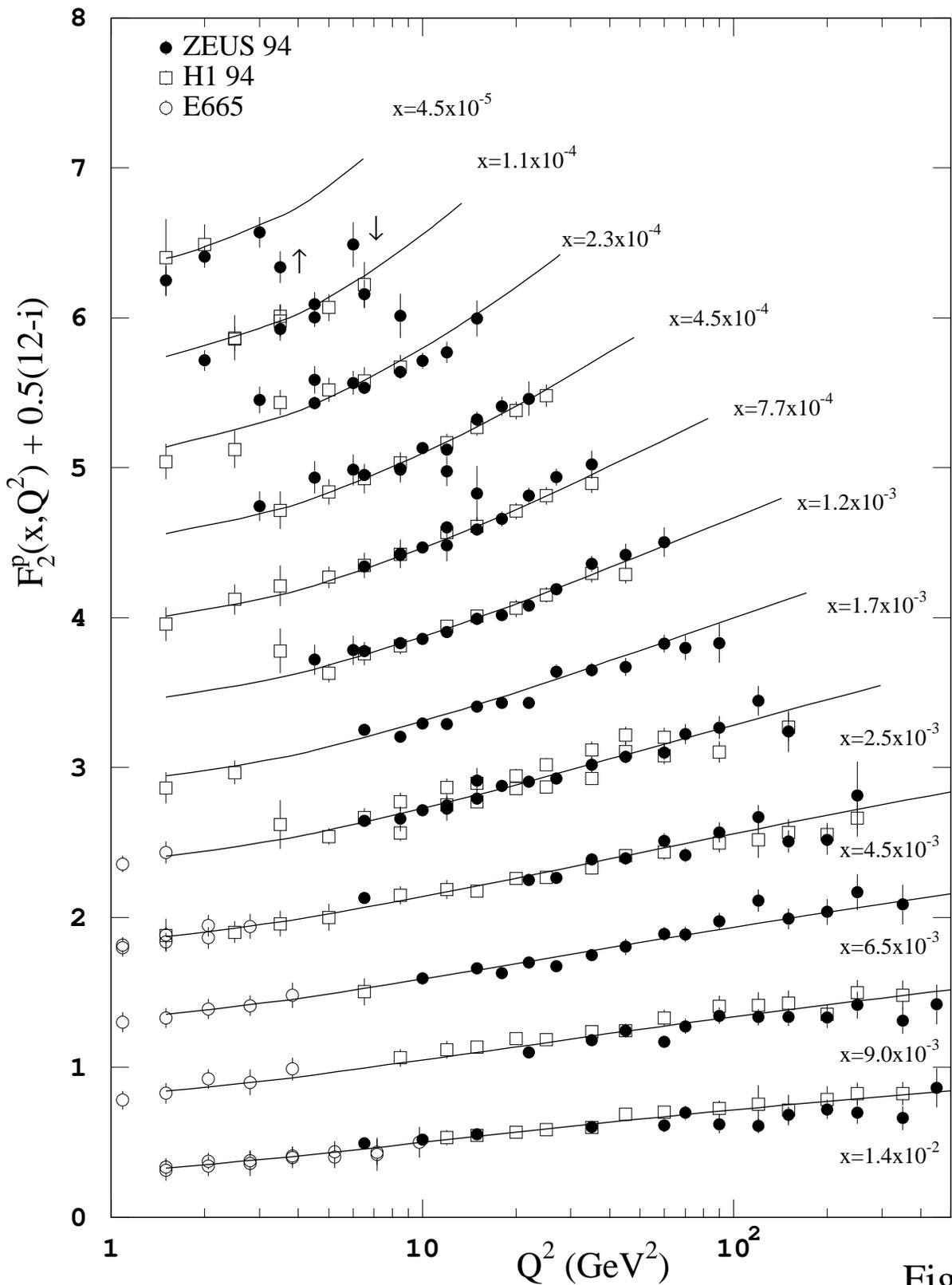
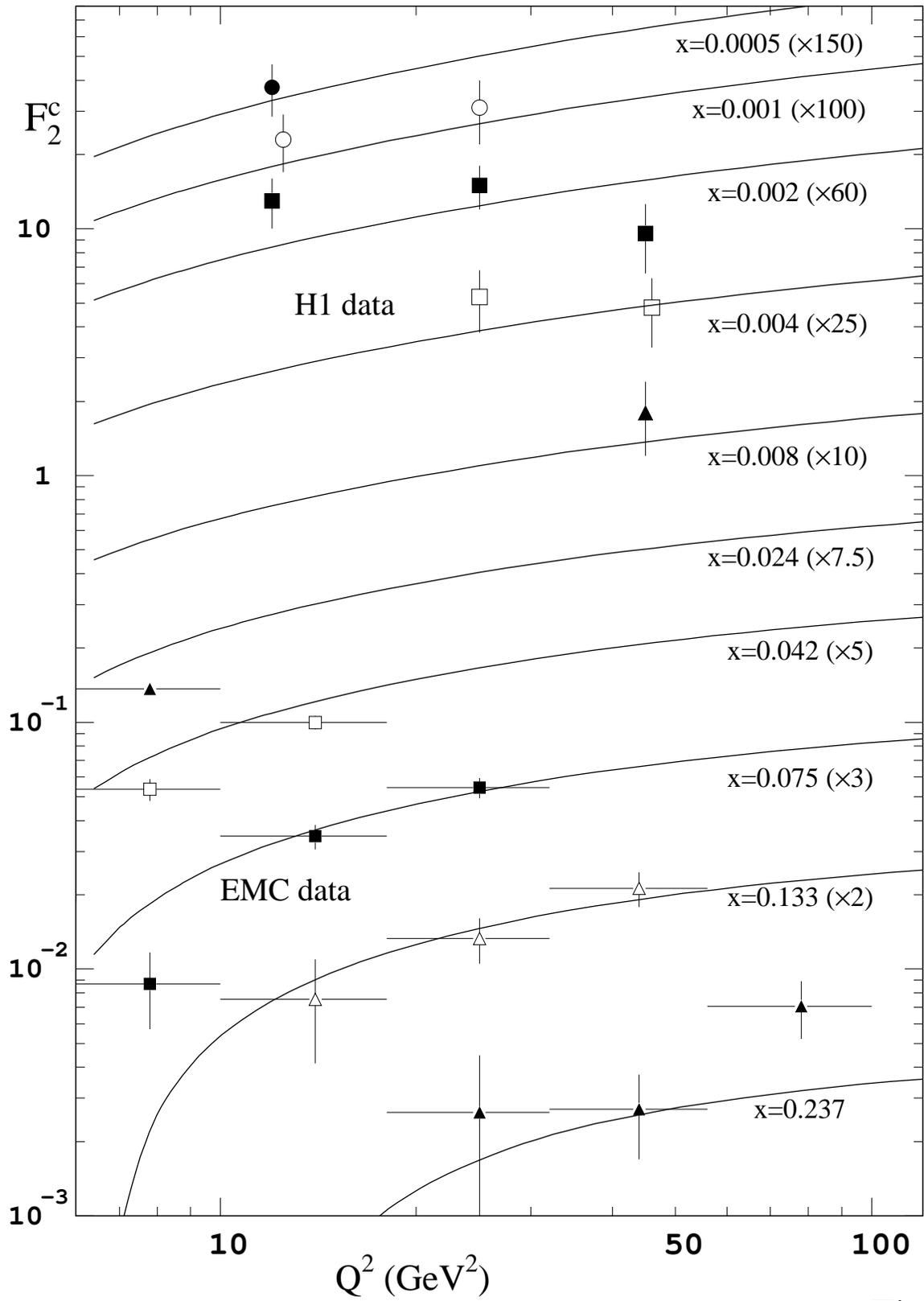
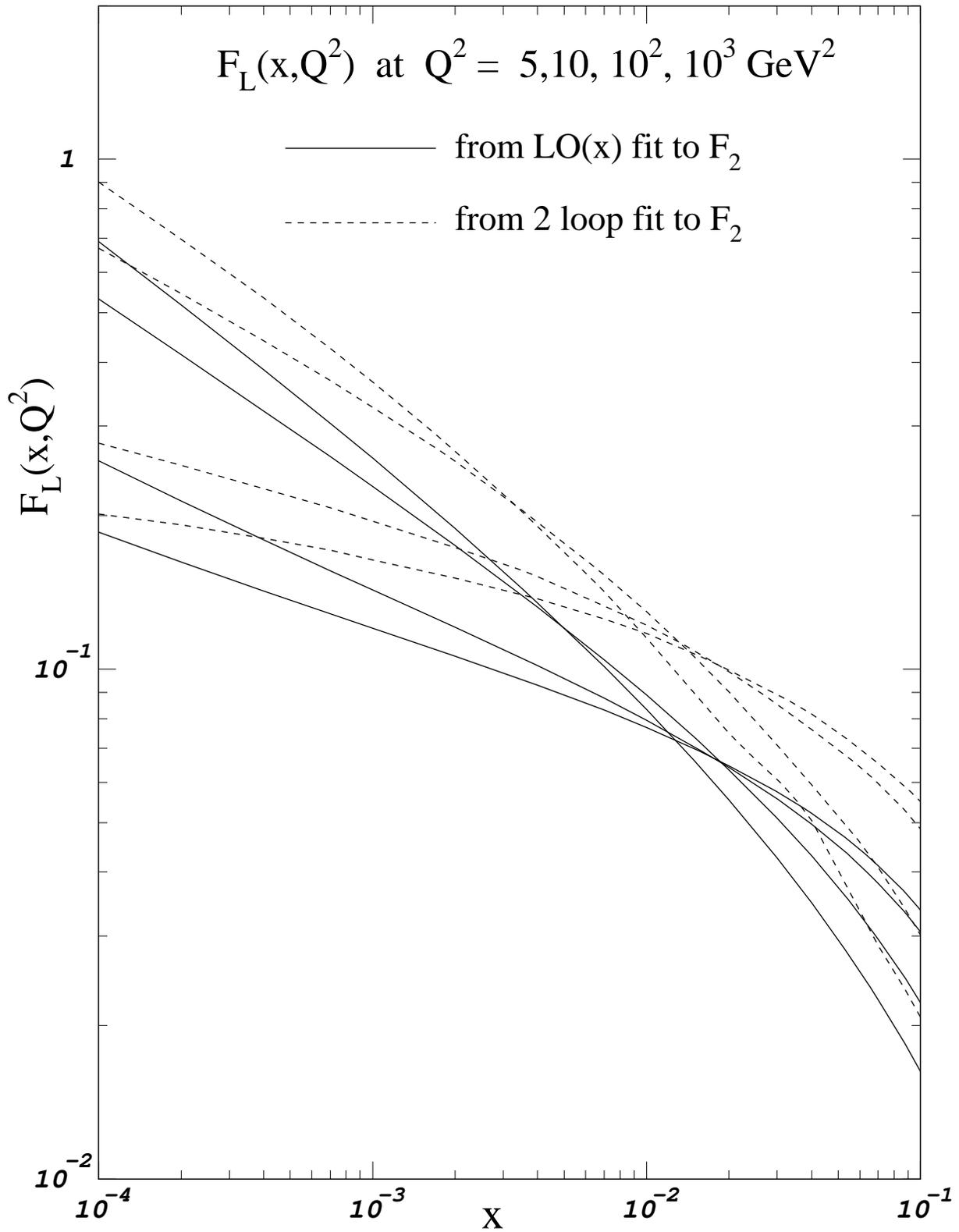
\end